\documentclass{aa}

\usepackage{graphicx}

\usepackage[varg]{txfonts}

\usepackage{natbib}
\bibpunct{(}{)}{;}{a}{}{,} 

\begin{document}

\title{How magnetic field and stellar radiative feedback influences the collapse and the stellar mass spectrum of a massive star forming clump}

\titlerunning{How magnetic field and stellar radiative feedback influences the stellar mass spectrum}

   \author{  Patrick Hennebelle \inst{\ref{inst1}} 
   \and Ugo Lebreuilly  \inst{\ref{inst1}} 
   \and Tine Colman  \inst{\ref{inst1}}    
   \and Davide Elia  \inst{\ref{inst2}}     
   \and Gary Fuller  \inst{\ref{inst3}, \ref{inst4}}      
   \and Silvia Leurini  \inst{\ref{inst5}}    
   \and Thomas Nony  \inst{\ref{inst6}}   
   \and Eugenio Schisano  \inst{\ref{inst2}}    
   \and Juan D. Soler  \inst{\ref{inst2}}  
   \and Alessio Traficante \inst{\ref{inst2}}   
   \and Ralf S. Klessen  \inst{\ref{inst7}}   
   \and Sergio Molinari \inst{\ref{inst2}} 
   \and Leonardo Testi  \inst{\ref{inst1}}     
   }

   \institute{ AIM, CEA, CNRS, Université Paris-Saclay, Université Paris Diderot, Sorbonne Paris Cité, F-91191 Gif-sur-Yvette, France,
      \label{inst1} 
      \and    
  INAF-IAPS, via del Fosso del Cavaliere 100, I-00133 Roma Italy,
      \label{inst2}
      \and
      Jodrell Bank Centre for Astrophysics
, Department of Physics and  Astronomy, The University of Manchester, Manchester M13 9PL, United Kingdom
\label{inst3}
\and
Physikalisches Institut, University of Cologne, Z\"ulpicher Str. 77, 50937 K\"oln, Germany
\label{inst4}
\and
INAF - Osservatorio Astronomico di Cagliari, Via della Scienza 5, I-09047 Selargius (CA), Italy
\label{inst5}
\and
Instituto de Radioastronomía y Astrofísica, Universidad Nacional Aut\'onoma de M\'exico, Apdo. Postal 3-72, 58089 Morelia, Michoac\'an, M\'exico
\label{inst6}
\and
Universit\"at Heidelberg, Zentrum f\"ur Astronomie, Institut f\"ur theoretische Astrophysik Albert-Ueberle-Str. 2, 69120, Heidelberg, Germany
\label{inst7}
      }

\abstract
{In spite of decades of theoretical efforts, 
the physical origin of the stellar initial mass function (IMF) is still debated. }
{ We aim at understanding the influence of various physical processes such as radiative stellar feedback,
magnetic field and non-ideal magneto-hydrodynamics on the IMF. }
{We present a series of numerical simulations of collapsing 1000 M$_\odot$ clumps
taking into account radiative feedback and magnetic field with spatial resolution down to 1 AU. Both  ideal
and non-ideal MHD runs are performed and various radiative feedback efficiencies are considered. 
We also develop analytical models that we confront to the numerical results.  }
{The sum of the luminosities produced by the stars in the calculations  is computed and it compares well 
with the bolometric luminosities reported in observations of massive star forming clumps.
The temperatures, velocities and densities are also found to be in good agreement with recent observations.
The stellar mass spectrum inferred for the simulations is, generally speaking, not strictly universal and in particular  
varies  with magnetic intensity. It is also influenced by the choice of the radiative feedback efficiency. 
In all simulations, a sharp drop in the stellar distribution is found at about $M_{min} \simeq$ 0.1 M$_\odot$, which 
is likely a consequence of the adiabatic behaviour induced by dust opacities at high densities. As a consequence, when 
the combination of magnetic and thermal support is not too large, the mass distribution presents a
 peak located at 0.3-0.5 M$_\odot$. When magnetic and thermal support are large, the mass distribution 
 is better described by a plateau, i.e.  $d N / d \log M \propto M^{-\Gamma}$, $\Gamma \simeq 0$. At higher masses the mass 
 distributions drop following  power-law behaviours until a maximum mass $M_{max}$ whose value 
 increases with field intensity and radiative feedback efficiency. 
 Between $M_{min}$ and $M_{max}$ the 
 distributions inferred from the simulations agree well with an analytical model inferred from gravo-turbulent
theory. Due to the density PDF $\propto \rho^{-3/2}$ relevant for collapsing clouds, values on the order 
of $\Gamma \simeq 3/4$ are inferred both analytically and numerically. 
 More precisely,
 after 150 $M_\odot$ of gas have been accreted, the most massive star has a mass of about 8 
M$_\odot$ when magnetic field is significant, and 3 M$_\odot$ only when both radiative feedback efficiency and magnetic field 
are low, respectively.  }   
{When both magnetic field and radiative feedback are taken into account, they are found to have a significant influence on the stellar mass spectrum. 
In particular  both reduce fragmentation and lead to the  formation of more massive stars.}

   \maketitle


\section{Introduction}

Star formation is a topic of fundamental importance in astrophysics.
In particular the mass distribution of stars, described by the initial mass function \citep[IMF][]{salpeter55,kroupa2001,chabrier2003,bastian2010,offner2014,lee2020} , plays a crucial role in setting the abundances of heavy elements and regulating stellar feedback, which in turn play major roles in the formation and evolution of galaxies and the interstellar medium. In the efforts to find a complete description of the IMF, it is sometimes overlooked that observed stellar masses span more than three orders of magnitudes: from 0.1 M$_\odot$ to more than 100 M$_\odot$. This facts likely implies the existence of several regimes of dominant physical processes and star formation conditions.
Clearly this problem requires a long standing community effort and during the last decades 
several teams have conducted systematic investigations with 
the help of numerical simulations,  introducing progressively 
more and more physical processes with increasingly higher numerical resolution. 

The first attempts to obtain stellar mass spectra from numerical simulations
in isothermal, self-gravitating, supersonic turbulent flows have been made 
by \citet{klessen2001} and \citet{Bate03}. Together with several
high-resolution  studies performed by various authors
 \citep[e.g.][]{Girichidis11,Bonnell11,BallesterosParedes15,leeh2018a}, 
 they find stellar mass spectra that present similarities with the 
 observationally inferred mass spectra. In particular, at high masses the distributions are compatible with powerlaws, i.e. 
 $dN / dlog M \propto M^{-\Gamma}$, although in many runs,  values of 
 $\Gamma=3/4$ to 1, seemingly 
 shallower than the canonical $\Gamma \simeq$1.3 value inferred by \citet{salpeter55} have been obtained \citep[see the discussion in][]{leeh2018a}. The inferred distributions also present a peak, which however, 
 when the simulations are strictly isothermal, is due to limited spatial resolution. A robust, numerically converged peak is obtained when an 
 effective equation of state  with an adiabatic index larger than $4/3$ is 
 taken into account \citep{leeh2018b}. 
 
 The influence of the magnetic field on the stellar mass spectrum
  has  been investigated 
 by \citet{haugbolle2018}, \citet{lee2019}, \citet{guszejnov2020}
 perfoming high spatial resolution simulations with various magnetisations. 
 The resulting mass 
 spectra have been found to be similar to those inferred from 
  simulations without magnetic field. In particular \citet{guszejnov2020} 
 stress that magnetic field cannot provide a characteristic mass 
 that may explain the peak of the IMF and that thermal processes have to be 
 considered.

 Several attempts have been made to study the IMF using  radiative transfer calculations. 
\citet{urban2010}  considered radiative feedback, i.e. stellar and accretion luminosity,
 by adding them onto the sink particles. They concluded that isothermal and radiative transfer calculations 
are significantly different, in particular the stars are much more massive in simulations with radiative feedback. 
\citet{bate2009} performed high resolution calculations, introducing the sink particles 
at very high density the released gravitational energy, i.e. 
$n > 10^{19}$ cm$^{-3}$ but stellar feedback  onto the sink particles is not explicitly included, which makes it much
weaker than it should. 
\citet{krumholz2012} performed adaptive mesh refinement calculations with a resolution of 20-40 AU. 
Both stellar and accretion luminosity are  added to the sinks. 
A relatively flat mass spectrum  that is to say such that $\Gamma \simeq 0$ is inferred
 when winds are not considered while 
in the presence of stellar winds the mass spectra present a peak around 0.3 $M _\odot$
and a power-law with $\Gamma \simeq 0.5-1$. 
Likely enough when winds are present, the radiation  escape along the cavities and the heating 
is reduced. 
 \citet{mathew2020} presented
simulations with a spatial resolution of 200 AU and perform calculations which use either a  polytropic equation 
of state or heating from stars.  They found that 
 when heating is included more massive stars would form. 
 \citet{hetal2020} conducted adaptive mesh simulations with a spatial resolution of 4 AU and down to 1 AU. 
 Both stellar and accretion luminosity are treated, with various efficiencies, $f_{acc}$,
  ranging from 0 to 50$\%$, as well as
 two sets of initial conditions, namely a very compact and  more standard clumps have been
 considered. For the most compact clumps and when $f_{acc}$ is high, a flat mass 
 spectrum develops. Otherwise all runs present mass spectra with a peak around 0.3-0.5 M$_\odot$ and a powerlaw 
 at higher masses, even when radiative feedback is not considered, i.e. $f_{acc}=0$, and  when a barotropic equation 
 of state is used instead. High efficiency radiative feedback runs however tend to present a broader distribution, both at the low mass and high mass end, with high mass stars up to 2 - 3 times more massive than in the barotropic and low feedback efficiency runs.

In the present paper we pursue the investigation of the origin of 
the stellar mass spectrum within a massive star forming clump. 
In particular, we focus on the role that magnetic field, in conjunction 
with radiative feedback may have. 
A number of studies performed calculations with both magnetic field and  radiative feedback although most of the time, without predicting
mass spectrum. As revealed by previous work 
\citep{peters2010,peters2011,commercon2011,myers2013} both these physical processes significantly influence the collapse and star formation, particularly by reducing the fragmentation. Moreover, their joint effect is not a mere superposition. These studies, however, did not present sufficient statistics to draw conclusions regarding the stellar mass spectrum.
A stellar mass spectrum has been obtained by \citet{lip2018} 
where a magnetized and radiative calculation is performed with a spatial 
resolution of about 30 AU. These statistics 
need to be expanded and various initial conditions must be systematically explored. 
To do so we perform high resolution simulations 
of  massive star forming clumps where both magnetic field and radiative 
feedback are accounted for. To get a good description of the small scales
which are mandatory  to describe the formation of low mass stars, 
we employ an adaptive mesh reffinement with a spatial resolution  
down to 1 AU. 
As the magnetic intensity is likely varying from clump to
clump and not many constraints from observations are available yet,
we explore three magnetisations.
Also the radiative feedback efficiency is subject to large uncertainties, so we
consider two different values.
 Importantly, we also perform a simulation in which non-ideal MHD effects, namely ambipolar diffusion \citep{mestelANDspitzer1956}, are explicitly 
taken into account.
We stress that these runs are the first for which both magnetic 
field and radiative feedback are taken into account, while considering a configuration which leads to sufficient statistics 
and spatial resolution to provide a reliable stellar distribution in the range 0.1 to 10 M$_\odot$.

The paper is structured as follows.
The second section presents the equations that are being solved, the relevant physical processes as well as the numerical methods used to solve these equations. It also presents the initial conditions and describes the various runs presented in the paper. In the third section,
we look at the evolution of the clump during its collapse and investigate the effect of the magnetisation and radiative feedback.
The global properties such as the total accreted mass and radiated 
energy, the temperature, magnetic field and mass distribution are studied. An analytical model, which is presented in an appendix is developed 
to understand the temperature distribution in the simulations.  Comparisons are made with observational results. The fourth section presents the stellar mass spectrum obtained in the simulations. They are quantitatively compared with an analytical model which gives more insight into the effect of the different physical processes and is also presented in an appendix.
In the fifth section a discussion is given while 
the sixth section concludes the paper.

\setlength{\unitlength}{1cm}
\begin{figure*}
\begin{picture} (0,12)
\put(0,6){\includegraphics[width=8cm]{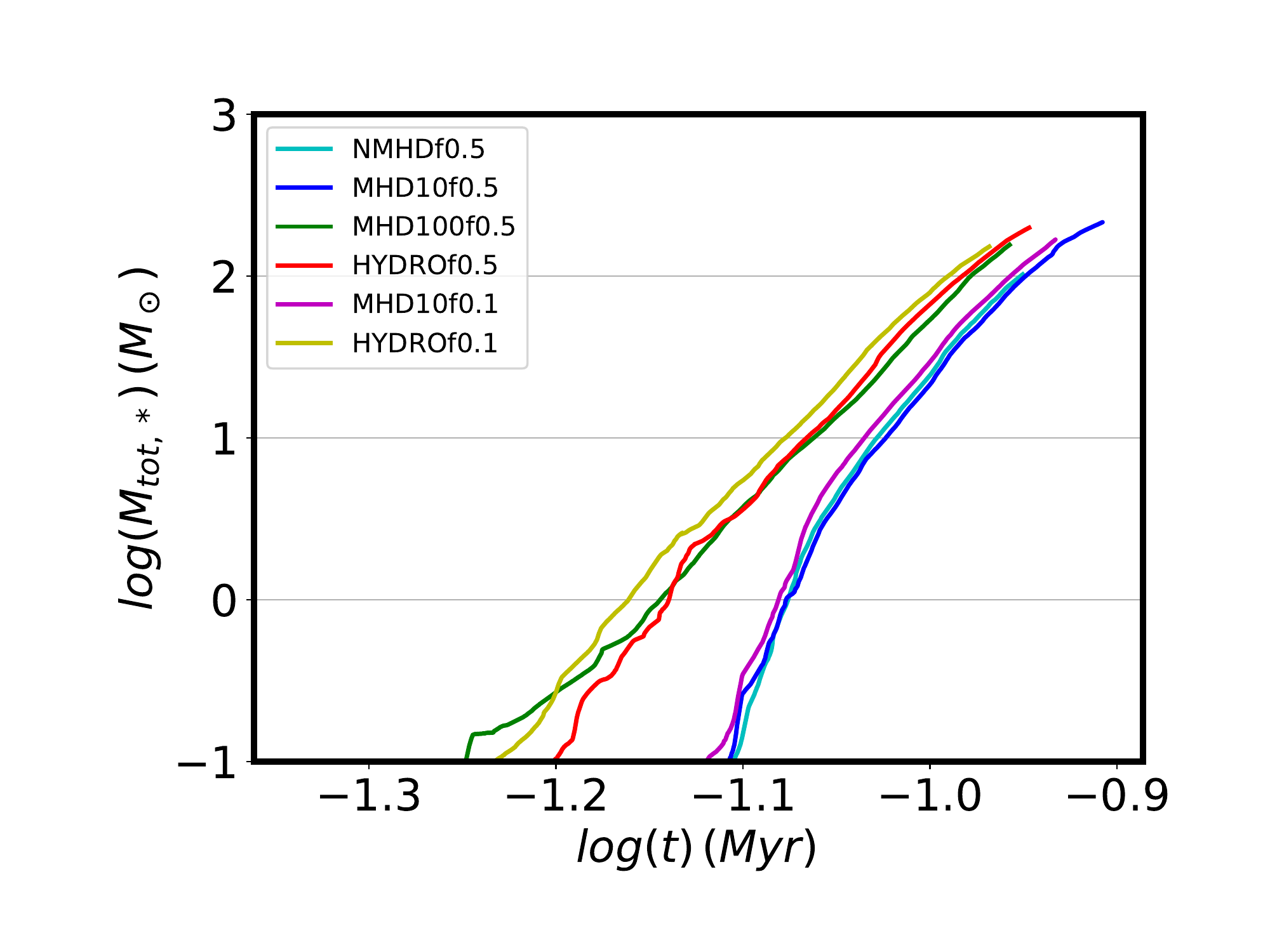}}  
\put(9,6){\includegraphics[width=8cm]{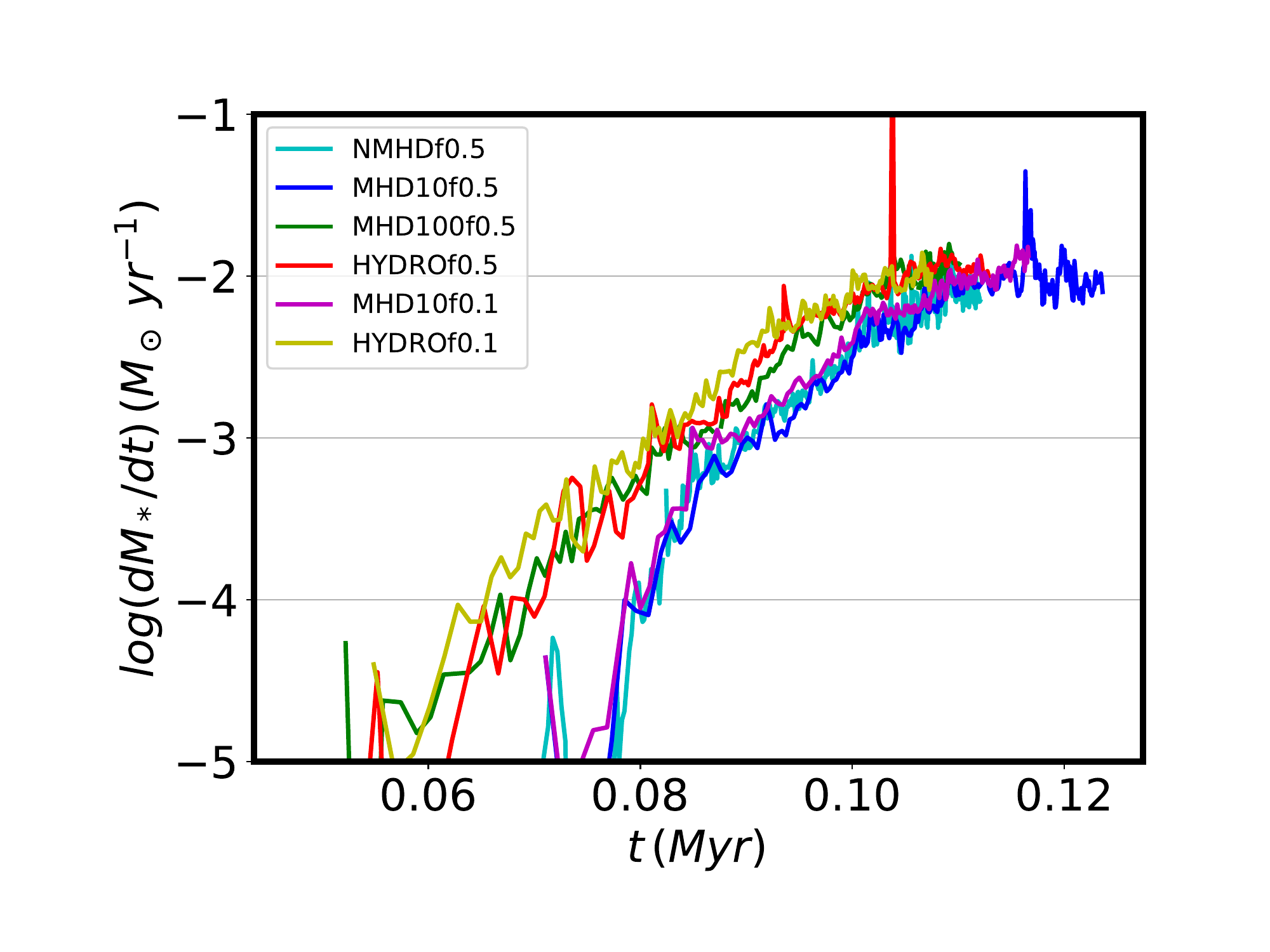}}  
\put(0,0){\includegraphics[width=8cm]{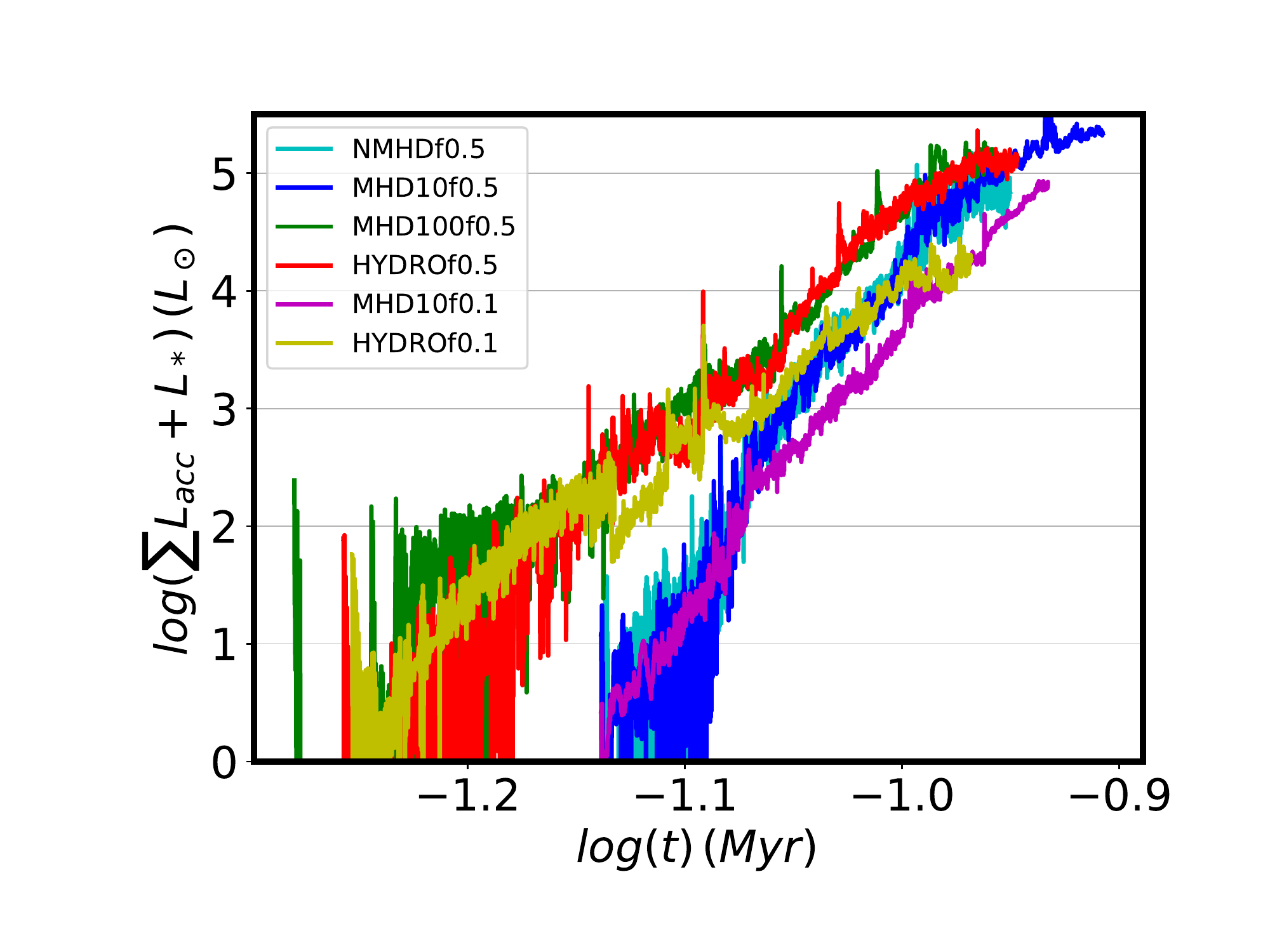}}  
\put(9,0){\includegraphics[width=8cm]{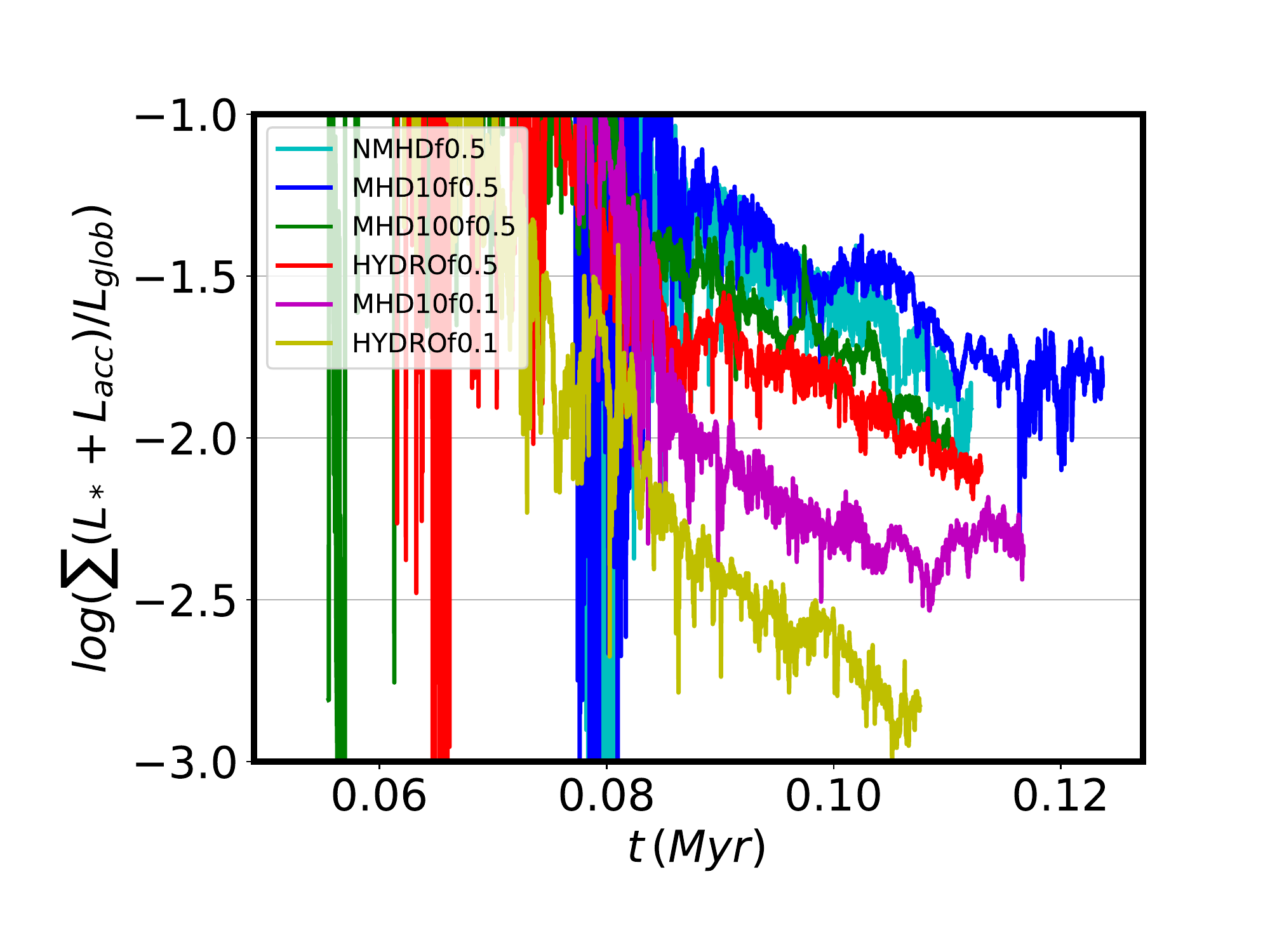}}  
\end{picture}
\caption{Top row shows the total accreted mass (top-left panel) and the total accretion rate (top-right panel) as a function of time for the various runs. Bottom-left panel portrays the total luminosity while bottom-right panel shows the  same quantity divided by $L_{glob} = 0.5 \times G M_{*,tot} \dot{M} _{*,tot} / (2 R_\odot)  $. 
}
\label{time_mass}
\end{figure*}

\setlength{\unitlength}{1cm}
\begin{figure*}
\begin{picture} (0,21)
\put(0,0){\includegraphics[width=15cm]{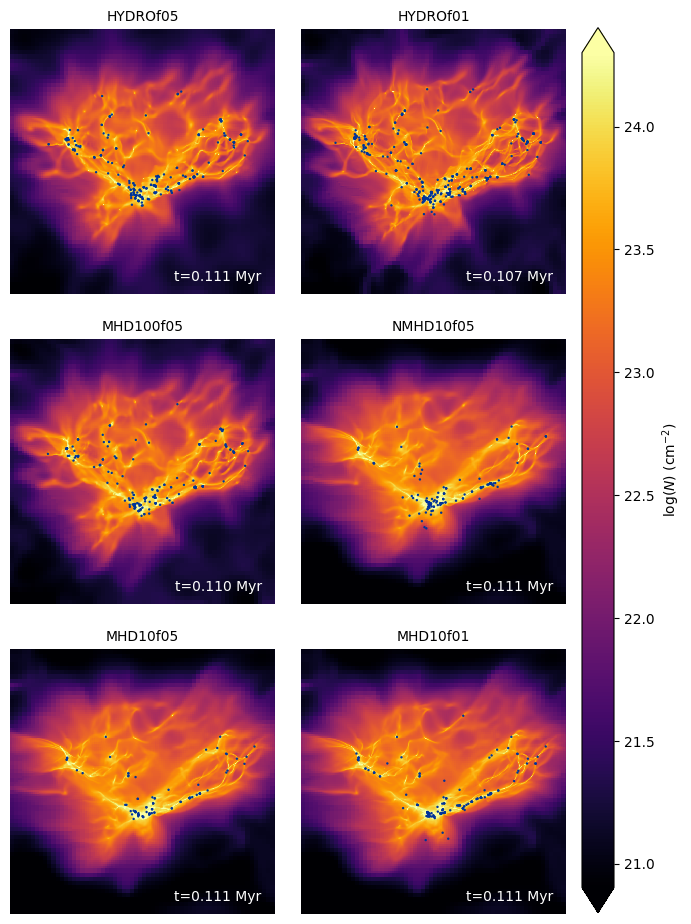}}
\end{picture}
\caption{Column density at $t=0.11$ Myr
for the six runs. The dark circles represent the sink particles aiming
at describing the stars.
}
\label{fig_coldens}
\end{figure*}


\setlength{\unitlength}{1cm}
\begin{figure*}
\begin{picture} (0,23)
\put(8,15.6){\includegraphics[width=7.5cm]{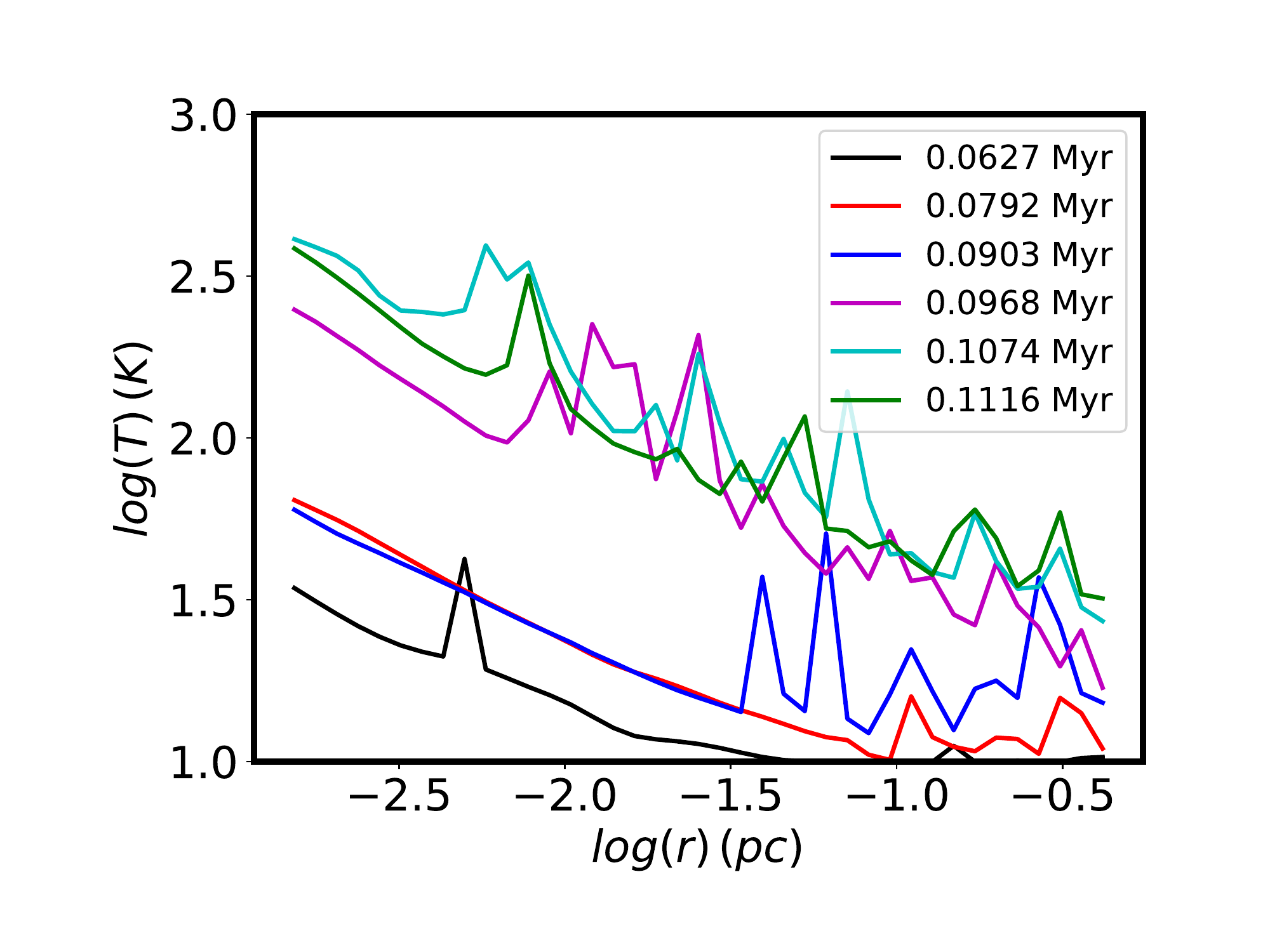}}
\put(10,20.7){HYDROf05}
\put(0,15.6){\includegraphics[width=7.5cm]{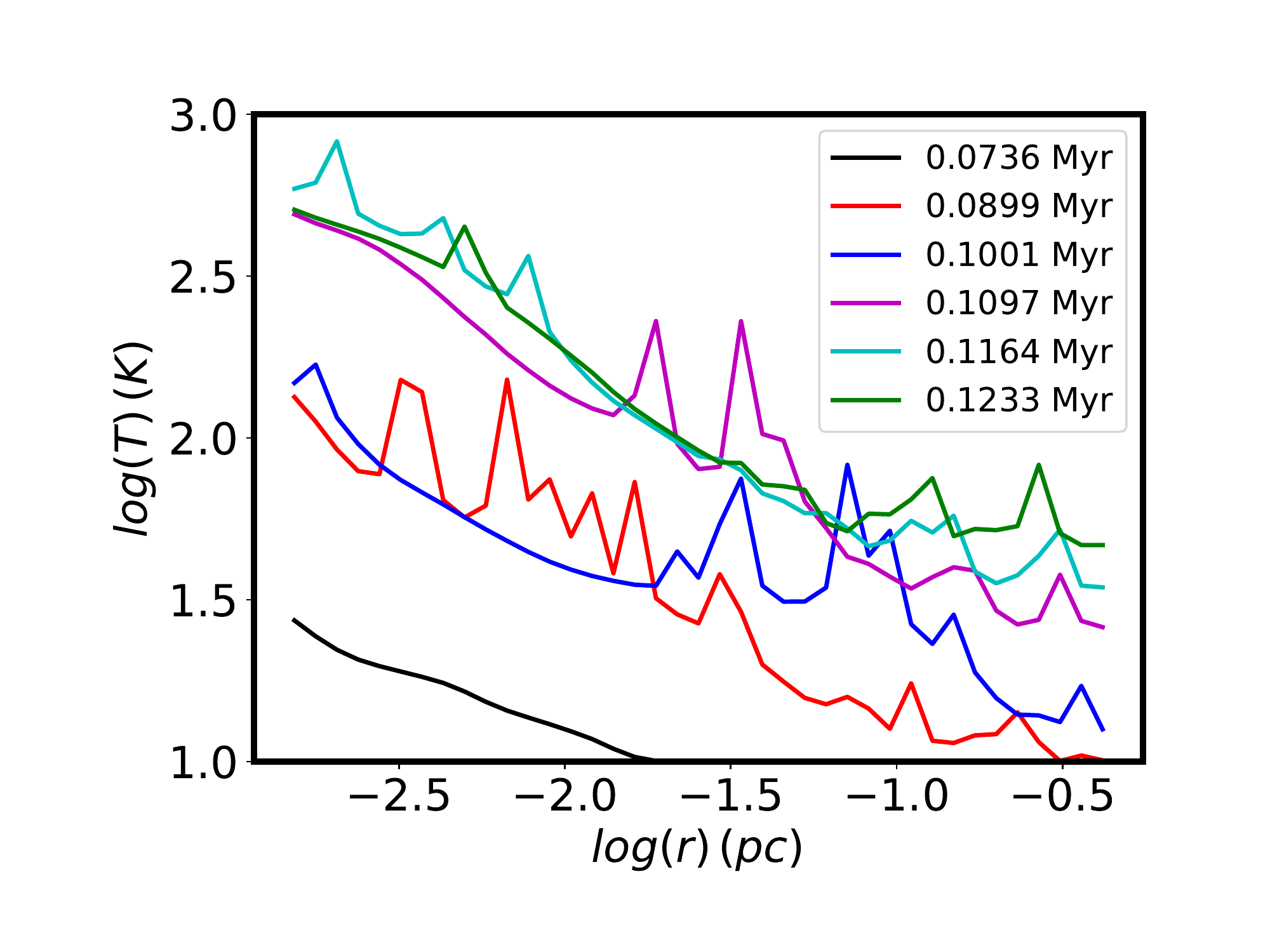}}  
\put(2,20.7){MHD10f05}
\put(8,10.4){\includegraphics[width=7.5cm]{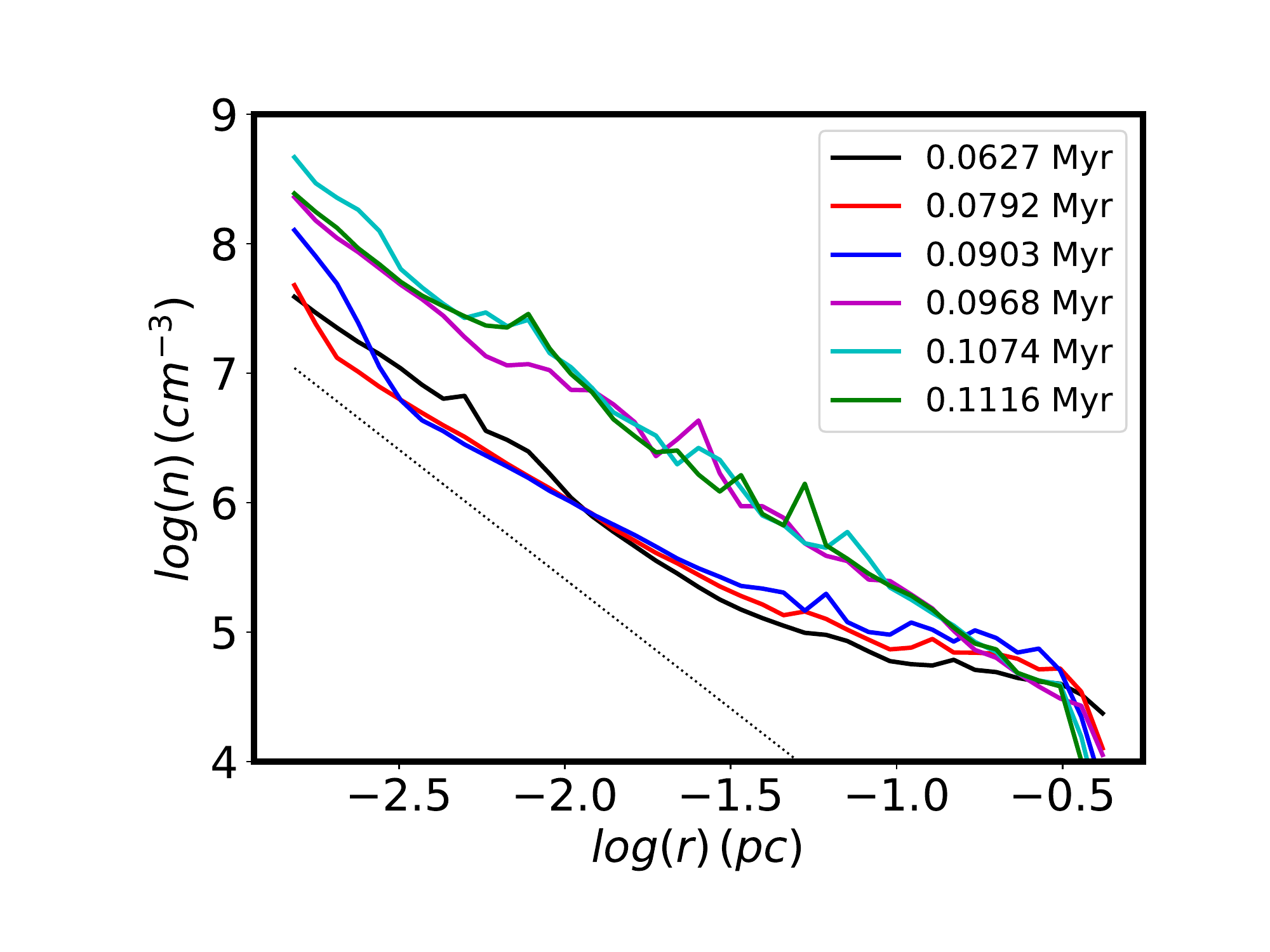}}
\put(10,15.5){HYDROf05}
\put(0,10.4){\includegraphics[width=7.5cm]{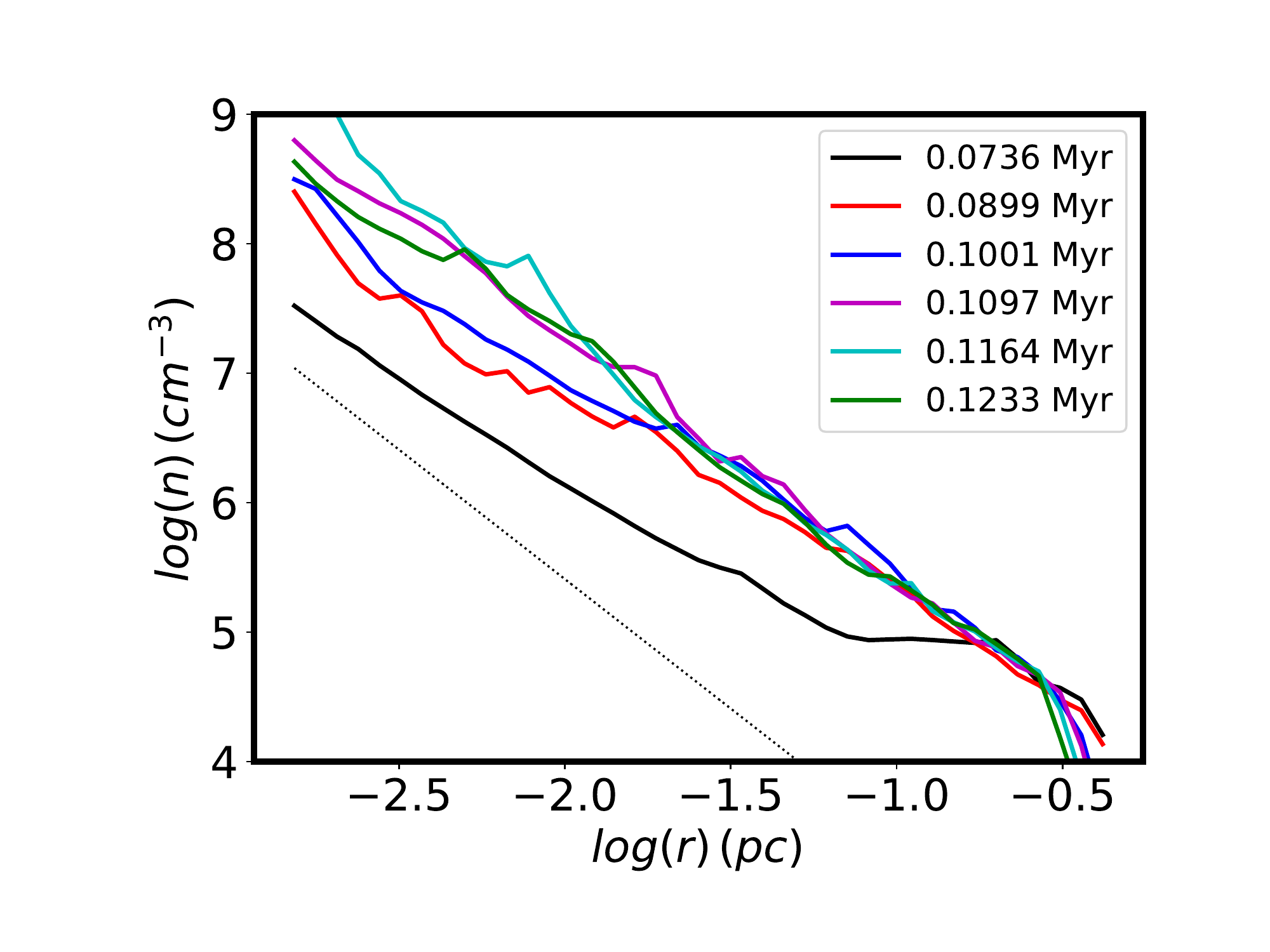}}  
\put(2,15.5){MHD10f05}
\put(8,5.2){\includegraphics[width=7.5cm]{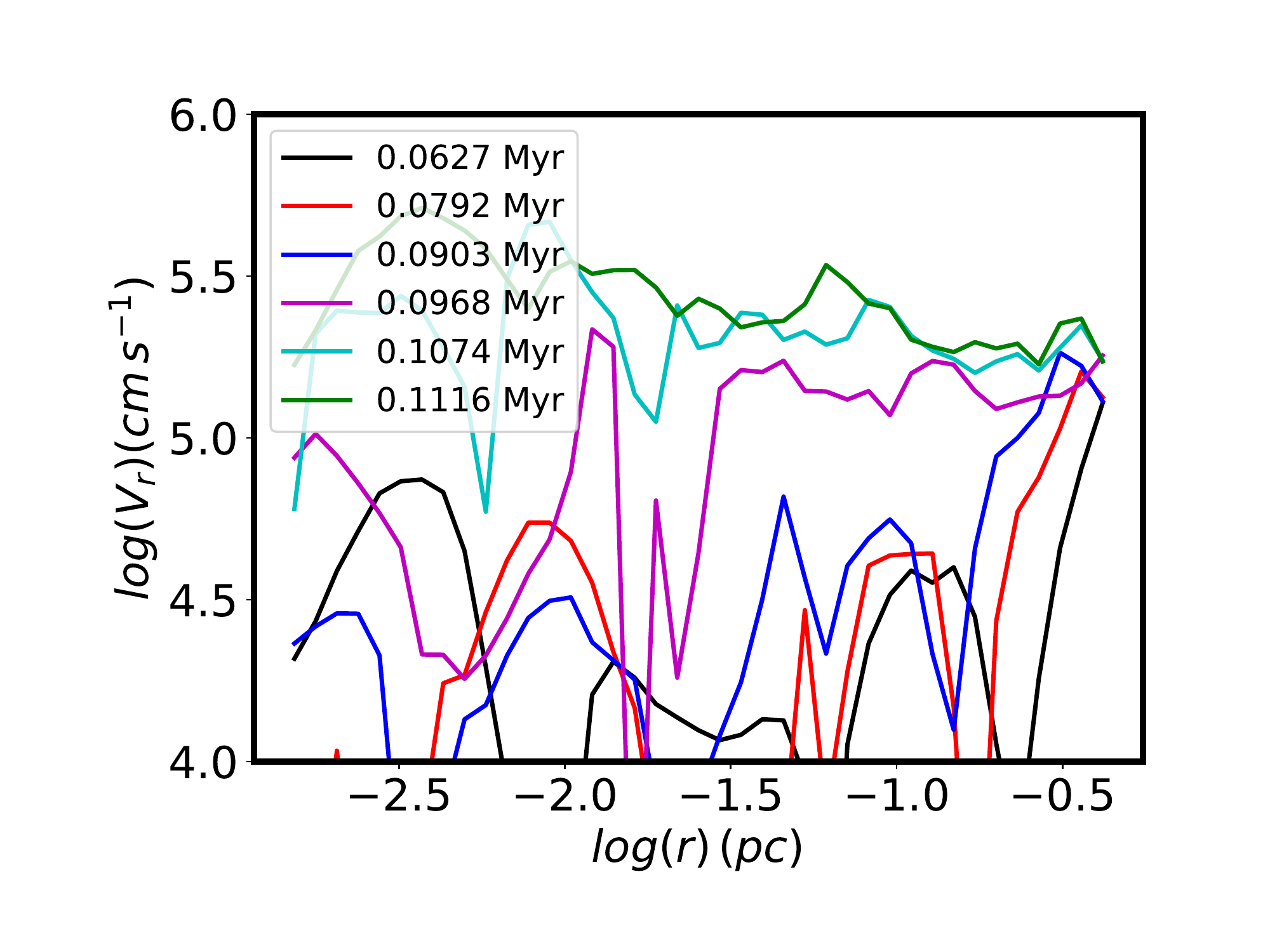}}
\put(10,10.3){HYDROf05}
\put(0,5.2){\includegraphics[width=7.5cm]{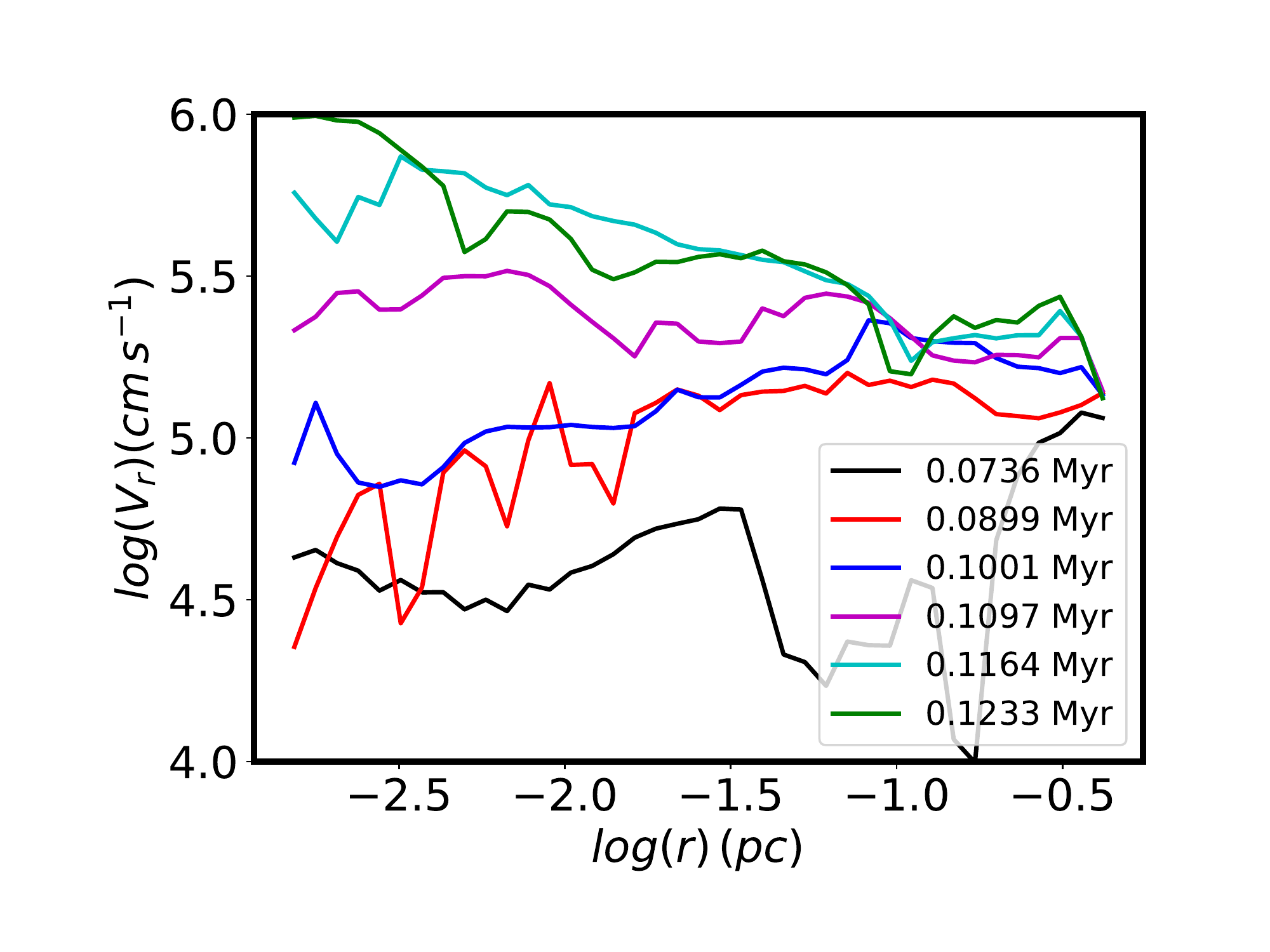}}  
\put(2,10.3){MHD10f05}
\put(8,0){\includegraphics[width=7.5cm]{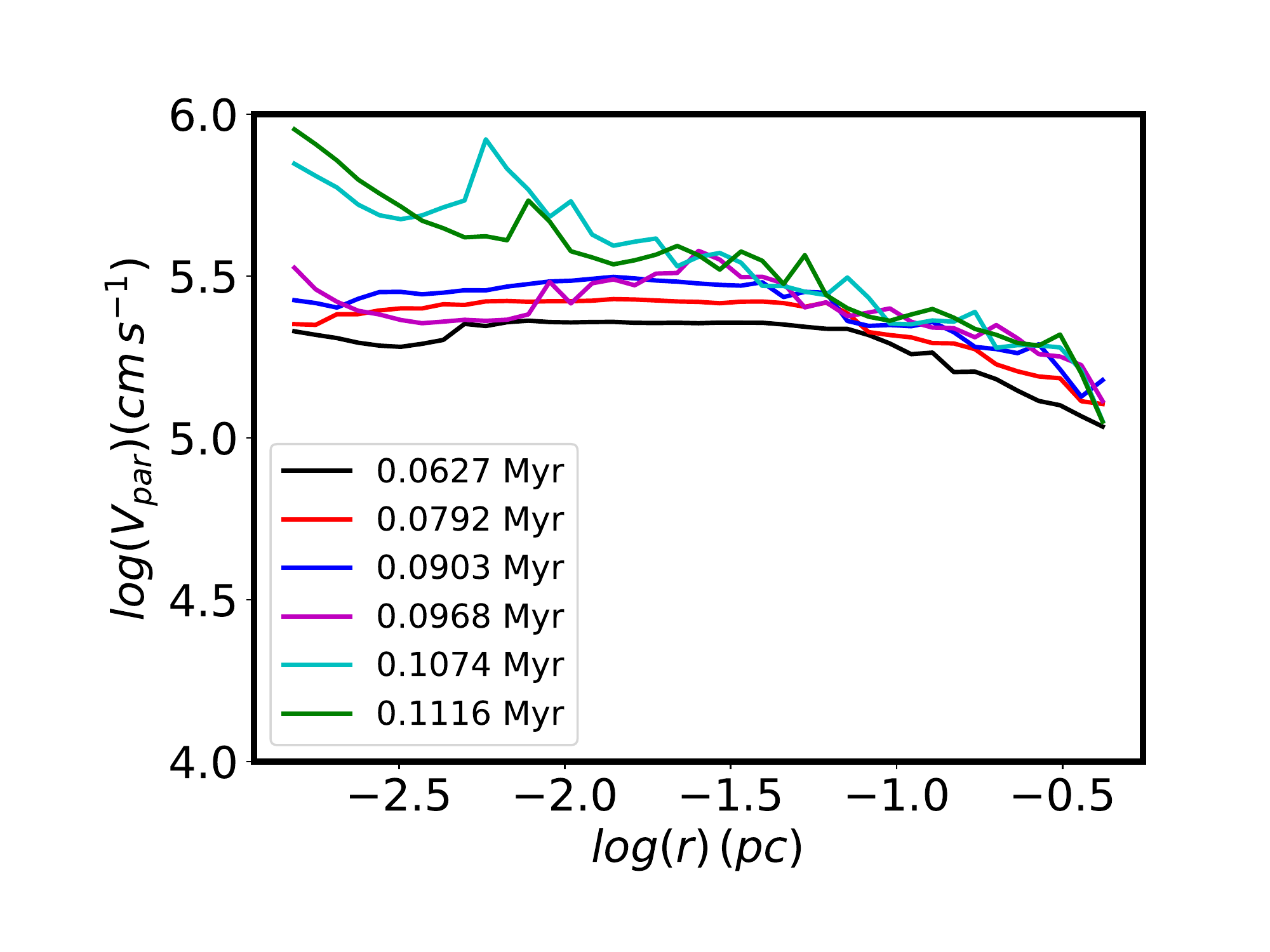}}  
\put(10,5.1){HYDROf05}
\put(0,0){\includegraphics[width=7.5cm]{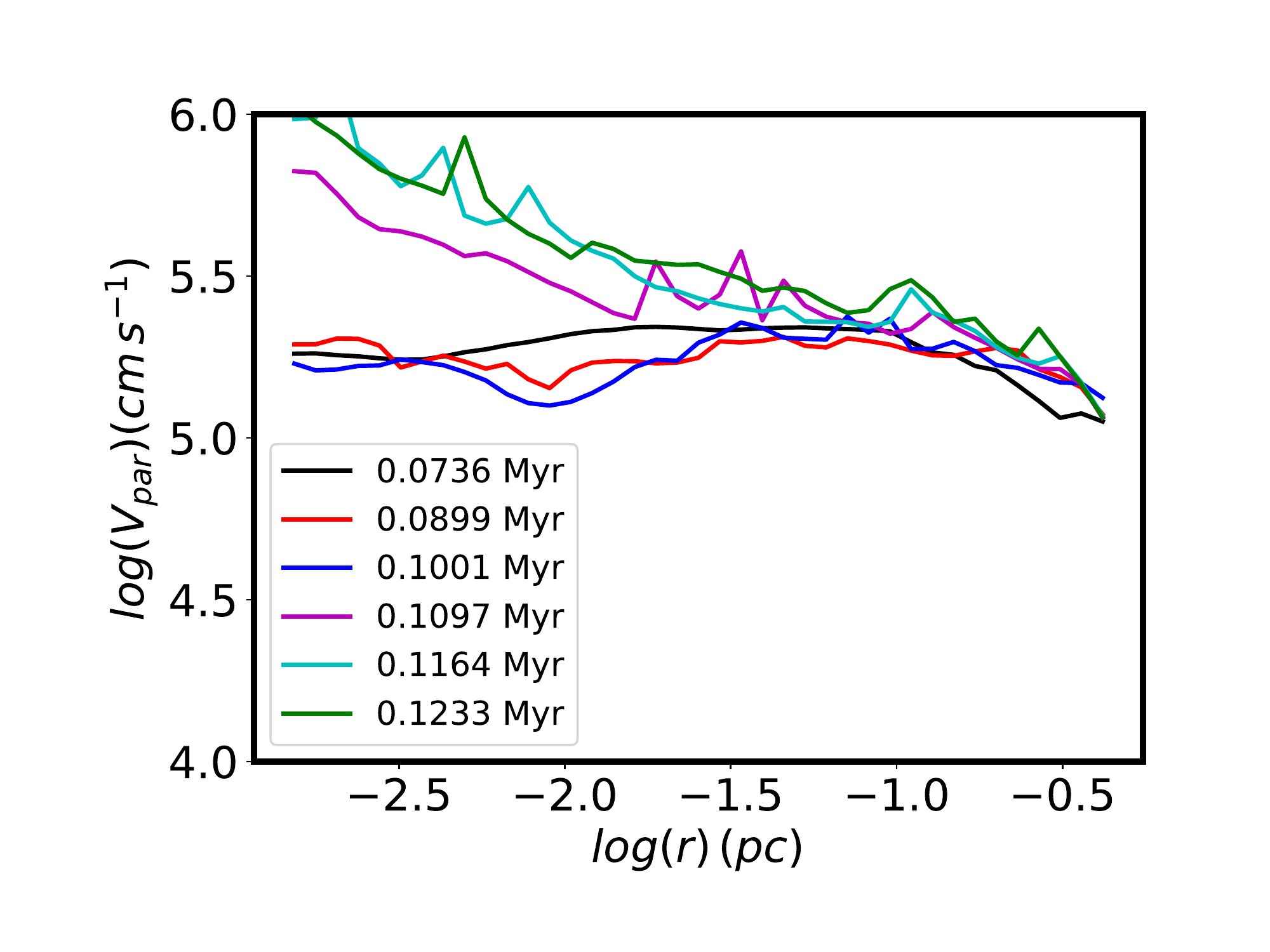}}  
\put(2,5.1){MHD10f05}
\end{picture}
\caption{Radial profiles at several timesteps in HYDROf05 and MHD10f05, respectively.
The black dashed line in the density panel shows the singular isothermal sphere density.}
\label{fig_radial}
\end{figure*}
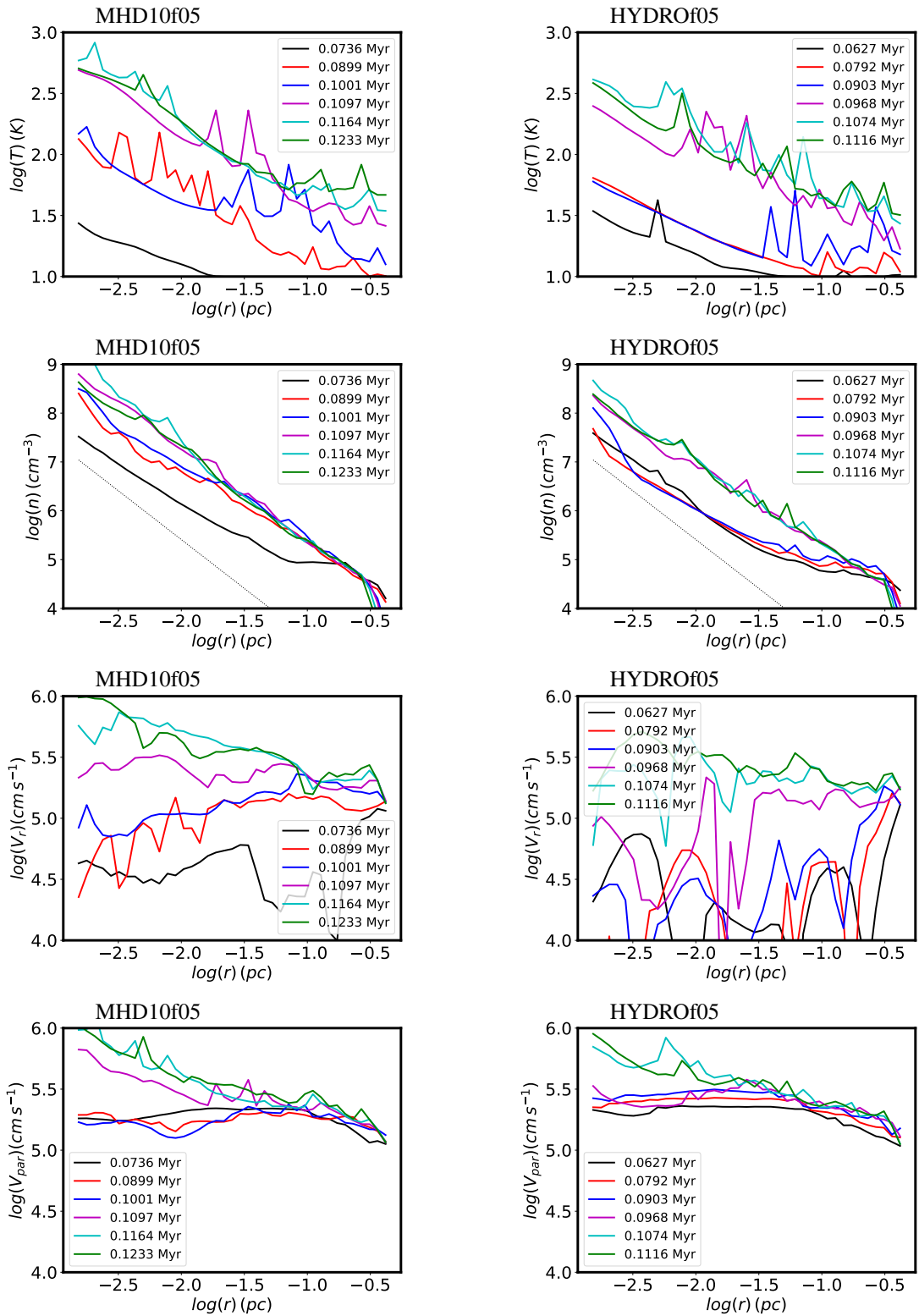

\setlength{\unitlength}{1cm}
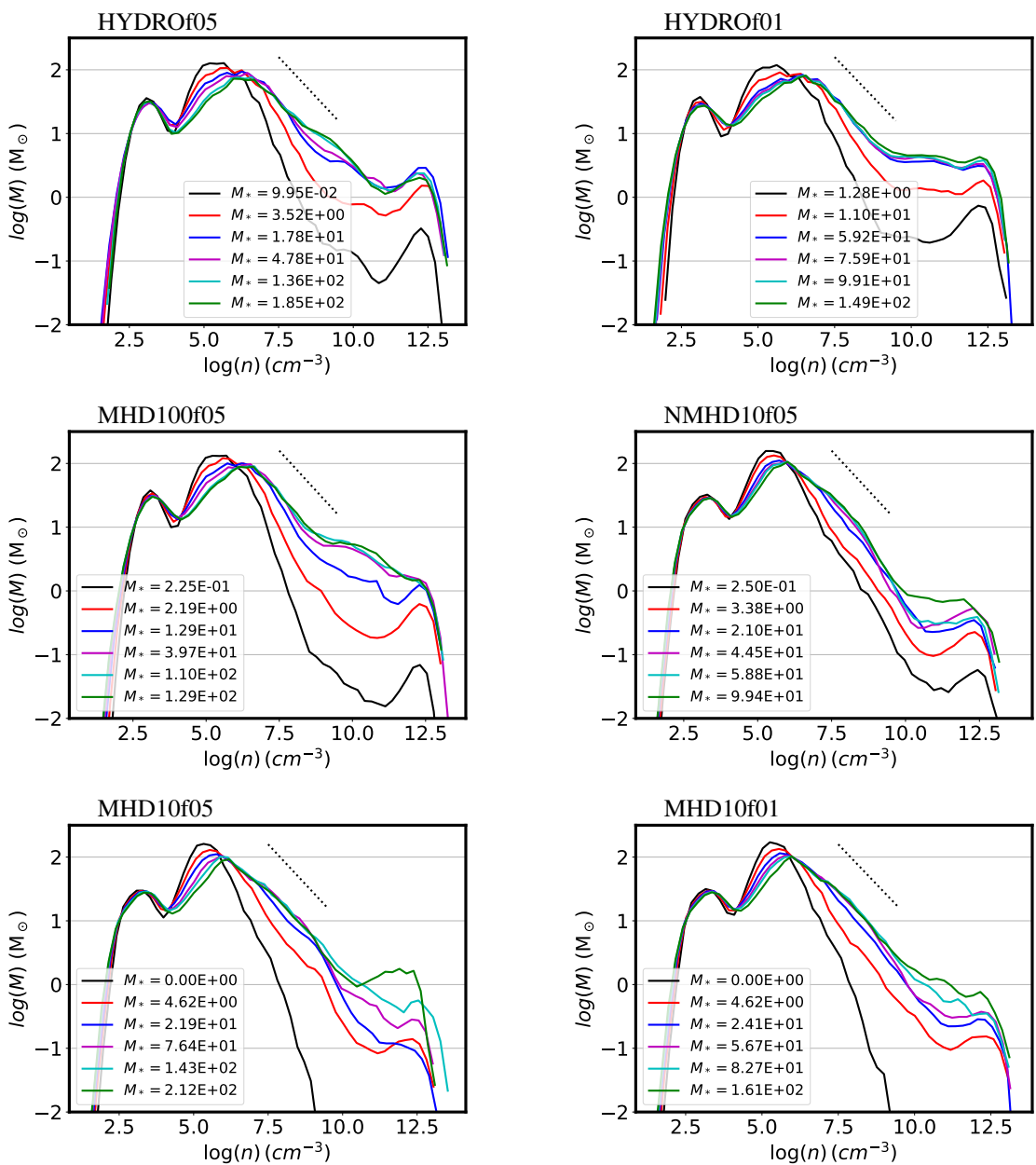
\begin{figure*}
\begin{picture} (0,17)
\put(8,11.2){\includegraphics[width=8cm]{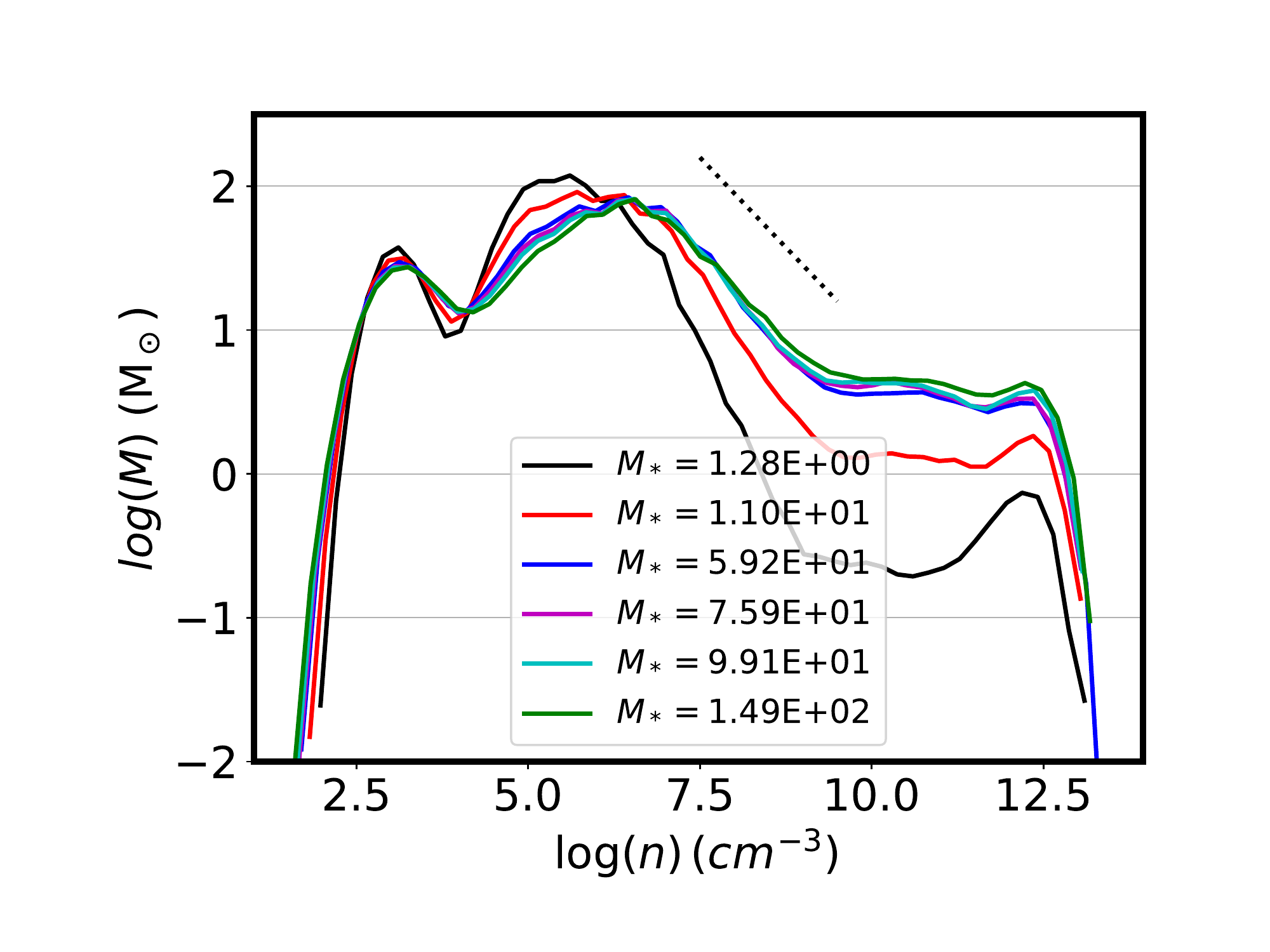}}
\put(10,16.6){HYDROf01}
\put(0,11.2){\includegraphics[width=8cm]{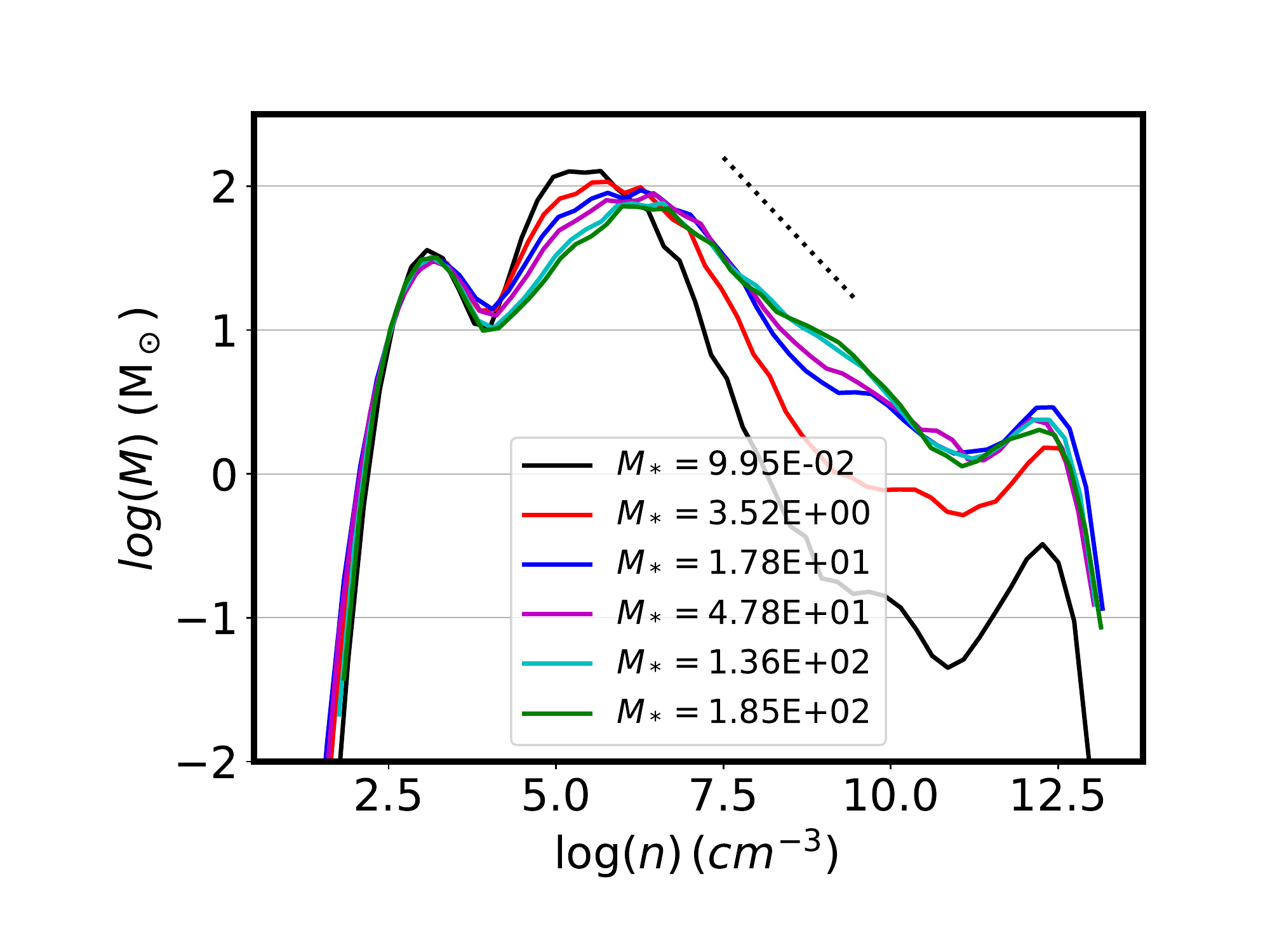}}  
\put(2,16.6){HYDROf05}
\put(8,5.6){\includegraphics[width=8cm]{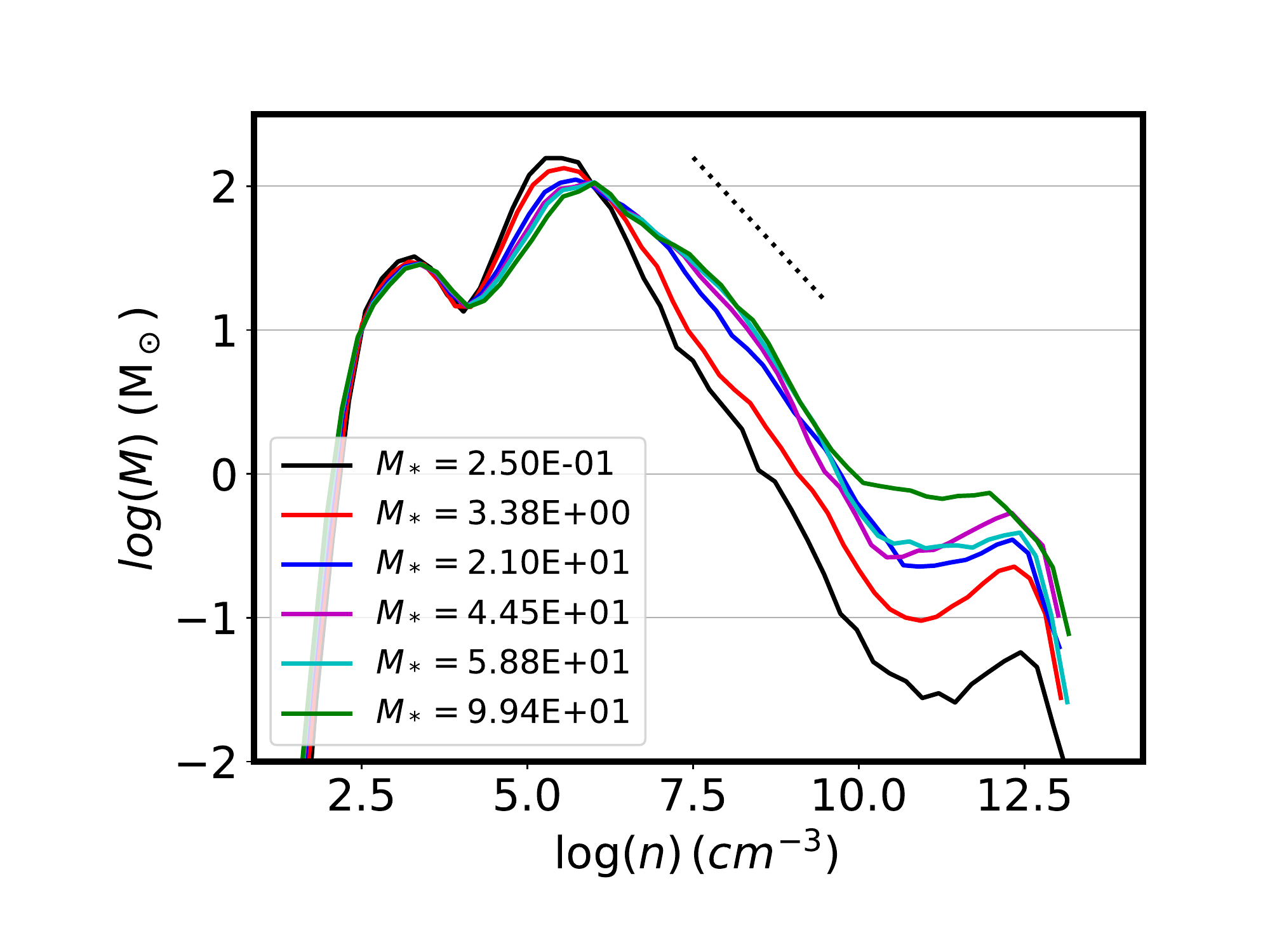}}
\put(10,11){NMHD10f05}
\put(0,5.6){\includegraphics[width=8cm]{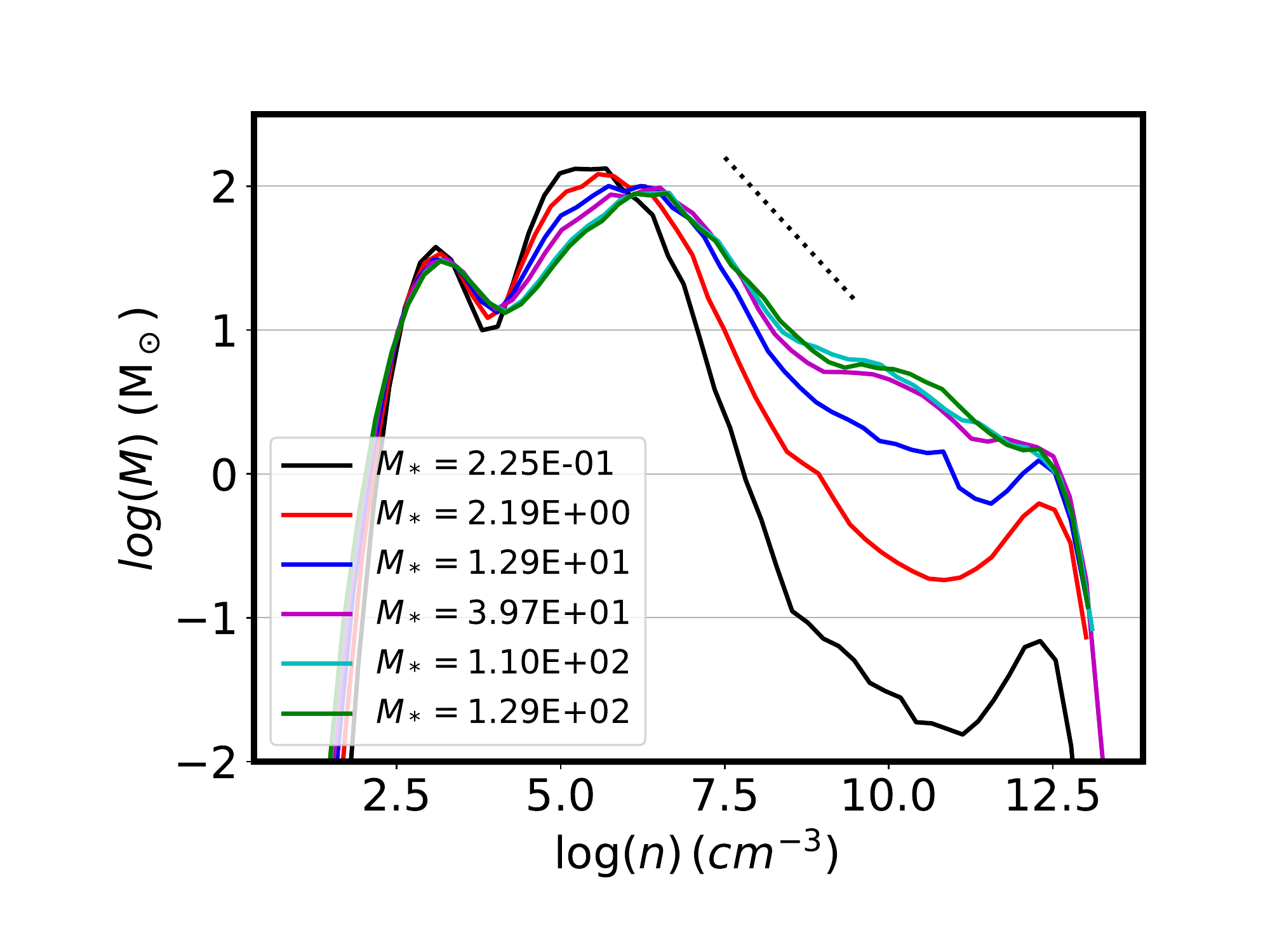}}  
\put(2,11){MHD100f05}
\put(8,0){\includegraphics[width=8cm]{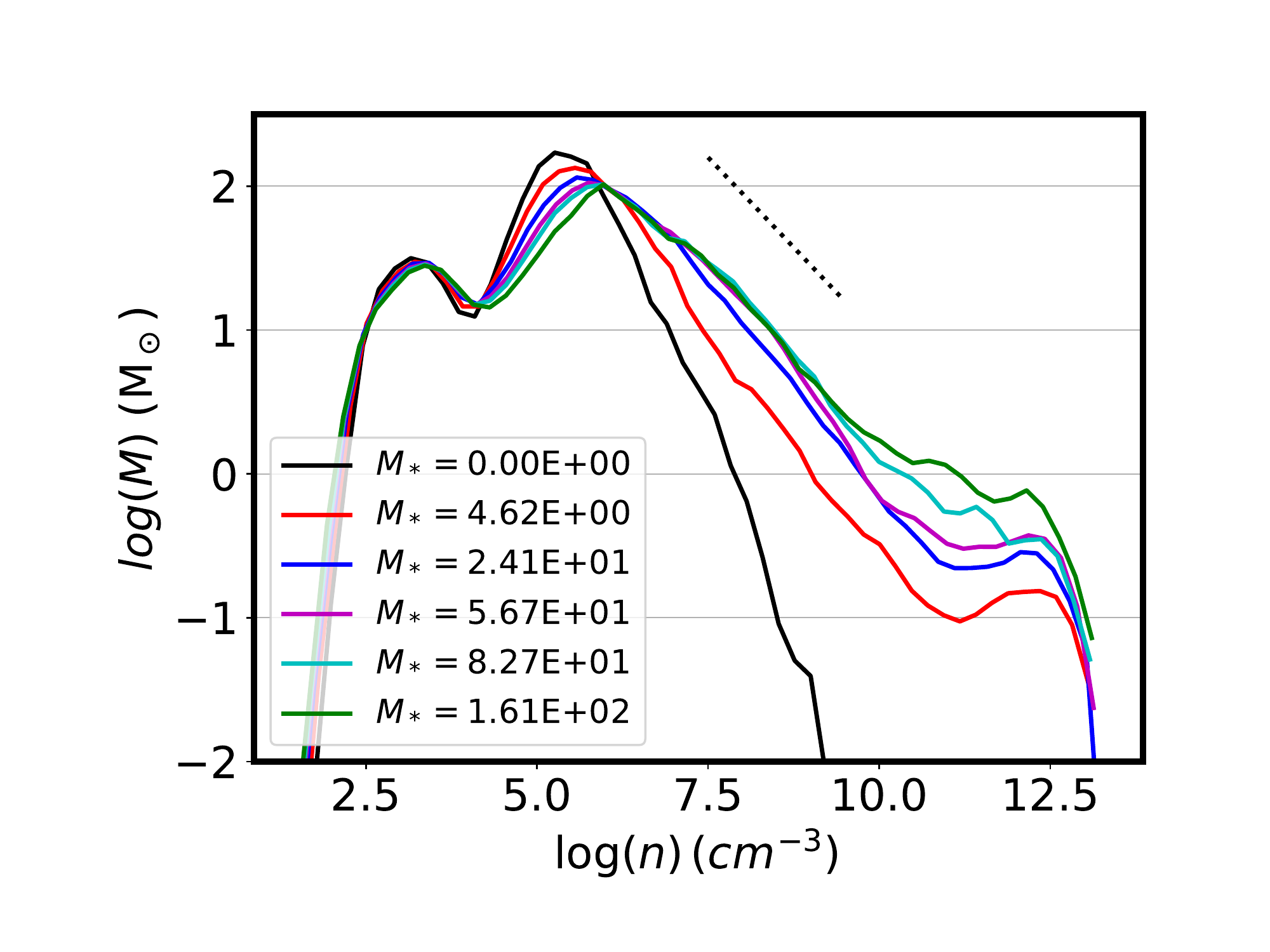}}  
\put(10,5.4){MHD10f01}
\put(0,0){\includegraphics[width=8cm]{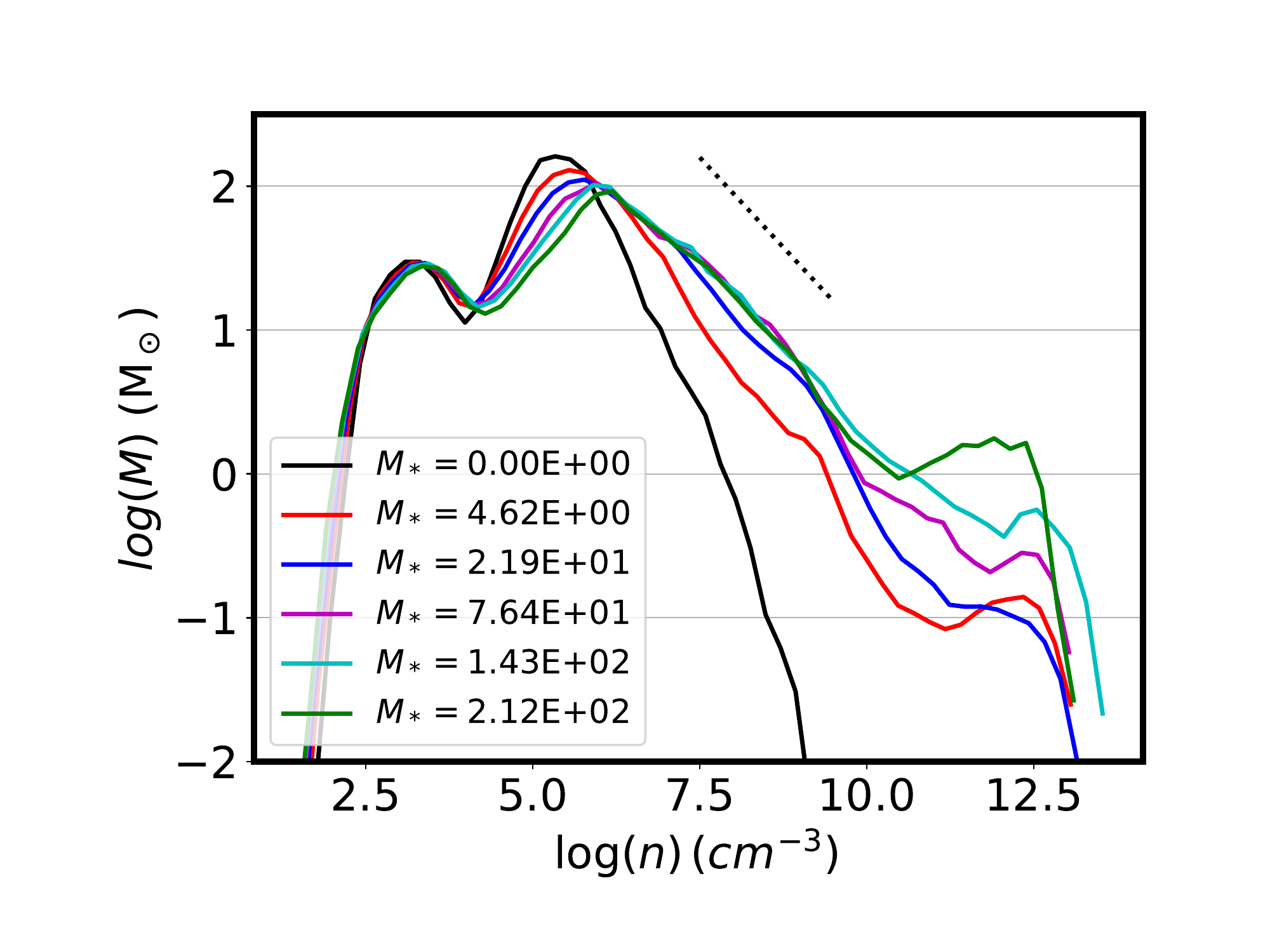}}  
\put(2,5.4){MHD10f05}
\end{picture}
\caption{Mass distribution as a function of density at several timesteps
for the six runs.  Section~\ref{mass_distrib} describes the corresponding physical regimes.
The dotted line shows a powerlaw behaviour $M \propto n^{-1/2}$ as expected for a density PDF $\propto n^{-3/2}$ (see Sect.~\ref{col_env}).
}
\label{fig_pdf_rho}
\end{figure*}

\setlength{\unitlength}{1cm}
\begin{figure*}
\begin{picture} (0,17)
\put(8,11.2){\includegraphics[width=8cm]{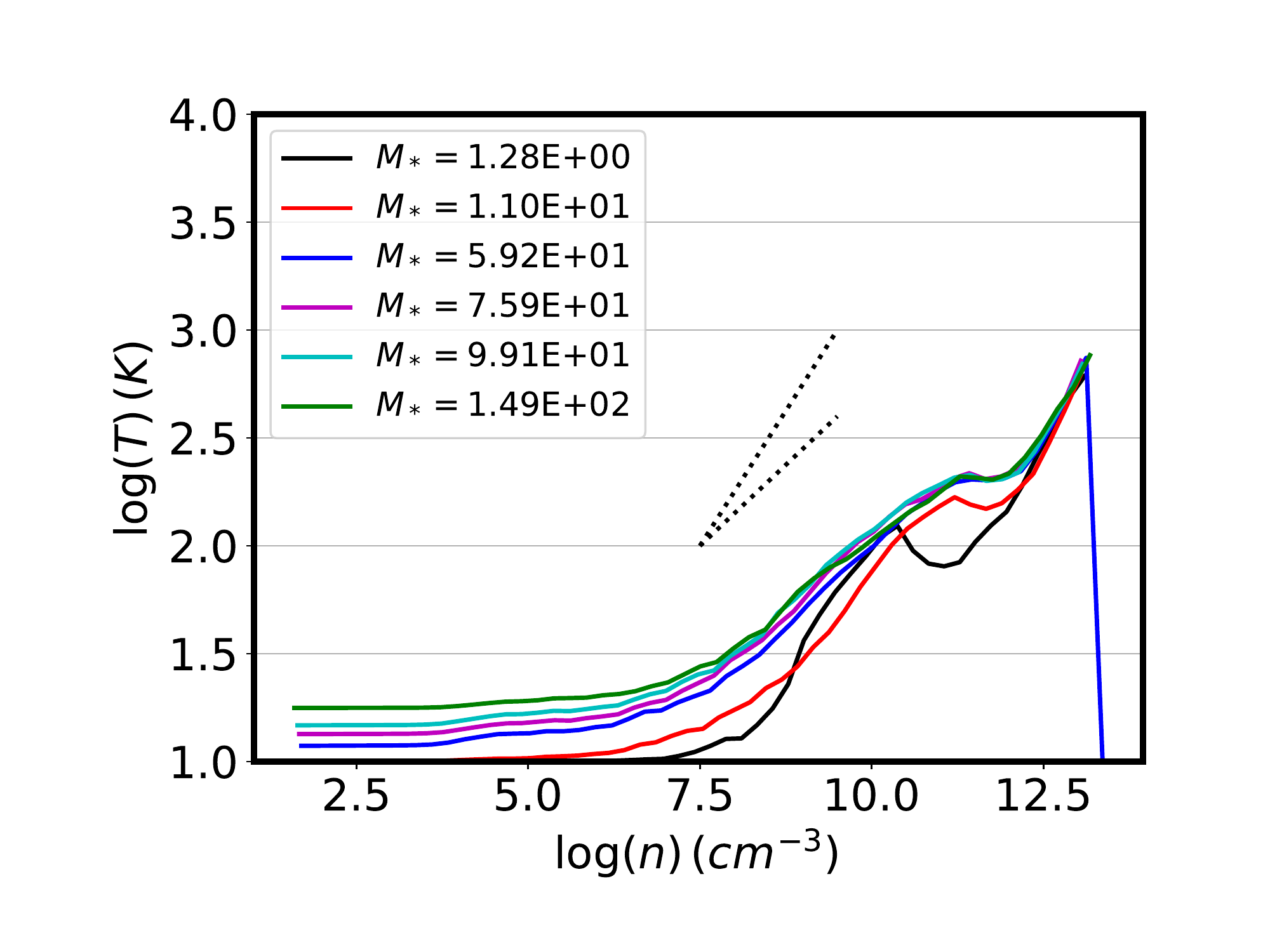}}
\put(10,16.6){HYDROf01}
\put(0,11.2){\includegraphics[width=8cm]{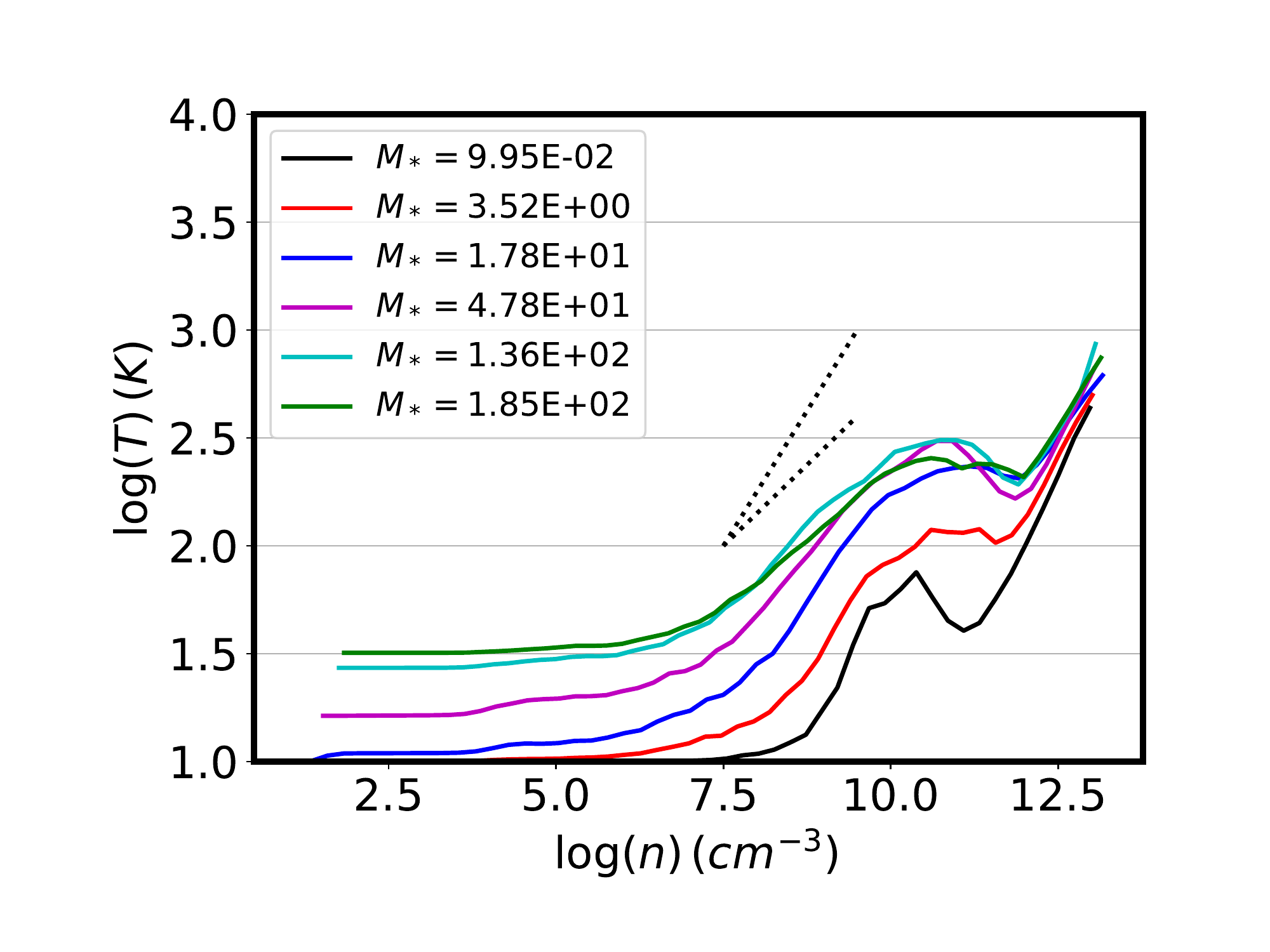}}  
\put(2,16.6){HYDROf05}
\put(8,5.6){\includegraphics[width=8cm]{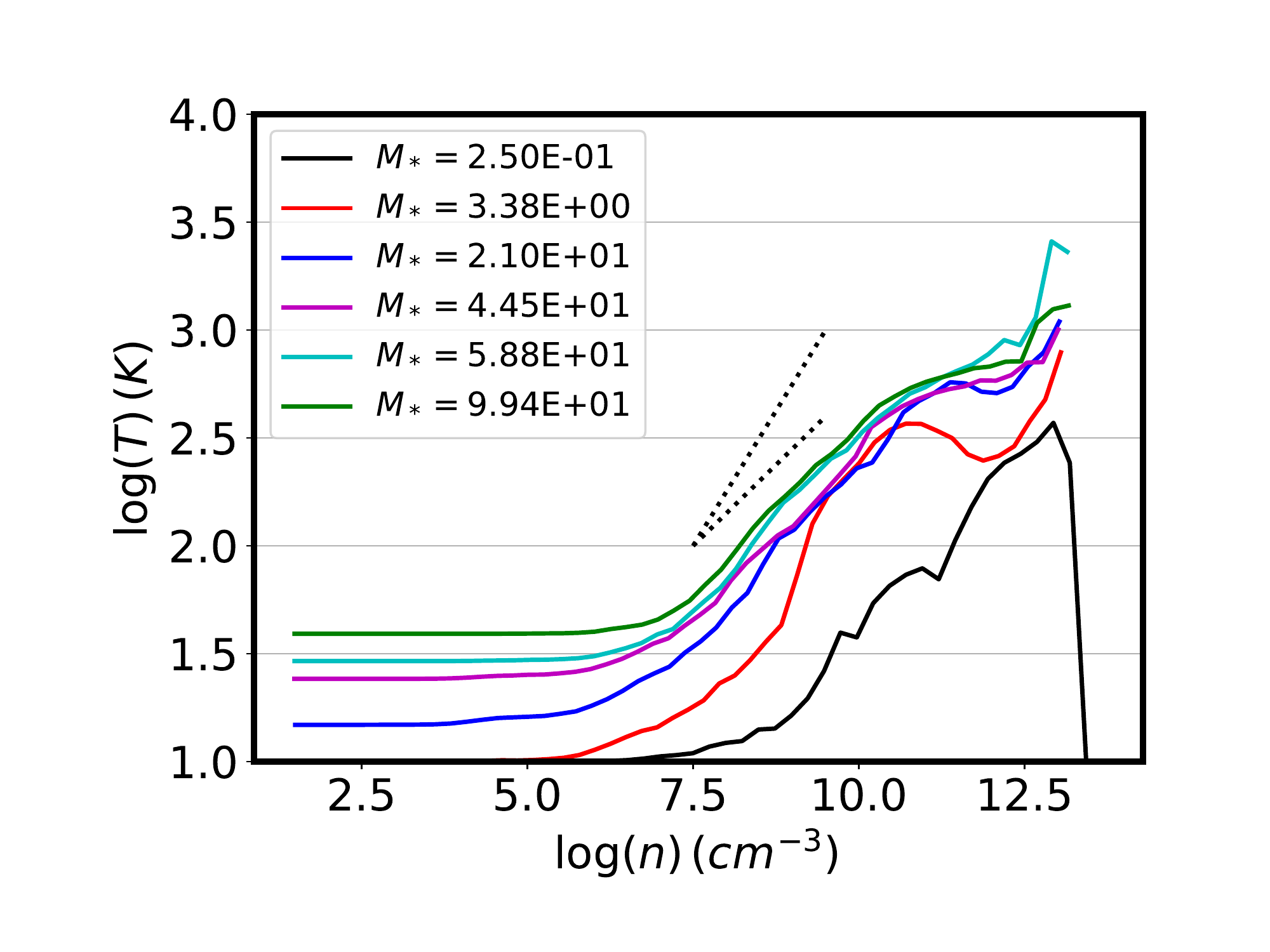}}
\put(10,11.){NMHD10f05}
\put(0,5.6){\includegraphics[width=8cm]{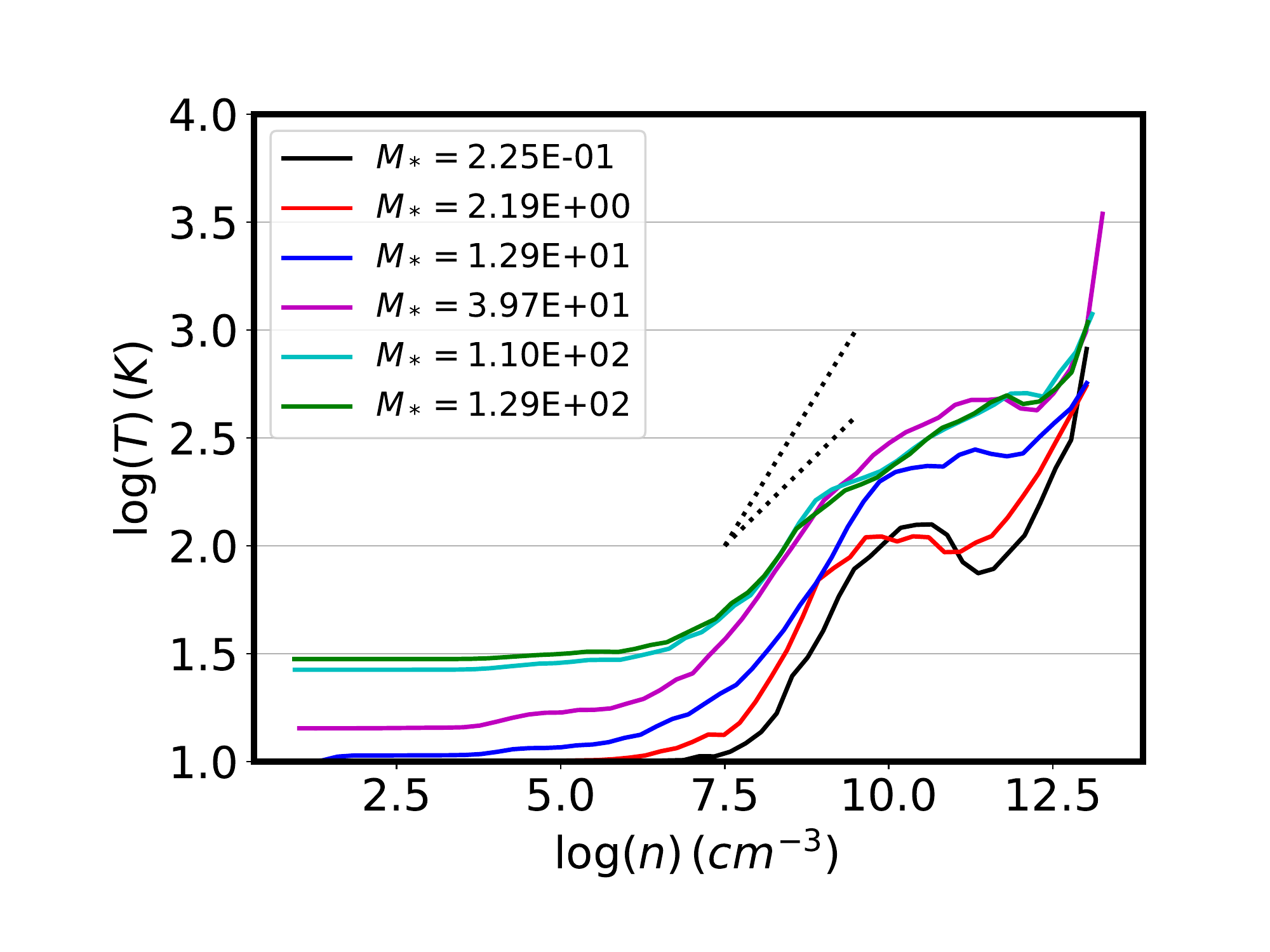}}  
\put(2,11.){MHD100f05}
\put(8,0){\includegraphics[width=8cm]{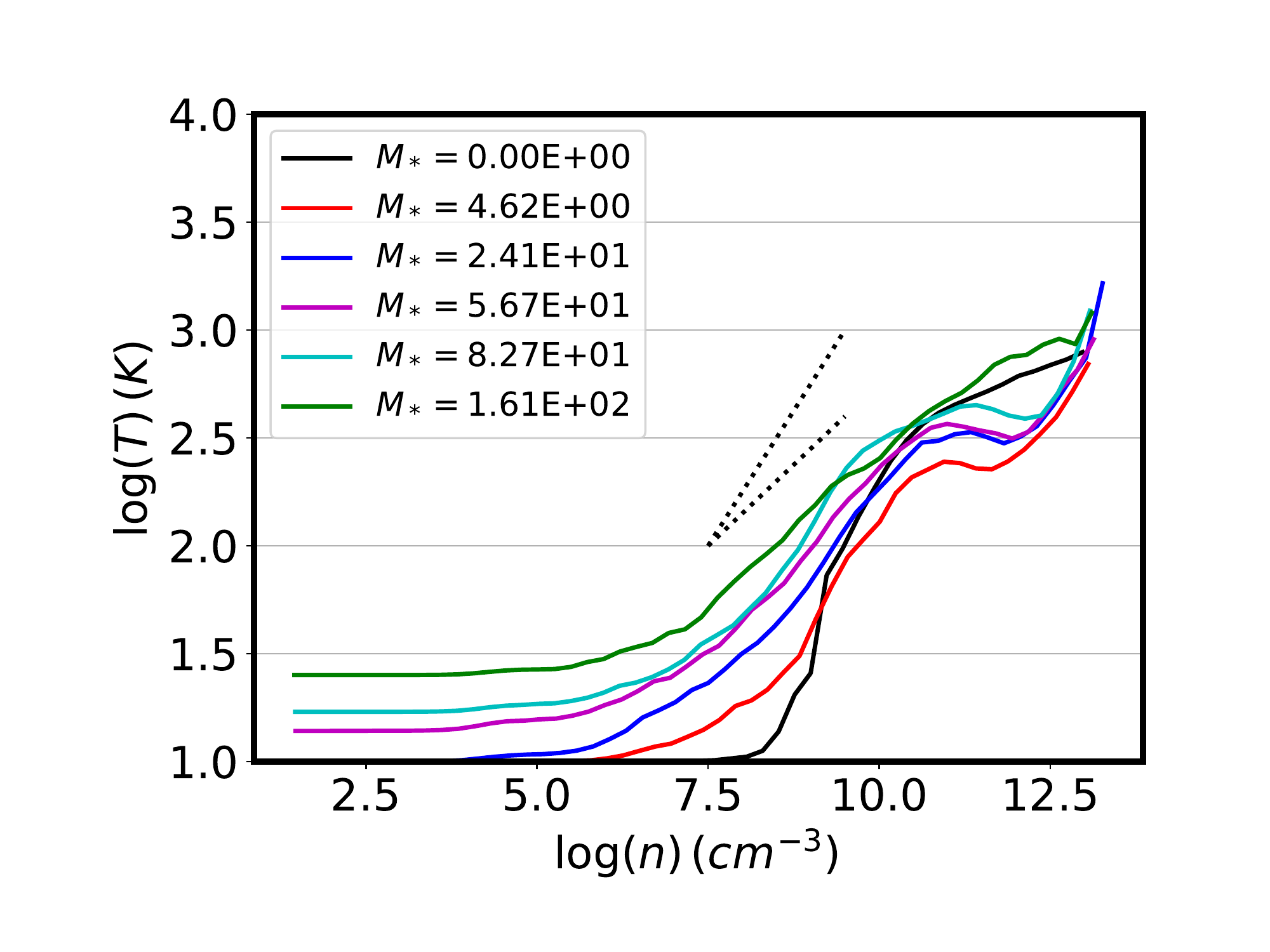}}  
\put(10,5.4){MHD10f01}
\put(0,0){\includegraphics[width=8cm]{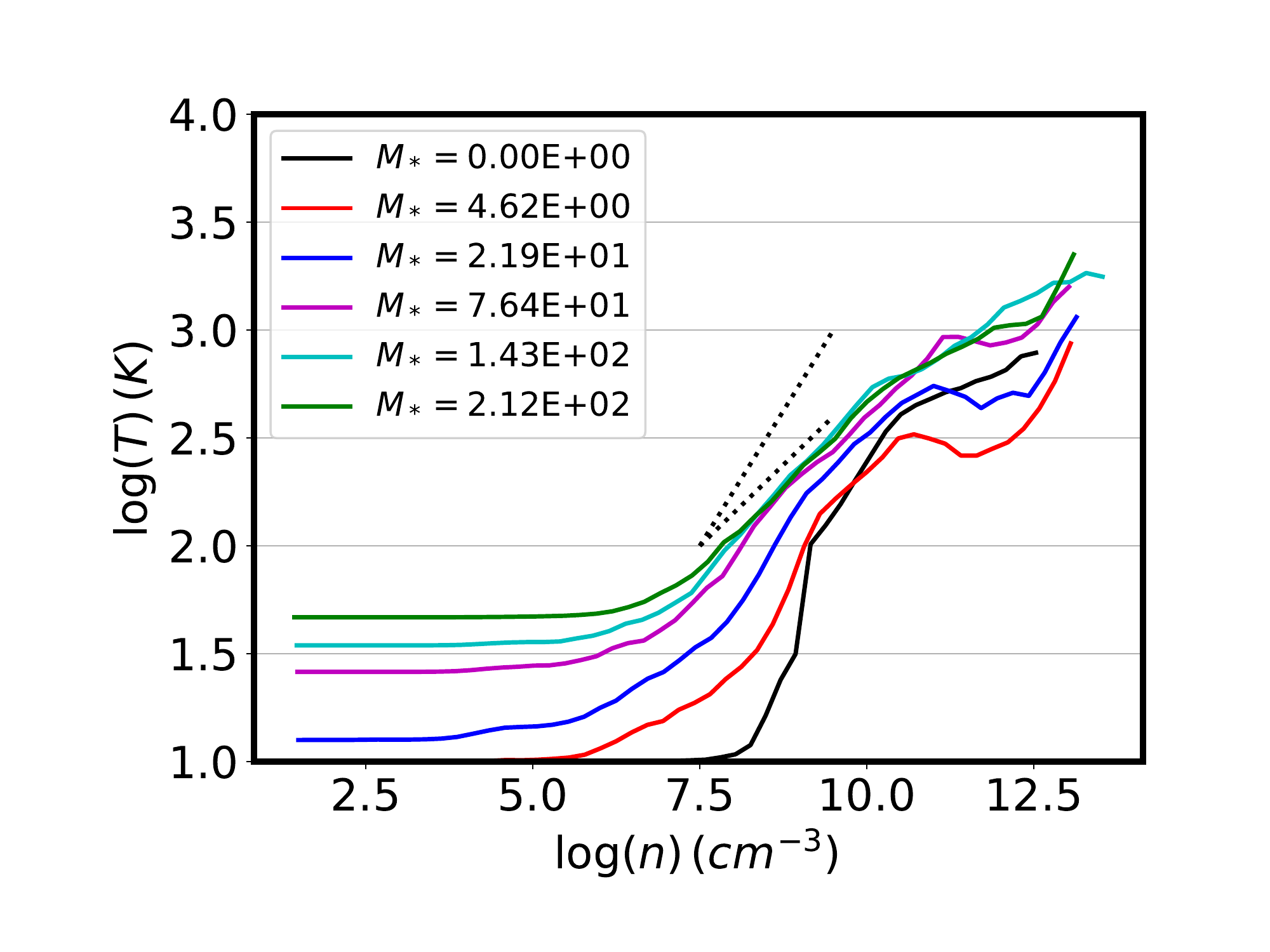}}  
\put(2,5.4){MHD10f05}
\end{picture}
\caption{Mass-weighted temperature in density intervals as a function of density at six time steps 
for the six runs. The dotted lines show a powerlaw behaviour of $T \propto n^{1/2}$ and $T \propto n^{0.3}$, respectively.
}
\label{fig_T}
\end{figure*}
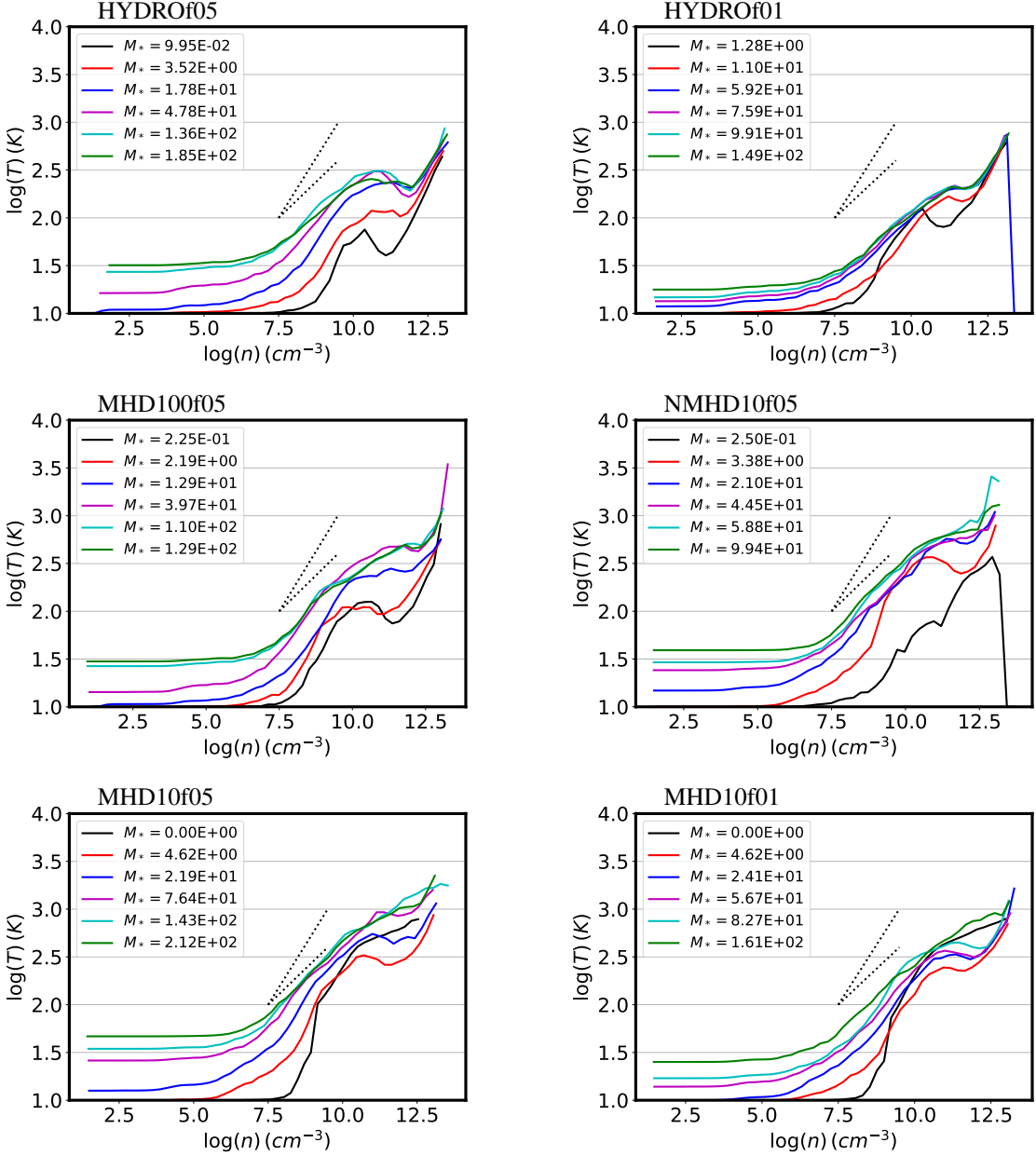

\setlength{\unitlength}{1cm}
\begin{figure*}
\begin{picture} (0,12)
\put(8,5.6){\includegraphics[width=8cm]{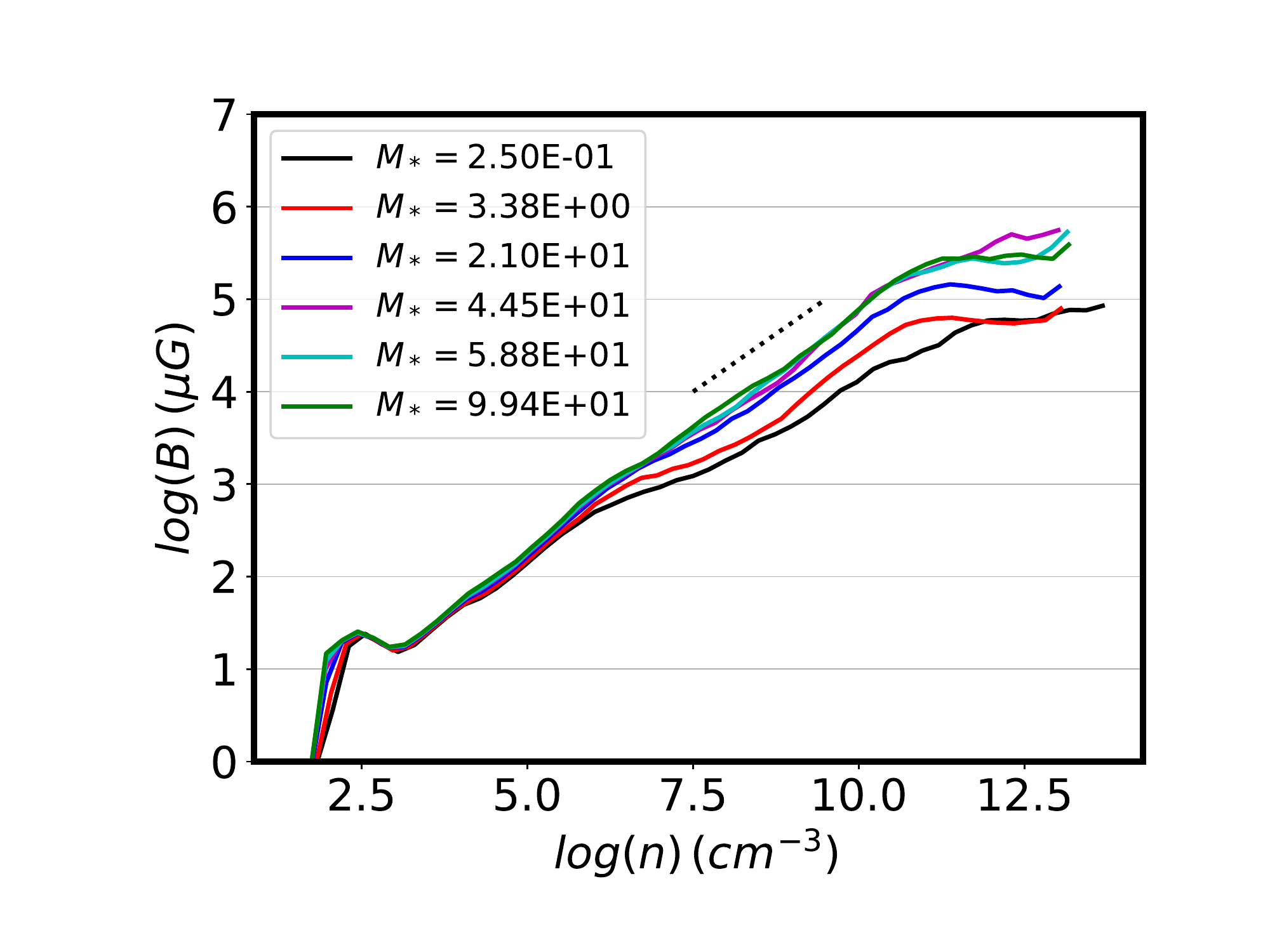}}
\put(10,11.){NMHD10f05}
\put(0,5.6){\includegraphics[width=8cm]{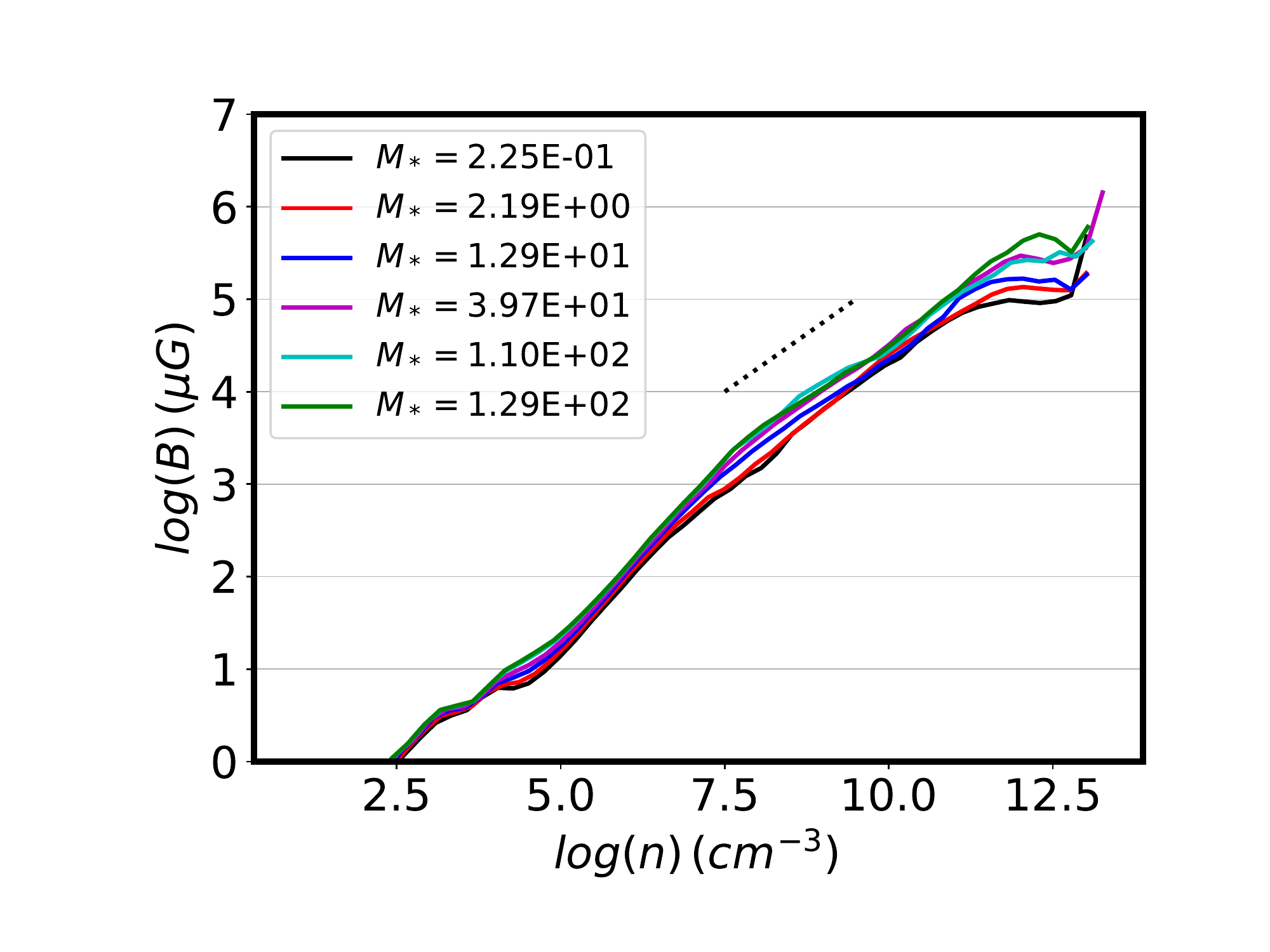}}  
\put(2,11.){MHD100f05}
\put(8,0){\includegraphics[width=8cm]{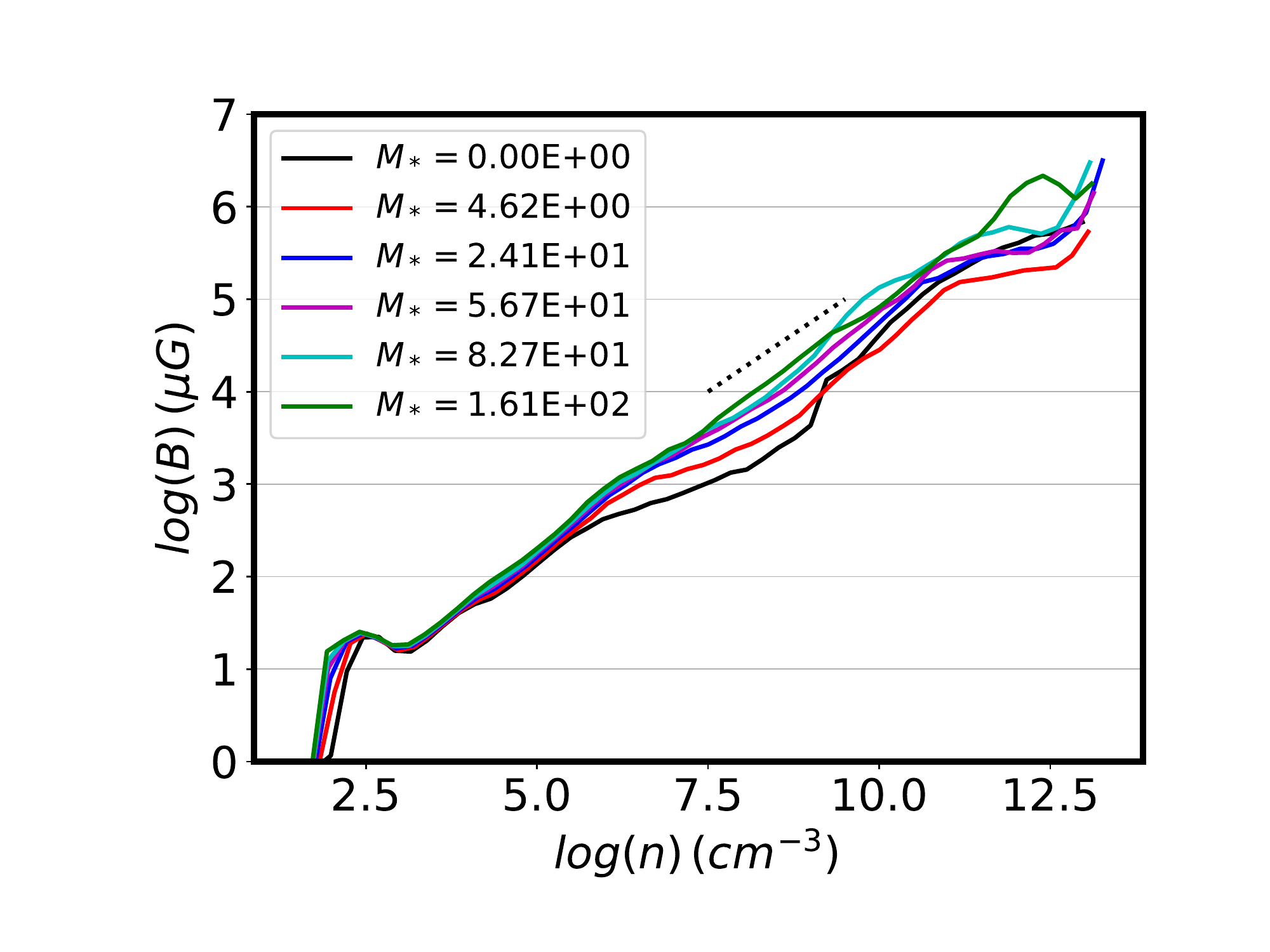}}  
\put(10,5.4){MHD10f01}
\put(0,0){\includegraphics[width=8cm]{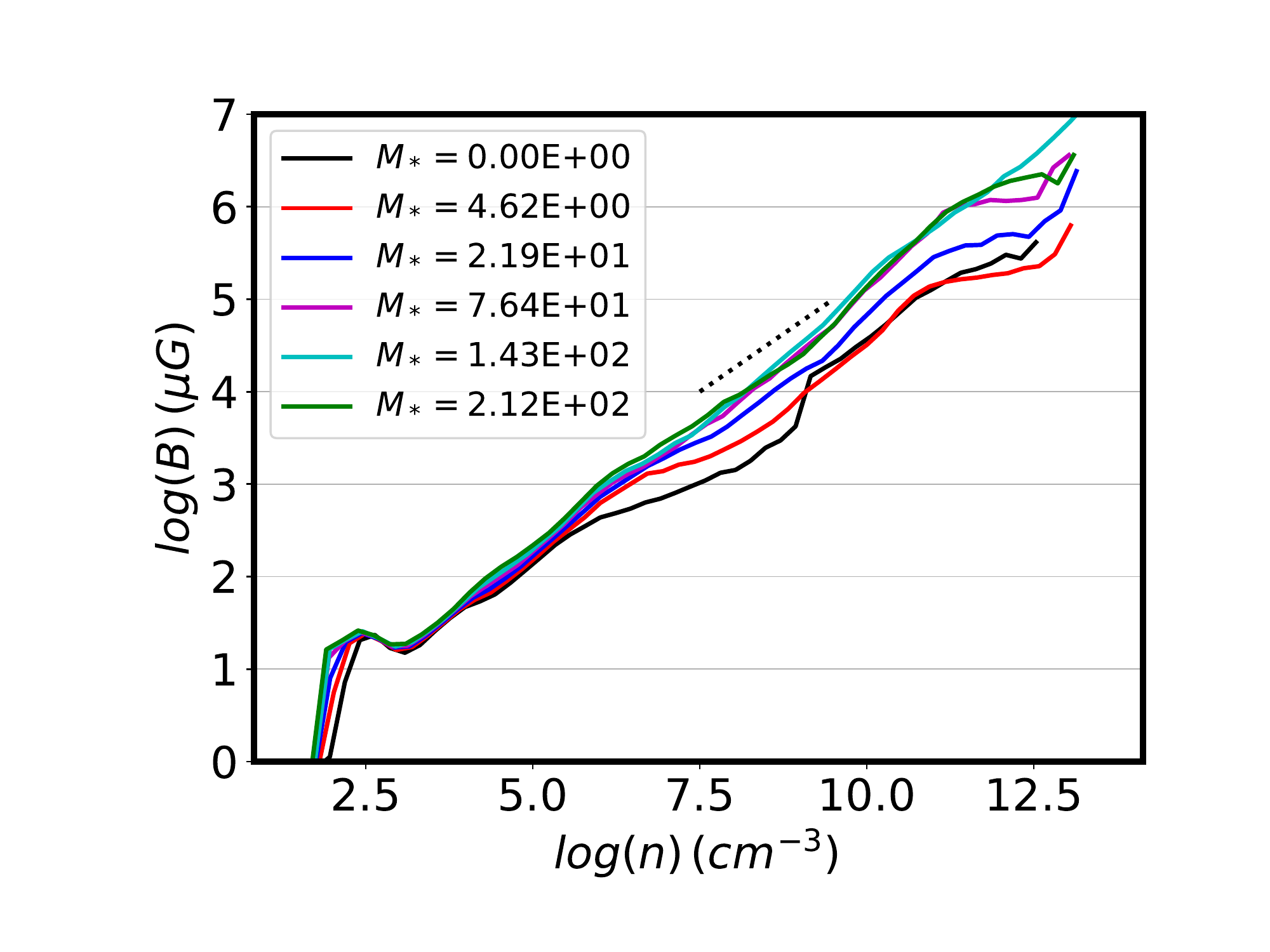}}  
\put(2,5.4){MHD10f05}
\end{picture}
\caption{Magnetic field as a function of density at several timesteps
for the four magnetized runs. The dotted line shows a powerlaw behaviour $B \propto n^{1/2}$.
}
\label{fig_B}
\end{figure*}

\section{Numerical simulations}

\subsection{Equations, numerical methods and setup}
In this paper we solve the equations of the radiative magneto-hydrodynamics. 
All the radiative quantities are estimated in the co-moving frame and assuming the grey approximation; 
that is to say  the radiative energies are integrated over the entire frequency spectrum \citep[e.g.][]{commercon2011}. The equations are
 \begin{equation}
\begin{array}{llll}
\partial_t \rho + \nabla\cdot \left[\rho\textbf{u} \right] & =  0, \\
\partial_t \rho \textbf{u} + \nabla \cdot\left[\rho \textbf{u}\otimes \textbf{u} + P \mathbb{I} \right]& = - \lambda\nabla E_\mathrm{r} +\textbf{F}_\mathrm{L} -\rho\nabla\phi,\\
\partial_t E_\mathrm{T} + \nabla\cdot \left[\textbf{u}\left( E_\mathrm{T} + P_\mathrm{} + \frac{B^2}{8 \pi} \right)  \right.&  \\
\hspace{30pt}\left.- {1 \over 4 \pi}\left(\textbf{u}\cdot\textbf{B}\right)\textbf{B}-\textbf{E}_\mathrm{AD}\times\textbf{B}\right] &=  - \mathbb{P}_\mathrm{r}\nabla:\textbf{u}  - \lambda \textbf{u} \nabla E_\mathrm{r} \\
 &   \hspace{9pt} +  \nabla \cdot\left(\frac{c\lambda}{\rho \kappa_\mathrm{R}} \nabla E_\mathrm{r}\right)+ S_\star\\&\hspace{9pt} -\rho\textbf{u}\cdot\nabla\phi, \\
\partial_t E_\mathrm{r} + \nabla\cdot \left[\textbf{u}E_\mathrm{r}\right]
&=
- \mathbb{P}_\mathrm{r}\nabla:\textbf{u}  +  \nabla \cdot\left(\frac{c\lambda}{\rho \kappa_\mathrm{R}} \nabla E_\mathrm{r}\right) \\
 &  \hspace{9pt} + \kappa_\mathrm{P}\rho c(a_\mathrm{R}T^4 - E_\mathrm{r})\\
 &\hspace{9pt} + S_\star,\\ 
\partial_t \textbf{B} - \nabla\times\left[\textbf{u}\times\textbf{B} +\textbf{E}_\mathrm{AD}\right]&=0,\\
\nabla\cdot\textbf{B}&=0,\\
\Delta\phi & = 4\pi G \rho,
\end{array}
\end{equation}
\noindent where $\rho$ is the material density, $\textbf{u}$ is the velocity, $P$ the thermal pressure,  $\lambda$ is the radiative flux limiter \citep{minerbo1978}, $E_\mathrm{r}$ is the radiative
 energy, $\textbf{F}_\mathrm{L}=  1/(4 \pi) (\nabla\times\textbf{B})\times\textbf{B}$ is the Lorentz force, $\phi$ is the gravitational potential, $E_\mathrm{T}$ the total energy $E_\mathrm{T}=\rho\epsilon +1/2\rho u^2 + B^2 /(8 \pi) +E_\mathrm{r}$ ($\epsilon$\
 is the gas specific internal  energy), $\textbf{B}$ is the magnetic field, $\textbf{E}_\mathrm{AD}$ is the ambipolar electromotor field (EMF), $\kappa_\mathrm{P}$ is the Planck mean opacity,  $\kappa_                                                               
\mathrm{R}$ is the Rosseland mean opacity, $\mathbb{P}_\mathrm{r}$ is the radiation pressure, $S _\star$ the luminosity source, 
and $T$ is the gas temperature.
The ambipolar EMF is given by
\begin{equation}
\textbf{E}_\mathrm{AD} = \frac{\eta_\mathrm{AD}}{B^2}\left[\left(\nabla\times\textbf{B}\right) \times \textbf{B} \right] \times \textbf{B},
\end{equation}
where $\eta_\mathrm{AD}$ is the ambipolar diffusion resistivity, calculated as a function of the density, temperature, and magnetic field amplitude.

The numerical method is overall very similar to the one used in \citet{hetal2020b}. 
The  simulations were performed with the adaptive mesh refinement (AMR) magnetohydrodynamics (MHD) code RAMSES \citep{Teyssier02,Fromang06}.
When non-ideal MHD, i.e. ambipolar diffusion, is included the scheme is the one described in 
\citet{masson2012} and used in previous studies \citep{masson2016,hetal2020,mignon2021,commercon2021,lebreuilly2021}. 
The resistivities are the ones calculated in \citet{marchand2016}. 

In all simulations presented here, 
radiative transfer is accounted for using  
 the flux limited diffusion method assuming  grey approximation  \citep[see][]{commercon2011a,commercon2014}. 
The flux limited diffusion method is known to present some restrictions for instance it does not 
 treat  shadows well due to its isotropic nature. 
  More accurate methods such as the M1 method \citep{gonzalez2007}, the hybrid method \citep{kuiper2010,mignon-risse2020}
 or the VETTAM method \citep{menon2022} have been developed and deal significantly better with anisotropic 
 radiative transfer \citep[see also][]{jaura2018,peter2022}. However, 
  they tend to be more costly than the flux limited diffusion method employed in this work, and are often limited in their 
  current implementations. This certainly represents line of future improvements.
 
At high density, the equation of state is taken from  \citet{saumon1992} and \citet{saumon1995} which takes into account
  H2, H, H+, He, He+, and He2+ (the He mass concentration is 0.27).
The opacities are as described in \citet{vaytet2013}. For the  range of temperatures and densities covered 
in this work, the opacities are  the ones calculated in \citet{semenov2003}.

 The boundary conditions are periodic.  The   cloud is initially spherical and has a radius 
four times lower than the computational domain size.
All simulations were run on a regular  grid of $256^3$ computing cells  and  10   AMR levels have been further added 
during the course of the calculation leading to 
a total number of  18 AMR levels. The resolution criterion is the Jeans length and it is resolved with at least 10 points. 
In the present paper, the issue of numerical resolution is not further discussed and we refer to the appendix of 
\citet{hetal2020b} for an investigation of the impact of numerical resolution. 

\subsection{Sink particles and stellar feedback} 
The sink particle algorithm is described in \citet[]{Bleuler14}.
Sink particles are formed at the highest refinement level at the peak of clumps whose maximum density 
is larger than $ n_{\rm acc}$. The 
sink particles are created if the parent clump has  a density $n > n_{\rm acc}$ and if it is sufficiently gravitationally
bound  \citep[see][]{Bleuler14}.
The value of  $ n_{\rm acc}$ is equal to 
   $ 10^{13}$ cm$^{-3}$. 
With this value of  $n_{\rm acc}$, the  computational cells having a density 
 equal to $n_{\rm acc}$ possess a  mass of roughly 1-2$\%$ of the mass of the first hydrostatic core, $M_L=0.03 M_\odot$. 
At each time step, 10$\%$ of the  gas mass inside the sink's accretion radius and 
with a density above  $n_{\rm acc}$ 
is retrieved from the grid and accreted by the sink.  
The sinks are not allowed to merge. 
The  impact of changing the value of $n_{\rm acc}$ has been discussed in \citet{hetal2020b}. It has been found that both the spatial resolution and 
 the value of $n_{\rm acc}$ may influence the peak of the stellar distribution. 
 However once the first hydrostatic core is sufficiently resolved, this 
 should not be the case.

Sink particles are also a source of radiation due to the stellar luminosity and gas accretion.
 The accretion luminosity is given by 
\begin{eqnarray}
L_{\rm acc} = {f_{\rm acc} G M_* \dot{M} \over R_*} . 
\label{accret_lum}
\end{eqnarray}
where $M_*$ and $R_*$ are respectivelly the star's mass and radius while $\dot{M}$ is the accretion rate.
If all the kinetic energy of the infalling gas  was radiated away, we would have $ f_{\rm acc} \simeq 1$. 
The accretion luminsosity  has been shown to be the dominant source of gas heating at early time and has 
important effects on the surrounding gas \citep[e.g.][]{krumholz2007,offner2009}.
The stellar luminosity of the protostars, $L_*$ and $R_*$ are 
taken from \citet{kuiper2013}
\citep[see also][]{hosokawa2009}.
As discussed in \citet{hetal2020b}, the value of $f_{\rm acc}$ that should be used is not
clearly established. In particular, the radiation is emitted at very small scale, i.e. few stellar radii
and is not expected to propagate uniformly  because of the highly anisotropic density distribution
\citep[e.g.][]{krumholz2012}. 
As in \citet{hetal2020b}, we perform simulations in which we use an {\it effective} accretion luminosity and explore the values $f_{\rm acc}=0.1$  and 0.5. By considering an effective luminosity smaller than the estimated total luminosity,  we envisage that 
the rest of the energy  either 
  escape preferentially along the cavities open by winds and jets
  or is converted into jet or a wind kinetic energy.
 This is obviously an important source of uncertainties which requires further investigations.

 We start considering accretion and stellar luminosities when the sink has a mass 
of about 2 $M_L$, i.e. 0.07 M$_\odot$. The reason is that due to the limited spatial resolution, when the sink 
is introduced the protostar is not truly formed yet. Since the size of the sink particles is not very different 
from the radius of the first hydrostatic core, it seems reasonable to assume that the protostar is formed only when the 
sink reaches a mass equal to a few $M_L$.
 Note that although reasonable, this assumption clearly requires further investigation. For instance, \citet{bhandare2020} who
have performed two-dimensional simulations of the second Larson cores, i.e. the young protostar, found that they grow with time 
far beyond the solar radii. This clearly suggests that at least some of the accretion energy is not fully radiated away but somehow stored
in the star for some time. Indeed, the accretion shock at the edge of the second Larson core is subcritical 
\citep{vaytet2013} meaning that most of the accretion energy is advected inside the protostar and not immediately radiated away. 
This constitutes an important source of uncertainty for calculations such as the ones performed in this work. 

\subsection{Initial conditions and runs performed}
Our initial conditions consist in  spherical clouds in which a turbulent velocity field has been added. 
 The velocity field  has a classical Kolmogorov power-spectrum  equal to 11/3 
with random  phases.  A fully self-consistent approach would require to also set up the 
density and magnetic field fluctuations. This is however not an easy task. In practice it requires running a large scale simulations and zooming-in
or  at least performing a preliminary run without self-gravity \citep[see for instance][]{lane2022}. Note that \citet{leeh2018a} have compared various approaches including starting from 
a previous phase during which the simulation is run without self-gravity and starting directly from a prescribed turbulent field as it is done 
here. They found very similar results. This suggests that, at least in the context of collapsing clumps, the choice of the initial turbulent field may
not be so important, probably because as the collapse proceeds, the fluctuations evolve and the initial perturbations are largely forgotten.


 The clump we consider has a
  mass of $10^3 M_\odot$ and an initial  radius of 0.4 pc corresponding to a uniform density of about 8$\times 10^4$ cm$^{-3}$ initially.
Observationally, this corresponds to relatively standard massive 
star forming clumps 
\citep[e.g.][]{Urquhart14,elia2017,elia2021,lin2022}.
With an initial temperature of 10 K, the ratio of the thermal over gravitational energy is about 0.008.
  The clump density leads to a freefall time of about 110 kyr.
The initial value of the Mach number is equal to  7 leading to 
a turbulent over gravitational energy ratio of about 0.4, i.e. the clumps are close to be initially virialised.

We set up the simulations with a uniform initial magnetic field through the cloud and intercloud medium. 
We considered two initial mean-field strengths, 
with  mass-to-flux ratios, $\mu$, of respectivelly 10 and 100 (corresponding 
to about 100\,$\mu$G and 10\,$\mu$G respectivelly).
These values are motivated by the observations of $\mu$ on the order of a few in dense cores \citep[e.g.][]{crutcher2012,myers2021}.
This selection also aims to account for the broad dispersions in the $\mu$ values of the massive clumps identified in the 1-kpc scale simulation presented in \citet{h2018}.
In these MHD simulations, which have a spatial resolution down to 400 AU, it has 
been  found that the mass-to-flux 
ratio, which presents a broad dispersion, is indeed on the order of a few for Solar mass cores but 
is lower for  the more massive ones.  More precisely, Fig.~5 of  \citet{h2018} shows the distribution 
of self-gravitating objects with density larger than $10^4$ cm$^{-3}$. It shows a clear trend that the mass-to-flux ratio, $\mu$, is increasing with mass in spite 
of a broad distribution. The most massive clumps displayed has a mass of about 100 M$_\odot$ and thus one needs to extrapolate to get a hint on 1000 
 M$_\odot$ clumps. Based on this figure, we would expect that the typical $\mu$ of a 1000  M$_\odot$ clumps is certainly larger than 10. 
 This may at first sight be surprising because low mass cores have been observed to present values of $\mu$ on the order of a few \citep{pattle2022}. 
 It should however be 
 remembered that the mass-to-flux ratio is the ratio of a volume over a surface weighted quantity. Thus considering objects with similar mean density, 
 the mass-to-flux is expected to increase with the object size.

Another fundamental aspect is  the physics of the magnetic field evolution. Whereas
many studies have assumed ideal MHD, it is however clear that this is a poor approximation at high density in 
star forming regions \citep[e.g.][]{zhao2020}. We performed one run with ambipolar diffusion and 
a magnetisation corresponding to $\mu=10$. Due to the small time steps induced by the second order derivative
in the ambipolar diffusion operator, this numerical simulation is quite challenging and has required about 500,000 cpu hours.

Table~\ref{table_param_num} summarizes the various runs performed.

      \begin{table*}
         \begin{center}
            \begin{tabular}{lcccccccc}
               \hline\hline
               Name & $R_c$ (pc) & $\mu$ &  $\mathcal{M}$ & $lmax$ & $dx$ (AU)   & $f_{\rm acc}$ & NMHD \\
               \hline
               NMHD10f05  &  0.4 & 10  & 7 & 18   & 1.15  & 0.5 & yes \\
               MHD10f05  &  0.4 & 10  & 7 & 18   & 1.15  & 0.5 & no \\ 
               MHD10f01  &  0.4 & 10  & 7 & 18   & 1.15  & 0.1 & no \\ 
               MHD100f05  &  0.4 & 100  & 7 & 18   & 1.15  & 0.5 & no \\ 
               HYDROf05  &  0.4 & $\infty$  & 7 & 18   & 1.15  & 0.5 & no \\ 
               HYDROf01  &  0.4 & $\infty$  & 7 & 18   & 1.15  & 0.1 & no \\ 
            \end{tabular}
         \end{center}
         \caption{Summary of the runs performed.  $R_c$ is the initial clump radius.
$\mu$ is the mass-to-flux over critical mass-to-flux ratio. 
$\mathcal{M}$ is the initial clump Mach number. 
$lmax$ is the maximum level of grid used and $dx$ corresponds
to the maximum resolution which is equal to 1.15 AU in these runs.
 $f_{\rm acc}$ gives the fraction of the accretion 
luminosity which is taken into account in the calculation. 
NMHD tells whether non-ideal MHD is accounted for. 
}
\label{table_param_num}
      \end{table*}

Two ideal MHD runs (MHD10f05 and MHD100f05) and one purely hydrodynamical runs (HYDROf05) allow us to investigate 
the role of the magnetic field during the collapse notably  on the initial mass function. Whereas the non-ideal MHD run
(NMHD10f05) is the most realistic simulation and to our knowledge is the first simulation of a massive star forming clump 
which includes both radiative feedback and non-ideal MHD. These four runs are complemented by two runs with a lower 
accretion luminosity parameter, $f_{acc}=0.1$, namely MHD10f01 and HYDROf01 which  allow to assess and discuss the influence of 
radiative feedback on our results.
All simulations are carried out until at least 150 $M_\odot$ of gas have been accreted onto the sink particles, except run NMHD10f05 for
which the final mass accreted by the sinks is 100  $M_\odot$. This is because, as explained above, this simulation is more computationally demanding.

\section{General clump description}
In this section, we look at the global evolution and final properties of the collapsing clump as a whole. 
We start by describing the general morphology, before proceeding to discuss the star formation and luminosity. We then study the gas density and temperature distribution inside the clump.

\subsection{Total accreted mass}
Figure~\ref{time_mass}  displays the total mass accreted (top-left panel) by sink particles, $M_{tot,*}$ as a function of time for the 
six simulations. The total accretion rate is also plotted (top-right panel). Since it is a heavily fluctuating
quantity, the latter is calculated by averaging its instantaneous  value
over 100 uniformly spaced time intervals. 
 As indicated above, all simulations are run until  about 150 $M_\odot$ have been accreted, corresponding to a star-formation efficiency 
 of 15$\%$.
 There are two  exceptions,  NMHDf05 for which at the end of the simulation
100 $M_\odot$ have been turned into the sinks, and MHD10f05 for which the final total mass of sinks is equal to 200 $M_\odot$. 
Two groups of simulations are easily distinguished. On one hand the hydrodynamical simulations and the low magnetized one, 
MHD100f05 and on the other hand the more magnetized ones, i.e. MHD10f01, MHD10f05 and NMHDf05. As expected, 
due to magnetic support, the latter group of simulations collapses a bit more slowly. Two points are worth mentioning, 
first the simulations with $f_{acc}=0.1$ and with $f_{acc}=0.5$ behave very similarly showing that in spite of 
strong heating, radiation does not significantly alter the large scale dynamics. Similar conclusion is also reached for the ambipolar diffusion. This is because i)  thermal support is rather weak and an increase of temperature even by a factor of several does not make thermal support sufficiently strong to provide a significant support at the clump scale; and also because ii) magnetic field is only significantly modified by non-ideal MHD processes, at 
high density (say  $n > 10^7$ cm$^{-3}$).   
Interestingly, we see that after a fast increase the accretion rate, $\dot{M}_{tot,*}$,  reaches 
 values, which are nearly identical for all simulations and equal to about $10^{-2}$M$_\odot$~yr$^{-1}$. This is because the accretion rate is controlled
by the largest scale, here the clump, which is globally collapsing. The magnetic intensity considered here is too weak to significantly modify 
this global dynamics. 

\subsection{Total  luminosity}
Bottom-left panel of Fig.~\ref{time_mass} portrays the total luminosities, $\sum (L_* + L_{acc})$ of the sink particles
(see also Fig.~\ref{time_mass2}). 
As for the accretion rate, after an increase which takes about 0.02 Myr, it reaches, in the case
with $f_{acc}=0.5$, a plateau 
at about $1-2 \times 10^5$ L$_\odot$. In the case $f=0.1$, the total luminosity is 10-20 times lower 
for run HYDROf01 and 3 times lower for run MHD10f01.
Note that
 Fig.~\ref{time_mass} shows that the clumps spend about 0.02 Myr in the protostellar phase with a
 luminosity several times lower than the peak values.

It is interesting to compare these values with observations. Although what is 
observationally available is the bolometric luminosities rather than the total source 
luminosities, they are obviously related and 
can be compared. In particular, it is expected that the radiation emitted by the star in the visible domain 
is quickly absorbed and reemitted in the infrared by the dust. 
 For instance Fig.~12 of \citet{elia2017} displays 
the bolometric luminosities as a function of mass for a sample of clumps. As can be seen
our values agree well with the luminosity distribution of the protostellar 1000 M$_\odot$ clumps, which 
range from 10$^2$ to few 10$^5$ L$_\odot$ although correspond to their upper values when $f_{acc}=0.5$. Note, however, that the bolometric luminosities
calculated in \citet{elia2017} correspond to wavelengths longer than 20$\mu m$.
Therefore these values represent themselves lower limits of the real luminosities.

\citet{lin2022} present detailed observations for several massive star forming clumps with 
comparable mass and radii (see their Table~6). Luminosities of a few $10^5$ L$_\odot$ are also reported.
 
Since the mass of the clump is 1000 $M_\odot$, this 
means that once the luminosity is about 10$^5$ L$_\odot$, the luminosity per solar mass is about $10-100$ L$_\odot$ / M$_\odot$. 
Again this is in good agreement with the values seen in Fig.~13 of \citet{elia2017}. 

In order to define a reference with which the luminosities can be compared, 
 we define a quantity $L_{glob} = 0.5 \times G M_{*,tot}
\dot{M} _{*,tot} / (2 R_\odot)$, which would correspond to the accretion luminosity of an object of 
mass $M_{*,tot}$ and radius 2 $R_\odot$, accreting at a rate $\dot{M} _{*,tot}$ with an efficiency $f_{acc}=0.5$.
The ratio $\sum L_* / L_{glob}$ is expected to be smaller than 1 because the luminosity is a non-linear quantity, 
which decreases with the number of stars. It gives a sense of how fragmented is the clump and how efficiently is the
 gravitational energy converted into radiation. 
Bottom-right panel shows $\sum L_* / L_{glob}$  for the six runs as a function of time. As expected, this quantity decreases with time and,  at later time,
 it reaches values as small as $10^{-3}$ and as high as  
$2 \times 10^{-2}$, depending on the run. Interestingly, there is a clear trend for the magnetized runs to have 
values 1.5-2 times larger than their hydrodynamical counterpart. As we show later this is because magnetic field tends to reduce fragmentation, therefore building more massive stars which present higher luminosities.

\subsection{General morphology}
Figure~\ref{fig_coldens} portrays the column density of the whole clump at time $t=0.11$ Myr, which as seen from 
Fig.~\ref{time_mass}, corresponds to a time where approximately 100-120 $M_\odot$ have been accreted. 
The dark circles show the sink particles, which represent individual stars.
The six simulations present a similar pattern. A complex network of intervowen and interconnected filaments 
have formed and three of them appear to be a little  more prominent. Their length is approximately $\simeq 0.3$ pc 
and is comparable to the whole clump size.
The three main filaments intersect, forming a hub located approximately at $y=0.75$ pc and $z=0.65$ pc. 
The stars are represented by the dark circles, are mainly, though not exclusively, located in the hub and in the filaments.
Let us recall that density filaments are naturally produced both by MHD turbulence 
\citep{hennebelle2013,federrath2016,xu2019}, shocks \citep{daisei2021},
and by gravity \citep[e.g.][]{smith2014,daisei2021}, which for different reasons tend both to amplify anisotropies. 

Beyond these general similarities, significant differences between the six simulations are clearly visible.
First of all, we see that radiative feedback has a clear influence on the cloud evolution and its fragmentation
\citep{krumholz2007,hetal2020b}. 
For instance there are more sinks in  run HYDROf01  than in run HYDROf05 and in run MHD10f01 than in 
MHD10f05 (this will be further quantified in \S~\ref{fragmentation}). This is a clear consequence of less heating when $f=0.1$ than when $f=0.5$. The impact on the 
gas structure appears to remain more limited.
Second of all, clearly magnetic field reduces  significantly the numbers of sinks. This is particularly obvious 
by comparing run HYDROf05 with run MHD10f05 as well as run HYDROf01 with run MHD10f01. It is also clear that 
sink particles tend to form in higher column density regions in the magnetized runs. This clearly is a consequence
of the support provided by the magnetic field which efficiently stabilizes the gas particularly when its column density 
is not too high. This happens obviously, only if magnetic field is strong enough. Indeed in run MHD100f05 which 
has an initial magnetic field 10 times lower than run MHD10f05, the sink distribution is very similar to 
run HYDROf05.
Interestingly, the sink distribution in run NMHDf05 is comparable to run MHD10f05 except near  the high column densities
areas where a small excess of sinks is sometimes visible.  This stems from the fact that ambipolar diffusion 
is efficient only at small scales and at high density.

\section{Gas and magnetic field distribution}
\label{gas_res}

\subsection{Density, velocity and temperature profiles}
\label{prof}
Figure~\ref{fig_radial} displays radial profiles of various density-weighted quantities for runs MHD10f05 (left) 
and HYDROf05 (right) and at several timesteps. For the sake of conciseness only 2 simulations are being displayed 
and discussed here.
The adopted center is the position  of the most massive 
sink particle, which is located in the hub at  $y=0.75$ pc and $z=0.65$ pc.

The first row displays the gas temperature that we remind is initially uniform and equal to 10 K. 
As time goes on, temperature increases by 2 to 3 orders of magnitudes in the center and about one order 
of magnitude in the clump's outer part. The  temperature profile in the clump inner part broadly 
behaves as $r^{-1}$ while it is almost flat in the clump outer part. 
We further note the presence of  many temperature peaks associated to sink particles distributed 
through the clouds. We also stress that the temperature is clearly larger by a factor $\simeq 1.5$ 
in run MHD10f05 than in run HYDROf05. We will come back on this particular point later but this effect 
is similar to what has been reported by \citet{commercon2011}, where radiative MHD calculations where also
performed and higher temperatures have been reported in the MHD case. This is a consequence of the non-linearity 
of the accretion luminosity proportional to $\dot{M} M$. By reducing fragmentation and extracting angular momentum, 
magnetic field increases both $M$ and $\dot{M}$ leading to higher accretion luminosity.
These temperatures appear to be in good agreement with the ones  presented 
in Fig.~10 of \citet{lin2022}. For instance, the temperature at few 0.01 pc is about 100-200 K while at 
0.1 pc it is typically 50-70 K. 

Second row shows the density profiles. The straight line represents the density of the singular isothermal sphere (SIS), 
i.e. $\rho = c_{s,0}^2 / (2 \pi G r^2)$, where $c_{s,0}$ is the sound speed taken here equal to 0.2 km s$^{-1}$. 
As collapse proceeds, the density increases from outside-in and after roughly 
0.1 Myr, it presents a powerlaw-like shape close to, but slightly shallower than, $r^{-2}$.
As we see the values evolve with times and also slightly depend on the radius. 
 We see however than 
it is nearly 2 orders of magnitude denser than the SIS, which is an expected consequence of the low initial thermal 
energy and the compactness of the cloud. The density in run HYDROf05 is slightly lower than in run MHD10f05, which is 
a consequence of the magnetic support. Let us remember that $r^{-2}$ density profile is the expected density structure
of a spherical collapsing cloud \citep[e.g.][]{larson1969,shu77} as, together with a uniform radial velocity, it leads 
to a roughly constant accretion rate through the cloud \citep{ligx2018,gomez2021}.
When turbulence is included, it is however common to find profiles  slightly shallower, for instance $\rho \propto
 r^{-1.5}$ is often reported \citep[e.g.][]{chang2015,lip2018}, though in the present case this value seems a little 
 too shallow.

The radial velocity through the cloud is presented in the third row. At 0.1 Myr, a constant radial velocity of $\simeq 1.5-2$ km s$^{-1}$
appears to reasonably represent the cloud radial velocity for radius between 0.03-0.3 pc. The radial velocity increases towards the cloud inner 
part where it reaches $\simeq$ 10 km s$^{-1}$. The parallel velocity, which represent both turbulent and rotation (i.e. the non-radial component) 
is displayed in the fourth row. Due to the chosen initial conditions, it is 
of the order of 
$\simeq 2$ km s$^{-1}$ in the cloud outer part. As the collapse proceeds and due to the increase of $v_r$, turbulence is further amplified toward the cloud center \citep{hennebelle2021} and this behaviour explains the density 
profile being shallower than $r^{-2}$. 
These velocity values  are in good agreement with the values presented in Fig.~22 of \citet{lin2022}.
For instance at 0.1 pc, values of about 3 km s$^{-1}$ are reported.

\subsection{Mass distribution} 
\label{mass_distrib}

The mass distribution, which we remind is equivalent to the mass weighted density PDF, 
 is displayed  in Fig.~\ref{fig_pdf_rho} for the six simulations at several timesteps.
 The distribution contains several features and going from low to high densities four domains can be identified: the interclump medium, the clump outer part, 
the collapsing envelopes and the high density material. We stress that the last two
do not correspond to a single physical region but rather develop around each individual collapse
center. 

\subsubsection{The interclump medium}
 At low density the  mass distribution
presents a roughly lognormal shape which peaks at about 500 cm$^{-3}$ \citep[e.g.][]{vazquez94,Federrath08,Kritsuk11} and remains 
stationary through time. It is due to the development of turbulence in the cloud outerpart. The latter has formed by the 
turbulent-driven expansion of the cloud external layer. Clearly it contains a small amount of mass. 

\subsubsection{The clump outer part}
 At higher density, i.e. $\rho \simeq 10^5-10^6$ cm$^{-3}$, a second peak of the mass distribution located at the cloud initial mean density, is visible. 
 It contains most of the mass of the cloud and shifts toward higher densities as collapse proceeds. Meanwhile as expected the mass it contains, 
 declines over time. Overall the mass distribution of this density range is similar for the six simulations. We can nevertheless note that the
 peak is a bit broader for the two hydro runs, than for the more magnetized runs MHD10f01, MHD10f05 and NMHDf05. This is likely 
 a consequence of the magnetic field which is known to reduce the turbulent dispersion of the density distribution 
 \citep[e.g.][]{molina2012}.

\subsubsection{The collapsing envelopes}
\label{col_env}
At densities higher than its peak value, the mass distribution is better described by a powerlaw behaviour up to 
densities of $10^9$ and even 10$^{10}$ cm$^{-3}$. This part of the mass distribution corresponds to the  $n \propto r^{-\alpha}$, 
$\alpha \simeq 2$,
envelope discussed in Fig.~\ref{fig_radial}.

 Let us remember that there is a simple correspondence between $\alpha$ and the 
index of the mass distribution. 
Let $d {\mathcal N}$ be the number of fluid particles located between radius $r$ and $r+dr$. We have $d{\mathcal N} \propto r^2 dr$. 
But since 
$n \propto r^{-\alpha}$, we have $d {\mathcal N} / d \log n \propto n^{-3 / \alpha} $ and the mass weighted density PDF is 

\begin{eqnarray}
n  {d \mathcal N \over d \log n  } \propto n^{-3 / \alpha+1}. 
\end{eqnarray}
For $\alpha \simeq 2$, we thus find that 
$n  {d \mathcal N} / d \log n \propto n^{-1/2} $, which indeed is close to the observed bevaviour of the mass distribution between 
$10^7$ and $10^9-10^{-10}$ cm$^{-3}$ as  shown by a comparison with the dotted lines. 

Several aspects are worth noticing. First at early stages (black and red curves), the mass distributions evolve with time.
More mass is gradually accumulated at high densities as collapse proceeds. Once the $n \propto r^{-2}$ envelope is 
fully developed, the mass distribution is stationary. This is all consistent with the stationarity observed in 
Fig.~\ref{fig_radial} illustrating that the accretion rate remains broadly constant with time.

\subsubsection{The high density material}
At  density larger than $10^9-10^{10}$ cm$^{-3}$,  the mass distribution becomes flatter, meaning that 
mass is pilling up. This is a consequence of rotational and thermal supports. Indeed protoplanetary
disks form \citep[see][for a description of disks in similar simulations]{lebreuilly2021}.
Clearly the amount of mass significantly varies with magnetisation and it is several times higher in the hydro runs than 
in the significantly magnetized ones (MHD10). This is a clear consequence of magnetic braking, which by extracting angular momentum 
leads to smaller and less massive disks.

\subsection{Temperature vs density distributions}

 Figure~\ref{fig_T} shows the mean temperature as a function of density in the six simulations. 
In each density bin, the mean temperature is simply the mass weighted temperature. 
The overall behaviour is as suggested by the temperature profiles shown in 
\S~\ref{prof}. 

The temperature associated to the  high density material is typically 
larger than $\simeq 300$ K and reaches values of few thousands K. As expected 
the temperature increases with $f$. 

For the lower density material (i.e. $n < 10^9$ cm$^{-3}$), we see first 
that the temperature decreases roughly as $T \propto n^{-0.3-0.5}$ (as indicated by the dotted line) 
and then at density
of about $\simeq 10^7$ cm$^{-3}$, it reaches a plateau and remains constant, $T=T_{ext}$, at lower densities.
Depending of the runs and the time the temperatures vary between 10 and up to $\simeq$ 30 K.

To interpret these temperatures, we developed a simple spherical model which is
presented in \S~\ref{temp_model}. Although we see from Fig.~\ref{fig_coldens} that the clouds 
are not spherical and that the sources are not clustered in the center as assumed in our model, 
this nevertheless allows us to get a deeper understanding of these temperatures.
The inferred powerlaw behaviours are as described by Eq.~(\ref{temp}) and Eq.~(\ref{Text}). 
More precisely, Eq.~(\ref{temp}) combined with Eq.~(\ref{rho_sis}) predicts that 
for $T > 100$ K, $T \propto n^{0.5}$ while for $T < 100$ K, $T \propto n^{0.3}$. 

To quantitatively estimate the values of $T_{\rm ext}$, we use Eq.~(\ref{Text2})

\begin{eqnarray}
T_{ext} =  33.5 K \; \left( {\tau_0 \over 0.5 }\right) ^{3/(4+2 \alpha)} \left(  {L_{tot} \over 10^5 L_\odot} \right)^{1/(4+2 \alpha)}
 \left( {\delta \over  100 } \right)^ {-1/(2+ \alpha)} 
 \label{Text_val}
\end{eqnarray}
where we remind that $\delta$ is as defined by Eq.~(\ref{rho_sis}) and $\tau_0$ is the optical depth at which the
radiation is free streaming. 

From Fig.~\ref{time_mass} and Fig.~\ref{fig_T}, we see that when $L_{tot} \simeq 10^5$ L$_\odot$, 
$T_{ext} \simeq 30$ K, whereas when $L_{tot} \simeq 10^4$ L$_\odot$ $T_{ext} \simeq 20$ K, which is close 
to what Eq.~(\ref{Text_val}) predicts.
Looking at Fig.~5 of \citet{elia2017}, we see that 20-30 K corresponds to the temperature of the warmest 
star forming clumps, which agrees well with the relatively high luminosities that we inferred.
The HiGAL-based temperature is the average termperature of the cold dust in a clump. They are derived from 160-to-500 (and 870, 1100, when available) $\mu m$ grey-body fit, so that probed temperatures cannot be higher than that. Since the mass in the outer part of the clump dominates, this cold component corresponds to that of the outer layers and of most of the volume  of the clump. This should therefore broadly correspond to what $T_{ext}$ is.

\subsection{Magnetic field distributions}
Figure~\ref{fig_B} portrays the volume weighted magnetic intensity  as a function of gas density
for the 4 magnetized simulations. Overall we see that, at least for $n$ between $10^7$ and $10^9$ cm$^{-3}$, 
the magnetic field scales with density broadly as 
$B \propto n^{1/2}$, a result observed in  previous works \citep[see for instance][for a review]{hinu2019}.
This is a consequence of the field amplification induced by field lines dragging by collapsing motion. 
Even more simply, this is likely a consequence of energy equipartition. As seen in Fig.~\ref{fig_radial}, 
$v^2$ depends weakly on $r$ while $n \propto r^{-2}$, therefore the kinetic energy scales as $r^{-2}$ 
and thus $B \propto r^{-1} \propto n^{1/2}$.
Interestingly, this implies that the Alfv\'en velocity, $V_a$, remains roughly constant in this range 
of density. 
We see however that its value is not identical for the four runs. We estimate that
for runs MHD10f05 and MHD10f01, $V_a \simeq 1-1.2$ km s$^{-1}$ while for run MHD100f05, $V_a$
is less than half this value. When non-ideal MHD is treated, the Alfv\'en velocity is reduced by 
tens of percents at $n \simeq 10^9$ cm$^{-3}$.

At lower densities, the behaviour depends on the field intensity. For run MHD100f05,
the dependence of $B$ on $n$, is a bit stiffer. This is expected as when the field is weak, 
the clump contraction tends to be spherical in which case $B \propto n^{2/3}$ \citep{li2015}. 
This explains why the magnetic field at high density in run MHD100f05 is larger than 
a tens of the $B$ values in run MHD10f05. Magnetic intensity is more vigorously amplified
when it is weaker. 

At high densities, i.e. $n > 10^9$ cm$^{-3}$, the magnetic field is further amplified up to density
values on the order of $10^{11}$ cm$^{-3}$. The highest magnetic intensities vary from one run to the other.
In the most magnetized runs, MHD10f05 and MHD10f01, it reaches $\simeq$100 G and about one third of this 
in run MHD100f05. 
Run NMHDf05 presents different behaviour. For $n > 10^{10}$ cm$^{-3}$, the intensity is nearly independent of $n$ and 
the largest intensities is about $\simeq$30 G. This behaviour, which has been discussed previously \citep[e.g.][]{masson2016,wurster2018} is 
a consequence of ambipolar diffusion, that tends to  diffuse the field. This implies that the influence of 
magnetic field on the high density gas is significantly reduced compared to ideal MHD runs.

\setlength{\unitlength}{1cm}
\begin{figure*}
\begin{picture} (0,6)
\put(0,0){\includegraphics[width=8cm]{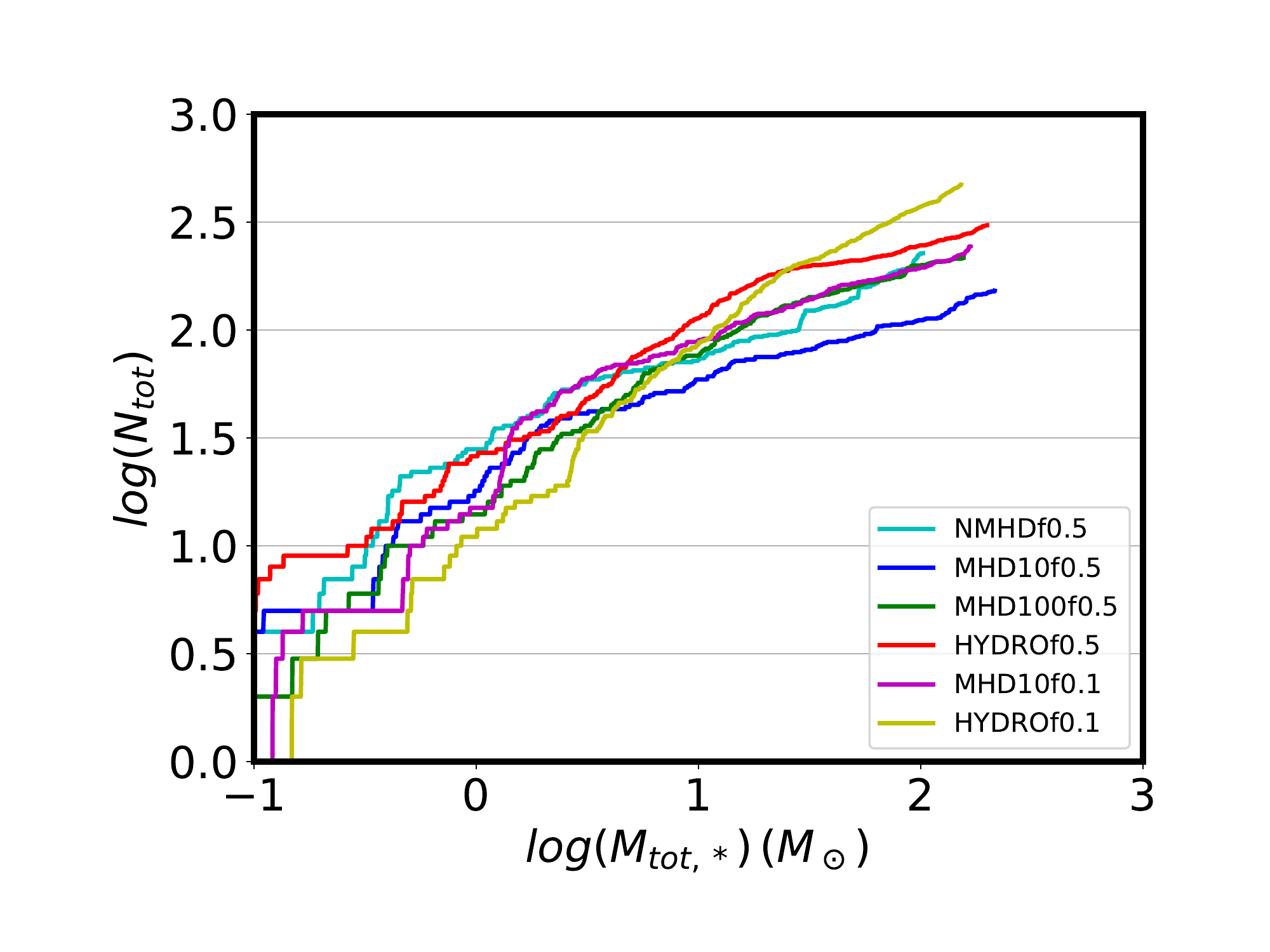}}  
\put(8,0){\includegraphics[width=8cm]{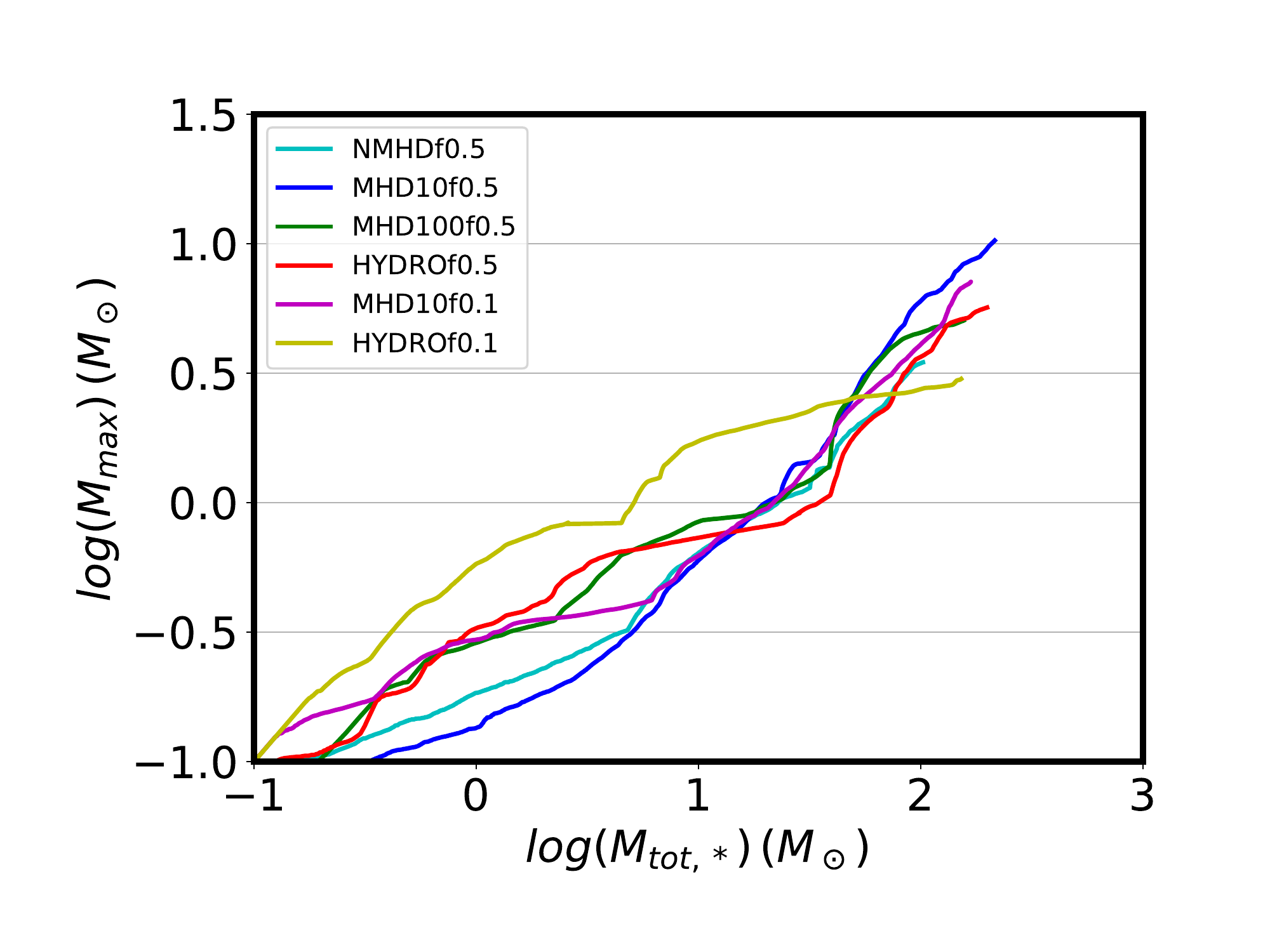}}  
\end{picture}
\caption{The  number of sinks (left panel), $N _{tot} $
and the largest sink mass (right panel)
as a function of the total accreted mass, $M_{tot}$, in several runs. 
}
\label{ntot_massmax}
\end{figure*}

\section{Stellar mass spectrum}

\subsection{Fragmentation and massive stars}
\label{fragmentation}
Figure~\ref{ntot_massmax} portrays the number of sink particles as a function of accreted
mass
(left panel) as well as the mass of the most massive star (right panel).

The number of sinks at the end of the simulations is typically between 100 and 300 depending 
of the runs. As anticipated from the clump images, both  magnetic field and radiative feedback 
reduce fragmentation. Here we see that the differences between runs HYDROf05 and MHD10f05 or 
between HYDROf01 and MHD10f01 is about a factor of 2, the difference being more pronounced 
for the two runs with $f_{acc}=0.5$. On the other hand  the differences between runs MHD10f05 and 
MHD10f01 is on the order of 50$\%$, showing that whereas radiative feedback contributes to reduce 
fragmentation, its effect is comparatively lower than magnetic field. 
Indeed, although the initial magnetisation of run MHD100f05 is quite weak, it nevertheless reduces 
the fragmentation by tens of percents compared to run HYDROf05. Interestingly run NMHDf05, that 
treats ambipolar diffusion and has the same magnetisation than run MHD10f05, presents a number 
of sinks similar to run MHD100f05. 

In all runs but HYDROf01, two phases can be distinguished. When $M_{tot,*}$ is smaller than 
$\simeq 3$ M$_\odot$ ($\simeq 10$ for run HYDROf05), the number of sinks increases fast and 
is nearly proportional to $M_{tot,*}^{m_{tot}}$ with $m_{tot} \simeq 0.7-1$. Beyond this value, the number of sinks increases 
much less  rapidly, and typically $m_{tot} \simeq 0.2-0.3$. For instance for run MHD10f05, the number of sinks has roughly doubled 
between the time when  $M_{tot,*}=10$  M$_\odot$  and $M_{tot,*}=100$ M$_\odot$. This is most certainly 
related to radiative feedback and to the global increase of temperature within the clumps. 
The consequence is obvioulsy that the sink particles, build their masses in this second phase 
after fast fragmentation has occured.

At the end of the runs, the mass of the most massive star is between $3$ and $10$ M$_\odot$. 
The observed trends are in good agreement with the sink numbers. The mass of 
the most  massive star
is higher when magnetic field and radiative feedback are larger and magnetisation 
is comparatively slightly more efficient than radiative feedback in producing massive stars. 
Two phases of growth can also be distinguished, typically below and above $M_{tot,*} \simeq 10$ M$_\odot$,
 where $M_{max}$ grows respectively slowly and fastly. We observe that $M_{max} \propto M_{tot,*}^{m_{max}}$
 with ${m_{max}} \simeq 0.5$ when $M_{tot,*} < 10$ M$_\odot$ while ${m_{max}} \simeq 1$ otherwise.

Note an important feature of the stellar mass distribution is that in a group of stars 
which in total contains about 100-120 $M_\odot$, 
a star more massive than 8 M$_\odot$ is expected. We see that in our simulations only runs 
MHD10f05 and MHD10f01 have reached this value. Runs HYDROf05 and MHD100f05 are slightly below
while run HYDROf01 is almost a factor 3 below. This may constitute a hint that magnetic field 
is playing a role regarding the building of the massive stars, essentially by reducing 
the cloud fragmentation.

\setlength{\unitlength}{1cm}
\begin{figure*}
\begin{picture} (0,17)
\put(8,11){\includegraphics[width=8cm]{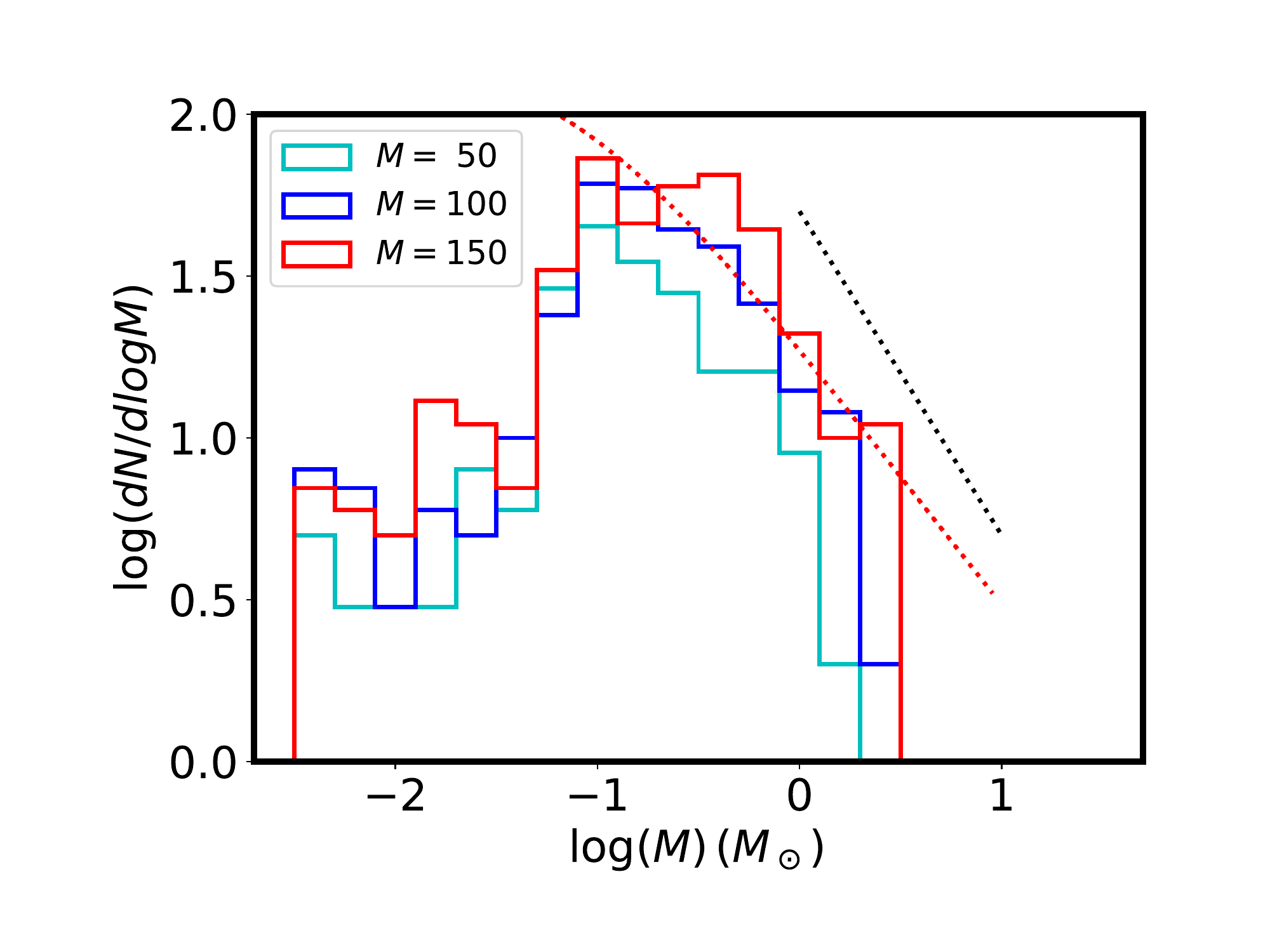}}
\put(10,16.6){HYDROf01}
\put(0,11){\includegraphics[width=8cm]{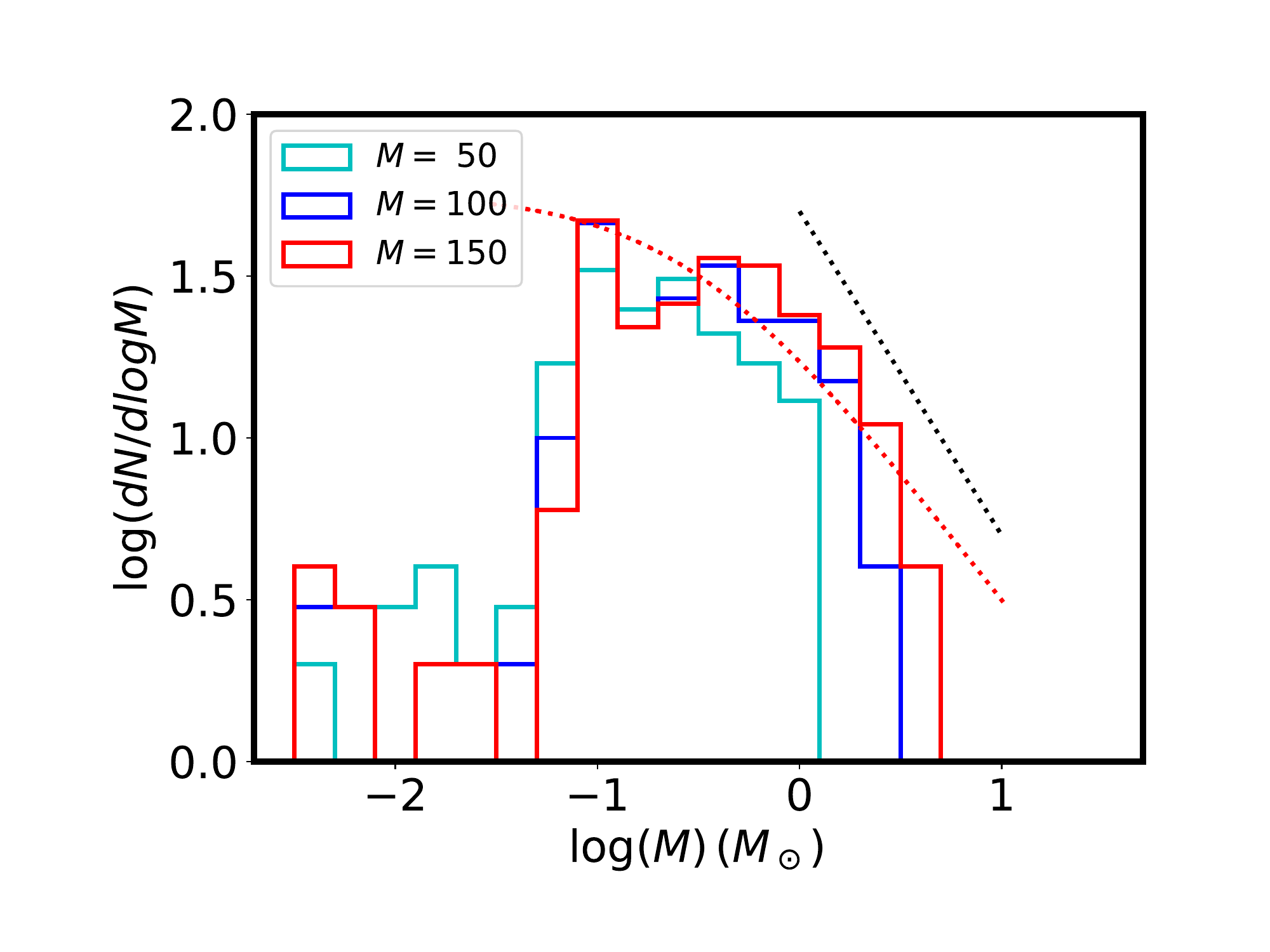}}  
\put(2,16.6){HYDROf05}
\put(8,5.5){\includegraphics[width=8cm]{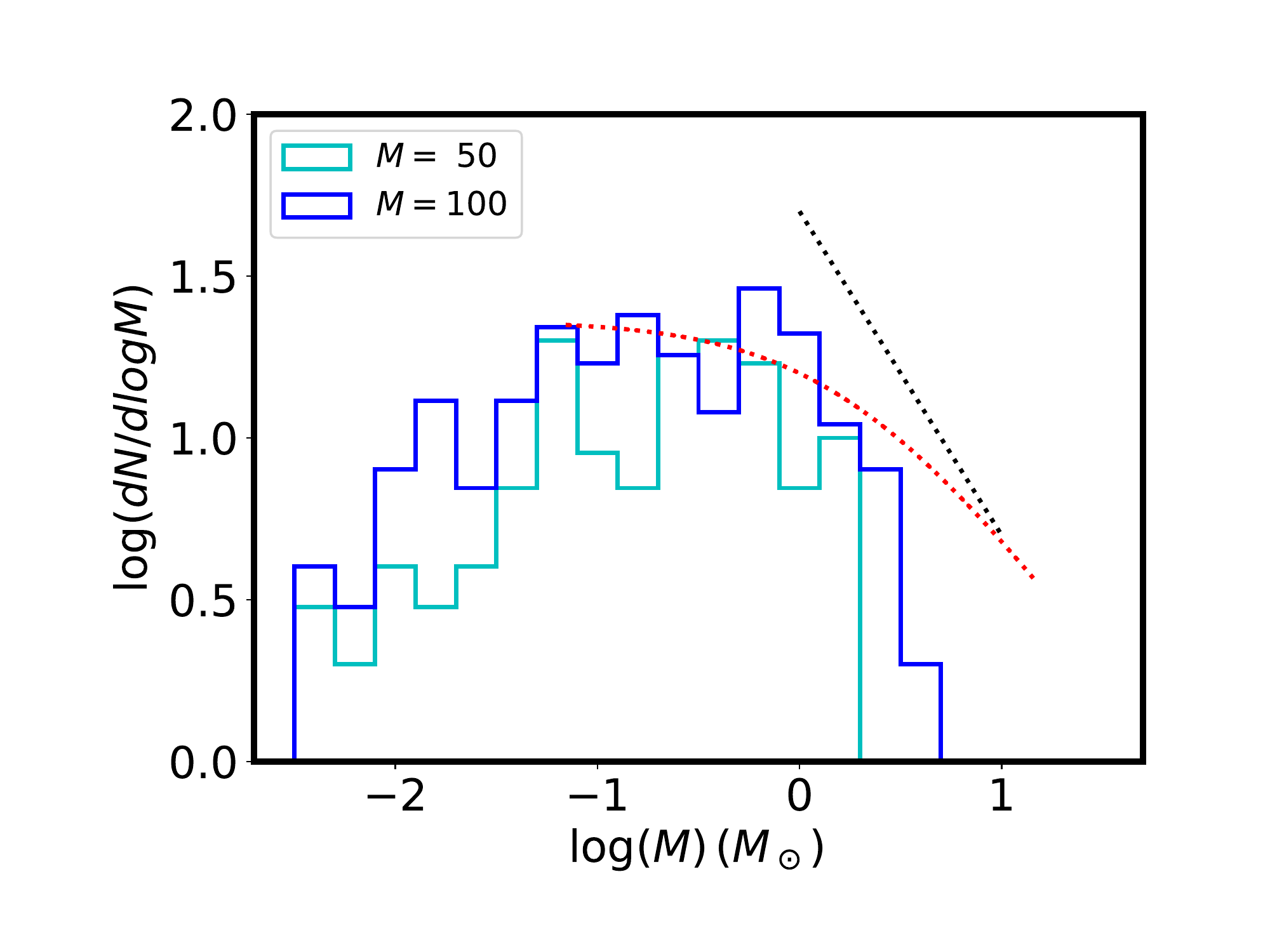}}
\put(10,11.1){NMHD10f05}
\put(0,5.5){\includegraphics[width=8cm]{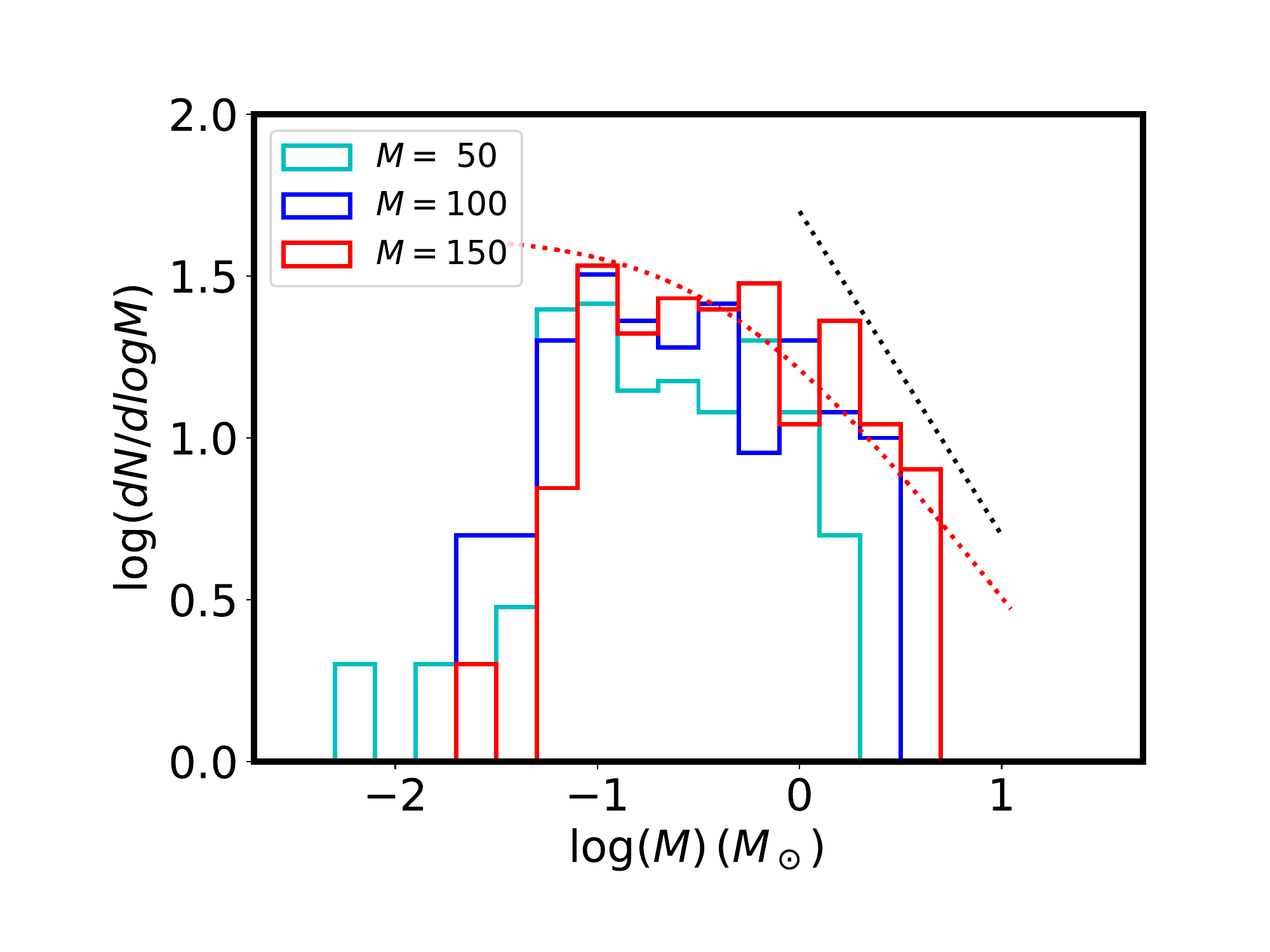}}  
\put(2,11.1){MHD100f05}
\put(8,0){\includegraphics[width=8cm]{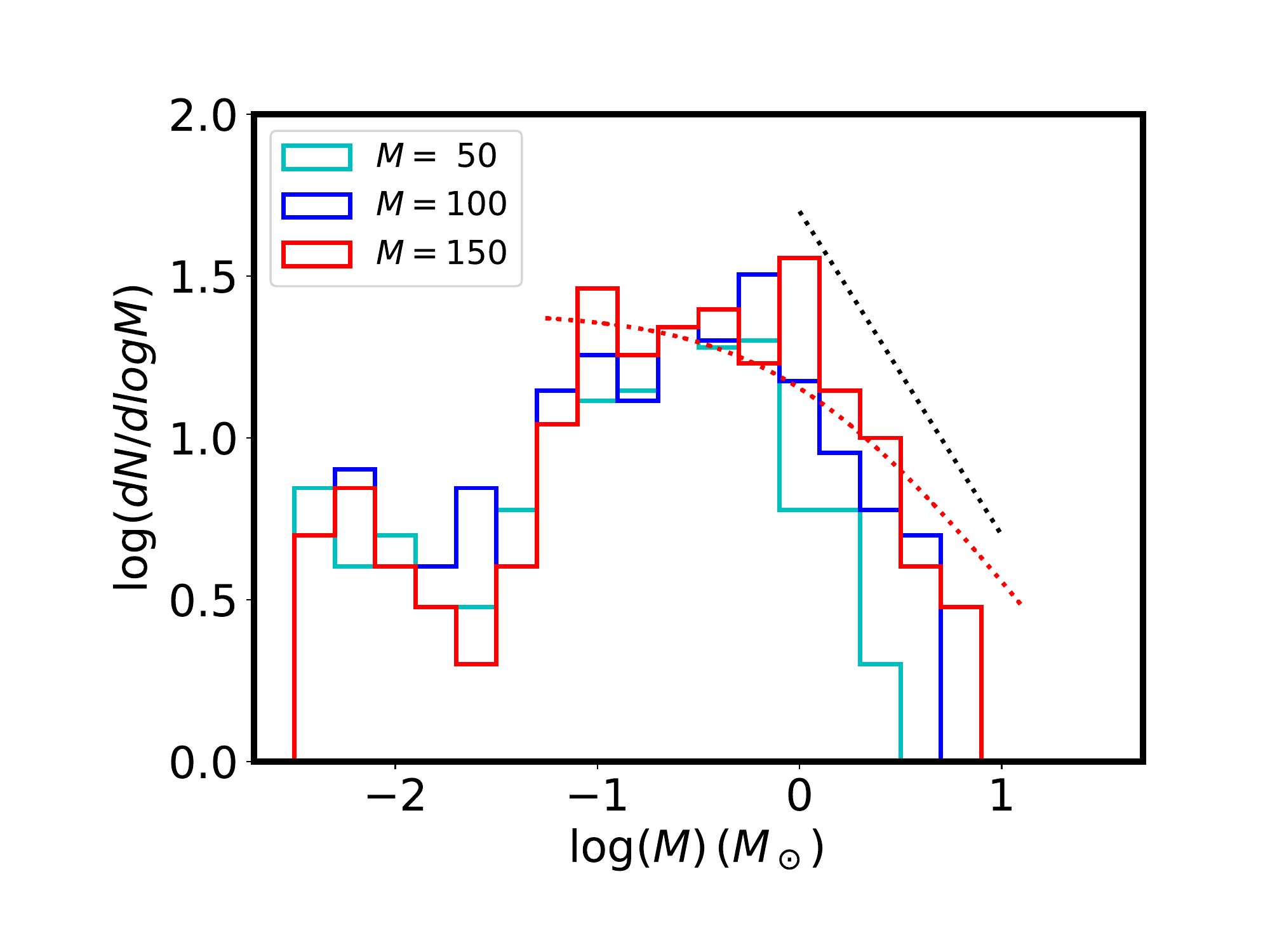}}  
\put(10,5.6){MHD10f01}
\put(0,0){\includegraphics[width=8cm]{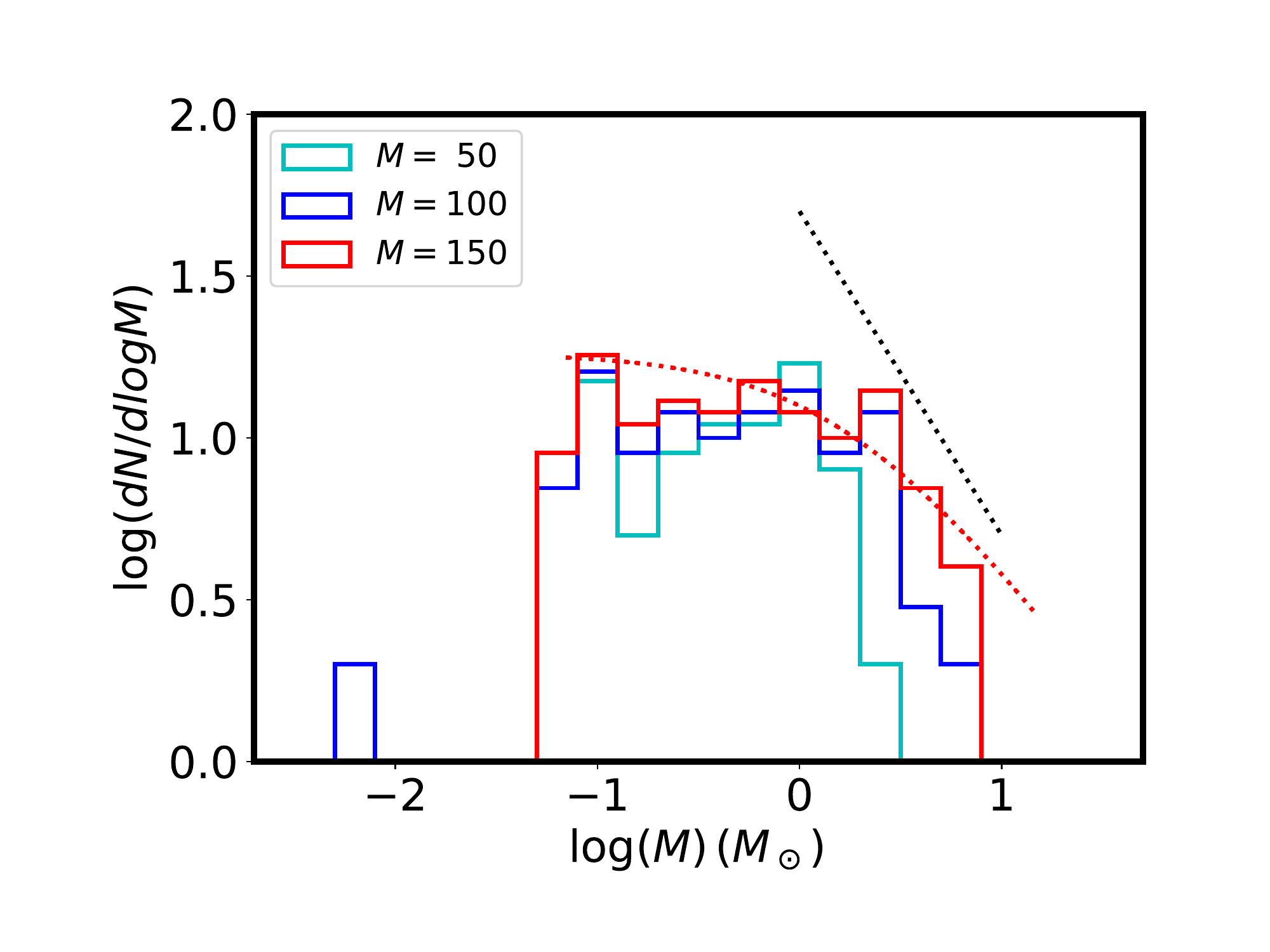}}  
\put(2,5.6){MHD10f05}
\end{picture}
\caption{Mass spectra at various times characterised by the total accreted mass, for the six runs and for three values of the 
accreted mass. The red dotted lines represent the analytical model presented in the paper for comparison. 
The black dotted one shows for reference a $M^{-1}$ power laws.
}
\label{fig_imf}
\end{figure*}

\subsection{The sink mass function}
Figure~\ref{fig_imf} displays the sink mass function, ought to represent the initial mass function, 
for the six runs and 3 values of $M_{tot,*}$. 

\subsubsection{Analytical model}
Before presenting the stellar mass spectra induced from the simulations, we discuss an analytical model that will be useful to interpret the results. It is in essence the model proposed in \citet{HC08} in which the 
density PDF is the one appropriated to the gravitational collapse and stated by Eq.~(\ref{fig_pdf_rho})
as proposed in \citet{leeh2018a}. For the sake of completeness, it is described in appendix~\ref{model_imf}.
Equations~(\ref{crit_Mtot}) and~(\ref{spec_mass2}) are the final equations to be used. 
Let us remember that the model predicts two asymptotic behaviours. At small mass, 
when thermal and/or magnetic support dominates, $\Gamma \rightarrow 0$, while at larger mass, when 
turbulent support dominates, $\Gamma \rightarrow 3/4$. The transition between these 
two regimes occurs at scales or equivalently masses (see Eq.~\ref{crit_Mtot} )  for which  thermal/magnetic and turbulent supports
are comparables.

In order to be compared with the numerical simulations, one needs to specify the values of the sound speed $c_s$, 
of the Alfv\'en speed, $V_a$, and of the turbulent velocity dispersion, $V_0$. All these values can be inferred from the results
presented in \S~\ref{gas_res}. Another important point when comparing simulations with the analytical model is the normalisation. For this purpose, we write 
\begin{eqnarray}
\nonumber
M_{tot,*} &=& \int _{M_{min}}  ^{M_{max}} M {\cal N}(M) dM \\
&=& \int _{\log_{10}(M_{min})}  ^{\log_{10}(M_{max})} 
M {  \cal N}(M)   {M \over \log 10} d \log_{10}M.
\end{eqnarray}
Thus, ${\cal N}_0$ as defined by Eq.~\ref{spec_mass2}, 
 is determined once $M_{tot,*}$, $M_{min}$ and  $M_{max}$ are specified. The various parameters are reported in table~\ref{param_mod}. Since 
$c_s$, $V_a$ and $V_0$ are all evolving with time and positions, the reported values are global estimates.

      \begin{table*}
         \begin{center}
            \begin{tabular}{lccccccc}
               \hline\hline
               Name & $R_c$ (pc) & $c_s$ (km s$^{-1}$) &  $V_a$ (km s$^{-1}$)& $V_0$ (km s$^{-1}$)& $M_{tot,*}$ (M$_\odot$) & $M_{max}$ (M$_\odot$)   & $ M_{min}$ (M$_\odot$)  \\
               \hline
               NMHD10f05  & 0.3  &  0.35     & 1 & 3     & 100   & 3  & 0.1   \\
               IMHD10f05  & 0.3   &  0.35    & 1 & 3     & 150  &  8  & 0.1  \\ 
               IMHD10f01  & 0.3  &  0.25    & 1 & 3     & 150   & 7  & 0.1   \\ 
               IMHD100f05  & 0.3  &  0.35   & 0.3 & 3     & 150   & 7  & 0.1   \\ 
               HYDROf05  & 0.3  &  0.35      & 0  & 3     & 150   & 7  & 0.1   \\ 
               HYDROf01   & 0.3   &  0.25      & 0  & 3     & 150   & 3  & 0.1  \\ 
            \end{tabular}
         \end{center}
         \caption{Parameters used to confront the stellar initial mass function inferred from 
         the simulations with the analytical model stated in \S~\ref{model_imf}.
         $R_c$ is the clump radius, $c_s$ is the typical sound speed, $V_a$ the Alfv\'en
         speed, $V_0$ the velocity dispersion, $M_{tot,*}$ is the total accreted mass,
         $M_{max}$ the mass of the most massive stars formed and $M_{min}$ the 
         smallest mass for which the comparison is meaningful.  }
\label{param_mod}
      \end{table*}

We recall that the model is isothermal in nature. The
  sound speed may vary for instance over time but remains
 uniform within the whole cloud.  
This has an important consequence, which is that the model does not predict a  minimum stellar mass.
One should however remember that the isothermal assumption becomes invalid when 
the density reaches density on the order of $10^{10}$ cm$^{-3}$ when the gas becomes 
progressively adiabatic. As discussed in \citet{hetal2019}, the change of thermal behaviour, which leads
to the formation of the first  hydrostatic core \citep{larson1969}, results in a peak/cut-off for 
the IMF at typically several times the mass of the first  hydrostatic cores, $M_L$ (that we recall is about 
$M_L \simeq$0.03 M$_\odot$). This implies that the analytical model is  valid for masses larger than a few times $M_L$ and this is why we choose $M_{min}=0.1$ M$_\odot$.
The values of $M_{max}$ are taken from Fig.~\ref{ntot_massmax}.

\subsubsection{The hydrodynamical runs}
The two top panels of Fig.~\ref{fig_imf} show results for run HYDROf05 and HYDROf01. 
The mass spectra are similar to those obtained in \citet{hetal2020b} with slightly different initial conditions and 
less spatial resolution. Essentially most of the sinks have their mass between few 10$^{-2}$ and few M$_\odot$. 
The distributions present a plateau that ranges between $\simeq$0.1 and $\simeq$0.5 M$_\odot$. A relatively sharp drop occurs around 0.1  M$_\odot$ and we get a small number of objects at lower mass, particularly in run 
HYDROf05. At mass larger than $\simeq$ 0.5 M$_\odot$, the distribution drops following a powerlaw-like 
behaviour, whose index cannot be reliably determined due to the lack of statistics. A tentative M$^{-1}$
distribution (dotted line) is represented for comparison. This value is similar to what previous authors
have inferred from simulations \citep{bonnell2011,girichidis2011,BallesterosParedes15,leeh2018a,leeh2018b,padoan2020}.
Overall we see that there is a good agreement between the analytical model 
(red dotted line), 
 and the sink mass distribution, or M$ > 0.1$ M$_\odot$.
We stress that the main effect of increasing radiative feedback is to broaden the distribution toward larger masses. From the analytical model, we see 
that this is compatible with this being a consequence of the
 mean cloud temperature increasing due to the radiative heating. 

The peak of the distribution however is barely affected. This confirms, as claimed in \citet{hetal2020b} that radiative feedback is not responsible of setting the peak of the IMF. 
In fact, at early time (total accreted mass of 50 M$_\odot$), the distribution is clearly peaked toward 
0.1-0.2 M$_\odot$ which is several time the mass of the first hydrostatic core. As time goes on, the mass
of the most massive stars increases while the number of low mass objects remains constant or increases 
moderately. This is entirely compatible with the idea that the stars inherite a minimum 
mass reservoir equal to a few times the mass of the first hydrostatic core \citep{hetal2019,colman2019}, which 
is fastly accreted. After this, the stars keep accreting from their mass reservoir which likely 
is set by gravo-turbulence \citep{padoan1997,HC08,hopkins2012}. 
While this process should largely be deterministic
in nature, it is also likely the case that stochastic processes modulate this accretion as well \citep{bonnell2001,basu2004,basu2015}.

\subsubsection{The influence of magnetic field on the stellar mass spectrum}
The influence of magnetic field can be seen by comparing on one hand runs HYDROf05, MHD100f05 and MHD10f05 
and on the other hand run HYDROf01 with run MHD10f01. 
Clearly,  magnetic field has a significant impact on the mass spectrum, that it tends to broaden towards
larger masses. In fact, the low mass distribution is almost unchanged. Again this provides further confirmation 
that radiative feedback has no significant impact on the low mass end
 of the stellar initial mass function since 
as discussed above magnetic field leads to stronger radiative feedback. This also obviously shows that magnetic 
field does not influence the low mass end of 
mass spectrum in good agreement with the idea that it is mainly linked to the 
mass of the hydrostatic core. 

Run MHD10f05 presents a plateau that extends from about  0.1 $M_\odot$ to $\simeq 2 M_\odot$. 
It is reminiscent of run A presented in Fig~6 of \citet{leeh2018a} and in the run presented
 in  Fig~2 bottom panel of \citet{jones2018}. These runs have in common to have a high thermal energy 
 initially, or equivalently a low Mach number. The analytical model 
suggests that when thermal support is high, a collapsing clump would indeed develop a
stellar mass spectrum $dN / d\log M \propto M^0$, while when turbulent support dominates the support of 
the mass reservoir,
$dN / d\log M \propto M^{-3/4}$ is expected.
 Likely enough run MHD10f05 falls in the regime where 
thermal and magnetic field dominates over turbulence at the scale of the mass reservoirs and this 
explains the flat mass spectrum. This is indeed what the good agreement with 
the MHD models and the simulations suggests since the broad plateau (where 
 $dN / d\log M \propto M^0$)
displayed by the analytical models is due to combination of a high 
Alfv\'en velocity and a high sound speed.

Compared to run MHD10f05,  the mass spectum of run MHD10f01
presents a plateau that is less broad. This is the case
both for the numerical and the analytical models, which are 
again in good agreement. This clearly is due to the lower 
temperatures in run  MHD10f01, which compared to run MHD10f05,
leads to weaker thermal support. 

\subsubsection{The impact of ambipolar diffusion}
The mass spectrum of run NMHDf05 presents similarities with 
the one of run MHD10f05 but also significant differences. Overall 
it is more similar to the mass spectrum of run MHD10f01.
First of all, unlike run MHD10f05, it does not present a plateau that 
extends up to $\simeq$3 M$_\odot$ but rather stops at 1 M$_\odot$ and the 
most massive stars are also less massive. This is in good agreement 
with the slightly lower magnetic field which is found for
run NMHDf05 (see Fig.~\ref{fig_B}) than for run MHD10f05. The similarity 
with run MHD10f01 likely comes from the total support due 
to both thermal and magnetic supports are closer because run MHD10f01
has stronger field but lower temperatures than run NMHDf05. 

A more surprising difference comes from the low mass objects. As can be seen 
there are more sink particles of masses lower than 0.1 M$_\odot$ in run NMHDf05 than 
in the  ideal MHD runs but also more than in the hydrodynamical runs.
The reason for this remains to be clarified. The most likely 
explanation is the relatively  weak magnetic field 
intensity at density above 10$^{10}$ cm$^{-3}$ in run NMHDf05
compared for instance to run MHD10f05. As seen from Fig.~\ref{fig_B}, the 
change of behaviour is relatively sharp, with $B$ being very comparable 
in runs NMHDf05 and MHD10f05 below 10$^{10}$ cm$^{-3}$. Thus while in both runs, high densities may develop due  to field support, the field support drops
at density above 10$^{10}$ cm$^{-3}$ for run  NMHDf05 and this may favor fragmentation. This may also be due to the difference in the disk 
populations that form in the various runs and presented in \citet{lebreuilly2021}.
The disks formed in non-ideal MHD runs are intermediate in mass and size
between the hydrodynamical disks and the the ones which form in ideal MHD runs. 
While the latter are usually very stable due to the fast growth of a toroidal 
magnetic component, the former fragments but since more mass is available in bigger disks, they tend to form bigger objects than in non ideal MHD disks.

\section{Discussions}

\subsection{Dependence of the high-mass slope of the stellar mass spectrum}

As discussed in the previous section, our numerical results suggest that 
from a few solar mass to at least 7-8 M$_\odot$, the stellar 
distribution presents a power law behaviour $d N / d \log M \propto M^{-\Gamma}$, 
with $\Gamma \simeq 3/4$. Analytically, this behaviour  is found when 
at the scale of the individual mass reservoir, $i)$ the dominant support against gravity 
is turbulence and $ii)$ when the density PDF is $\propto \rho^{-3/2}$ which 
is a consequence of gravitational collapse.  On the other-hand, when the density PDF is close to a lognormal 
distribution, we do expect $\Gamma \simeq 1.3$ as discussed in \citet{leeh2018a}. 
 In essence, the density PDF is a direct estimate of how the gas mass is distributed amongst 
densities and therefore controls the number of density fluctuations at a given density. Typically a 
log-normal distribution has less dense gas than a PDF $\propto \rho^{-3/2}$ and therefore 
less small mass objects are produced with the former than with the latter.
The transition between 
the two exponents, $\Gamma=3/4$ and $\Gamma=1.3$, is expected to occur at the density, $n_{trans}$, which typically 
connects the turbulent log-normal PDF to the  power law  $\propto \rho^{-3/2}$ gravitational  PDF. 
In the present simulations this occurs around $\simeq 10^6$ cm$^{-3}$.
Combining Eqs.~(\ref{masse}) and~(\ref{crit_Mtot}), we can estimate the mass, $M_{trans}$ it corresponds to
\begin{eqnarray}
\nonumber
M_{trans} \simeq  15 \; M_\odot \left( { V_0 \over 3 \; {\rm km \, s^{-1} } }\right)^6 \left( { R_c \over 0.3 {\rm pc} }\right)^{-3}
\left( { n_{trans} \over 10^6 \; {\rm cm^{-3}} }\right)^{-2}  \\
\simeq 15 \; M_\odot \left( { M \over 10^3 \; {M_\odot } }\right)^3 \left( { R_c \over 0.3 {\rm pc} }\right)^{-6}
\left( { n_{trans} \over 10^6 \; {\rm cm^{-3}} }\right)^{-2}.
\end{eqnarray}
 where for simplicity, we have assumed $\eta=0.5$ and $V_0^2 \simeq G M / R_c$.
Because of the sixth power which appears for $V_0$ or $R_c$, the clump radius, 
the value of $M_{trans}$ is clearly not accurate and likely can abruptly change from 
one environment to another. Typically we expect a fast transition around $R_c \simeq 0.3$ pc.  
It is however 
illustrative and shows that for our simulations, at high mass, the mass spectra are expected to be mostly if not 
exclusively described by the $\Gamma=3/4$ exponent since our stellar masses are smaller than 15 M$_\odot$.
It also shows that in a less dense and compact clump, the transition should occur at smaller masses since the 
value of $V_0$, or equivalently the value of $R_c$, should be smaller. 
While most of the studies which have started from  massive clumps, comparable to the ones studied here, tend
to present $\Gamma$ lower than the canonical Salpeter exponent \citep[see the discussion in][]{leeh2018a}, works in which the IMF is obtained from
larger scale clouds 
studies which have attempted to obtain the IMF in larger scale simulations with initial conditions that correspond to 
more standard giant molecular clouds, generally report $\Gamma$ values that are closer to 1.3. 
This is the case for instance for the run XL-F presented in Fig.~4 of \citet{chong2019} and the run presented 
in Fig.~3 of \citet{padoan2020} for masses between 10 and 50 M$_\odot$, respectively. 
This is also the case for the runs presented 
in \citet{ntormousi2019} and the core mass function extracted from these simulations \citep{louvet2021}.

\subsection{Observationally inferred mass distribution in actively star forming  regions} 

While it may sound at first surprising not to find $\Gamma \simeq 1.3$, which is 
the slope inferred by \citet{salpeter55}, it should be stressed that recent observations 
have been inferring that in some actively star forming regions, the IMF may indeed be top-heavy \citep{zhang2018,lee2020}. More precisely, in the Arches cluster \citet{hosek2019}
inferred $\Gamma \simeq 0.8$. On the other hand, recent studies of the core 
mass function also obtained within massive star forming regions, have also inferred 
power law behaviours with indices $\Gamma \simeq 0.95$ \citep{motte2018,pouteau2022}. 
As cores are widely assumed to be the progenitors of stars out of which they build their mass, the  inferred $\Gamma$ are compatible with the idea that the shape of the 
IMF in massive star forming regions is inherited from the shape of the CMF, at least at high masses, although eventually it should be compared with the IMF 
of the very same region. 

While more detailed investigations, including careful comparisons between simulations and observations must be carried out before firm 
conclusion can be drawn, there is a clear suggestion coming from both observations and theories that systematic variations of the IMF 
may occur, particularly in very compact star forming regions.

\subsection{Limits of the present work and the {\it universality} of the IMF }
Our work presents several important limits that need to be discussed. 
Indeed, one of the conclusion is that the combination of 
magnetic field and radiative transfer possibly leads to more 
variability that what observational inferences of the IMF 
may have led to conclude. Admittedly, this question even for our own 
Galaxy remains difficult to address, particularly because of the relatively limited samples
that are often available but it seems nevertheless unavoidable that at least 
some level of fluctuations should be present \citep[see for instance the comprehensive discussion provided in][]{dib2022}.

Determining whether the variations observed in the present work are compatible with the galactic fluctuations of the IMF,  is beyond the scope of the present paper but 
it is worth to remind that an important source of variations is due to the efficiency of the accretion 
luminosity expressed by the parameter, $f_{acc}$.  Whereas there may be some variability of 
$f_{acc}$, likely enough it is not a factor of 5 as we have been exploring here. The other 
possibly extreme variations we have considered is magnetic intensity since we 
have explored a factor of 10 (and even go to pure hydrodynamical cases). This is not 
well constraint yet but a 1000 M$_\odot$ clump is a relatively large ensemble and 
it is unclear what are the variations of the magnetisation in the galactic populations. 

Finally, we stress that in this work a possibly important process has been omitted, namely 
the protostellar jets. Recently, \citet{gus2021} have been exploring their impact 
in simulations comparable to the ones presented here (with a resolution of few tens of AU).
They concluded that protostellar jets may be playing a significant role in setting the IMF
in particular for the formation of low mass objects in the presence of a
significant initial magnetic field. Whether this process may help explaining 
the universality of the IMF is however not clear yet.

\section{Conclusions}

With the goal of understanding how magnetic field and radiative feedback 
influence the collapse and the fragmentation of a massive star forming clump, 
we have performed high resolution adaptive mesh calculations with a 
spatial resolution down to about 1 AU. 
Six runs in which  2 radiative feedback efficiencies, 3 magnetic intensities
as well as the impact of non-ideal MHD are explored. 
We show that the physical characteristics of the simulated star forming clumps 
compare well with various observations. This is for instance the case for the 
observational bolometric luminosities that we compared with the total 
luminosities of the sink particles produced in the simulations as well as for the 
gas temperatures. For the latter, we develop an analytical model which 
agrees well with the temperatures inferred from the simulations.

The stellar mass spectra of the six runs are analysed in detail and compared with 
an analytical model in which thermal, magnetic and turbulent supports are playing a major role. Overall the analytical model reproduces well the numerical 
mass spectra for masses above $\simeq 0.1$ M$_\odot$. At this mass 
which corresponds to a few times the mass of the first hydrostatic core
the underlying gas thermodynamics is nearly adiabatic and specific 
models should be considered \citep[e.g.][]{hetal2019}.
The combination between 
simulations and analytical results allows us to clearly assess the role and influence of each physical process which are as follows:
\begin{itemize}
\item[-] in the density range at which the gas is not adiabatic, the density PDF which is $\propto \rho^{-3/2}$ is deeply shaping the stellar mass spectrum and leads to two physical distinct regimes for the mass spectra. 
\item[-] at masses larger than $\simeq$ 0.1 M$_\odot$, thermal pressure and magnetic field may lead to a flat mass spectrum, i.e. $d N / d \log M \propto M^{-\Gamma}$ with $\Gamma \simeq 0$ if they are strong enough compared to turbulence.
\item[-] at larger scales, turbulence dominates and may lead to a mass 
spectrum with $\Gamma \simeq 3/4$. At even larger scales and lower density, the PDF is expected to be log-normal in shape and stiffer mass spectra, with larger $\Gamma$ are expected.
\item[-] the transition between the regime with 
$\Gamma \simeq 0$ and $\Gamma \simeq 3/4$ is not universal and depends
on the local physical processes such as thermal support, magnetic field and Mach number.
\end{itemize}

 Generally speaking, we find that the main effect of magnetic field and radiative transfer is  to reduce the 
total number of fragments and to increase the mass of the most massive stars. These latter have been found to  increase with the 
magnetic intensity and the radiation feedback efficiency. 
For instance, in the present work we found that for the hydrodynamical simulation with the lowest efficiency, 
the most massive star produced after 150 M$_\odot$ have been accreted, is about 3 M$_\odot$. With a higher radiative feedback efficiency or a sufficiently strong initial field, stars  of masses 7-8 M$_\odot$ are produced. 
We therefore conclude that whereas magnetic field and radiative feedback may not be essential to explain the peak or the various slope values of the IMF, they may be essential 
to reproduce the exact shape (like the transition between the various regimes), the level of fragmentation i.e. the number of stars formed,  and  the mass of the most massive stars.



\begin{acknowledgements}
We thank the anonymous referee for a useful report. 
This work was granted access to HPC
   resources of CINES and CCRT under the allocation x2014047023 made by GENCI (Grand
   Equipement National de Calcul Intensif). 
   This research has received funding from the European Research Council
synergy grant ECOGAL (Grant : 855130).
G.A.F also acknowledges support from the Collaborative Research Centre 956, funded by the Deutsche Forschungsgemeinschaft (DFG) project ID 184018867.
\end{acknowledgements}

\bibliography{lars}{}
\bibliographystyle{aa} 

\appendix

\section{A simple analytical model for the temperature}
\label{temp_model}
To better understand the  temperature profils through the clump, we make use of the simple model
discussed in \citet{hetal2020b}, improving on various aspects. 
The model assumes that the cloud is spherically symmetric and that all sources are located 
in the clump center. 
As seen in   \S~\ref{prof}, the radial profil of the density field in  collapsing envelopes is given by
\begin{eqnarray}
\label{rho_sis}
\rho (r) = {\delta _\rho  c_{s,0}^2 \over 2 \pi G r^2},
\end{eqnarray}
where $\delta _\rho$ is a dimensionless factor which typically is equal to $\simeq$30-100.
We assume that gas and dust have the same temperature and are
stationary. The grey body approximation that is being used,
 leads when  the medium is optically thick to
\begin{eqnarray}
\label{transfert}
- 4 \pi r^2  {c \over 3 \kappa (T) \rho (r) } \partial _r (a T^4) = \sum (L_* + L_{acc}).
\end{eqnarray}
where $a$ is the radiation density constant. 
In this expression we assume that all the emitting sources are located in the clump center.
The opacity temperature  dependences  \citep[e.g.][]{semenov2003}, suggest that we can distinguish two regimes of temperature,
\begin{eqnarray}
\label{kappa}
\kappa (T) = \kappa _0 \simeq 5 \, {\rm cm}^2 {\rm g}^{-1} \; {\rm for} \; T > T_{\rm crit} \simeq 100 \,  {\rm K} , \\
\kappa (T) \simeq \kappa _0 \left( { T \over T_{\rm crit}} \right)^{\alpha} \; {\rm for} \; T < T_{\rm crit}.
\nonumber
\end{eqnarray}
where $\alpha$ is typically between 1 and 2. In this work we adopted $\alpha=1.5$.
Combining Eqs.~(\ref{rho_sis}),~(\ref{transfert}) and~(\ref{kappa}), we get

\begin{eqnarray}
\label{temp}
\nonumber
 T(r) = \left( T_{\rm crit}^4 + K \left( {1 \over r^3} - {1 \over r_{\rm crit}^3}  \right)  \right)^{1/4} \; {\rm for} \; T > T_{\rm crit}, \\
 T(r) = \left(  K T_{\rm crit}^{-\alpha} {4 - \alpha \over 4}   {1 \over r^3}     \right)^{1/(4-\alpha)} \; {\rm for} 
 \; T < T_{\rm crit} \; {\rm and} \; r < r_{\rm ext},
\end{eqnarray}
\begin{eqnarray}
K = f_{\rm acc} \delta _\rho    {3 \kappa _0  c_{s,0}^2  \over 24 \pi ^2 a c G  }   \sum (L_*+L_{acc}),
\label{Brad}
\end{eqnarray}
where $r_{\rm crit}$ is the radius at which $T=T_{\rm crit}$ and is given by
\begin{eqnarray}
\label{rcrit}
r_{\rm crit} ^3 =   K   {4 - \alpha \over 4}   { T_{\rm crit}^{-\alpha} \over   T_{\rm crit}^{4-\alpha}  -  T_{\rm ext}^{4-\alpha}  }.
\end{eqnarray}
Finally, $T_{ext}$ is  the temperature where  the optical depth is about 1 and $r_{ext}$ the corresponding radii.
We therefore have $\kappa (T_{\rm ext})  \rho(r_{\rm ext}) r_{\rm ext} \simeq \tau _0$ where $\tau_0$ should be 
on the order of 1.
Combining Eqs.~(\ref{rho_sis}),~(\ref{kappa}),~(\ref{temp}), we obtain for $r_{\rm ext}$
\begin{eqnarray}
\label{rext}
r_{\rm ext} =  \left( {  \kappa_0 \delta c_{s,0}^2 \over 2 \pi G \tau_0 } \right) ^{(4-\alpha)/(4+2 \alpha)}  \left( {  K \over T_{\rm crit}^4 } {4 - \alpha \over 4} \right) ^{\alpha/(4+2 \alpha)}.
\end{eqnarray}
At this point, the radiative flux becomes simply equal to the term $c E_R$ and Eq.~(\ref{transfert}) becomes invalid.
Under the assumption that the temperature remains the one of a blackbody, it then remains constant at larger radii and thus
\begin{eqnarray}
 T(r) = T_{\rm ext} = \left(  K T_{\rm crit}^{-\alpha} {4 - \alpha \over 4}   {1 \over r_{\rm ext}^3}     \right)^{1/(4-\alpha)}  \; \rm{for} 
 \; r > r_{\rm ext}
\label{Text}
\end{eqnarray}
The expression for $T_{\rm ext}$ is obtained by continuity at $r_{\rm ext}$.
By combining Eq.~(\ref{Text}) and Eq.~(\ref{rext}), we find that 
\begin{eqnarray}
T_{\rm ext} = \left(  \pi (1 - {\alpha \over 4}) {L_{tot} \over a c}
 {T_{crit}^{2 \alpha} G^2 \over \kappa_0 ^2 \delta^2 c_s^4  } \tau_0^3  \right)^{1/(4+2 \alpha)}
 \label{Text2}
\end{eqnarray}

\section{Analytical model of the mass spectrum}
\label{model_imf}
For completeness, we describe here the analytical model developed in 
\citet{HC08} and \citet{leeh2018a} that we use in the paper to interpret 
the numerical results.

It is based on the equality of mass of the density fluctuation which are unstable at scale $R$ (left-hand term) 
and the mass that ends up into the structures, i.e. the stars:
\begin{eqnarray}\label{bal_mass}
{ M_{\rm tot}(R) \over V_\mathrm{c}} =
 \int\limits^{\infty}_{\delta_R^\mathrm{c}} \overline{\rho} \exp(\delta)   {\cal P}_R(\delta)  d\delta
= \int\limits_0^{M_R^\mathrm{c}} M'  {\cal N}\! (M')  P(R,\!M')\, dM',
\end{eqnarray}
where $\delta = \ln(\rho / \bar{\rho})$, ${\cal P}_R$ is the density PDF, 
$P(R,M)$ is the probability of finding a self-gravitating clump of mass $M'$  embedded into 
a self-gravitating clump of mass $M_R$ unstable at scale $R$. It is assumed to be 1. 

Taking the derivative with respect to $R$, we get
\begin{eqnarray}\label{spec_mass1}
 {\cal N} (M_R^c)  =
  { \overline{\rho} \over M_R^c}
{dR \over dM_R^c} \,
\left( -{d \delta_R ^c \over dR} \exp(\delta_R^c) {\cal P}_R( \delta_R^c) \right).
\end{eqnarray}

The mass of the density fluctuations is given by 
\begin{eqnarray}
M=C_m \rho R^3, ~\text{where typically $C_m = 4 \pi/3$.}
\label{masse}
\end{eqnarray}

Here we assume that the density PDF is given by
\begin{eqnarray}
{\cal P}_R (\rho) =  {\cal P}_0 \left( { \rho \over \rho_0 } \right)^{-1.5},
\label{PL}
\end{eqnarray}

The gravitational instability criterion for a clump of mass $M$ at scale $R$ is 
\begin{eqnarray}
M > M_J = a_J { \Bigl[ c_s^2  + {V_a^2 \over 6} + {V_0^2 \over  3} \left({R \over R_c } \right)^{2 \eta} \Bigr]^{3 \over 2}   \over \sqrt{G^3 \overline{\rho} \exp(\delta) }  },
\label{cond_tot}
\end{eqnarray}
where $c_s$ is the sound speed, $V_a$ the Alfv\'en speed, $V_0$ the rms velocity dispersion at the 
cloud scale, $R_c$ is the cloud radius and $\eta$ an exponent to describe the turbulent scale dependence.
Typically $\eta=0.3-0.5$ and in this work the value $\eta=0.5$ is assumed for simplicity. Equation~(\ref{cond_tot})
is the standard Jeans mass expression in which the support is assumed to be as suggested by the virial theorem.
Note that the surface terms are not taken into account, they would typically modify this expression by a factor 
of 2.
Taking the standard definition of the Jeans mass,
the mass enclosed in a sphere of diameter equal to the Jeans length, we get $a_J=\pi^{5/2}/6$.
With Eq.~(\ref{masse}), this implies
\begin{eqnarray}\label{crit_Mtot}
M_R^c = {a_J^{2 \over 3} C_m^{1 \over 3} \over G    }
\left(   c_s^2  R + {V_a^2 \over 6}   R + {V_0^2  \over 3\, G  } \left({R \over R_c }\right)^{2\eta} R \right),
\end{eqnarray}
where $M_R^c$ is the critical mass at scale $R$.

With Eq.~(\ref{PL}), Eq.~(\ref{spec_mass1}) leads to 
\begin{eqnarray}\label{spec_mass2}
 {\cal N} (M_R^c)  = {\cal N}_0
  \left( { R \over M_R^c} \right)^{3/2}
{dR \over dM_R^c} \,
\left( - {1 \over M_R^c} {d M_R ^c \over dR} + {3 \over R}  \right).
\end{eqnarray}
Knowing the cloud physical conditions, $c_s$, $V_a$, $V_0$, together
with Eq.~(\ref{crit_Mtot}), Eq.~(\ref{spec_mass2}) allow to predict the stellar mass spectrum. 
The normalisation coefficient ${\cal N}_0$ is determined by specifying  the total mass within stars.

It is useful to see that
\begin{eqnarray}
\label{asymptot_ther}
M \rightarrow 0 \Leftrightarrow {\cal N} \rightarrow M^{-1} \Leftrightarrow {d N \over d \log M} \rightarrow M^{0}.
\end{eqnarray}
In this limit, the mass reservoir is thermally supported and the 
mass spectra present a plateau, i.e. ${d N \over d \log M} \propto M^{0}$.

On the other hand, in the limit
\begin{eqnarray}
\label{asymptot_turb}
M \rightarrow \infty  \Leftrightarrow {\cal N} \rightarrow M^{3/(4 \eta +2) -5/2} \Leftrightarrow {d N \over d \log M} \rightarrow M^{3/(4 \eta +2) -3/2}.
\end{eqnarray}
As revealed by Eq.~(\ref{crit_Mtot}), in this limit the mass reservoir is 
dominated by the turbulent dispersion. For $\eta=0.5$, the mass spectrum 
is  ${d N \over d \log M} \propto M^{-3/4}$.

We recall that at small masses, the asymptotic behaviour will eventually break
down when the gas becomes adiabatic due to the dust opacity and the formation of the first hydrostatic core, while  at large masses, the assumption of the density PDF being $\propto \rho^{-3/2}$, is 
eventually invalid (typically it eventually turns into a log-normal distribution).
Therefore while useful, these asymptotic behaviours must be handled with care.

\section{Accretion and stellar luminosities}
To get a better understanding of the origins of the luminosities, we investigate the stellar and 
accretion luminosities separately.  
The two panels of Fig.~\ref{time_mass2} show the sum of the stellar luminosities (left panel) and the sum of the accretion luminosities
(right panel). In a first phase, up to time $\simeq 0.1$ Myr, the accretion luminosity largely dominates.
Then as stars of few solar masses have formed, the stellar luminosities increase steeply and then reach 
values comparable to the accretion luminosities.

\setlength{\unitlength}{1cm}
\begin{figure*}
\begin{picture} (0,6)
\put(0,0){\includegraphics[width=8cm]{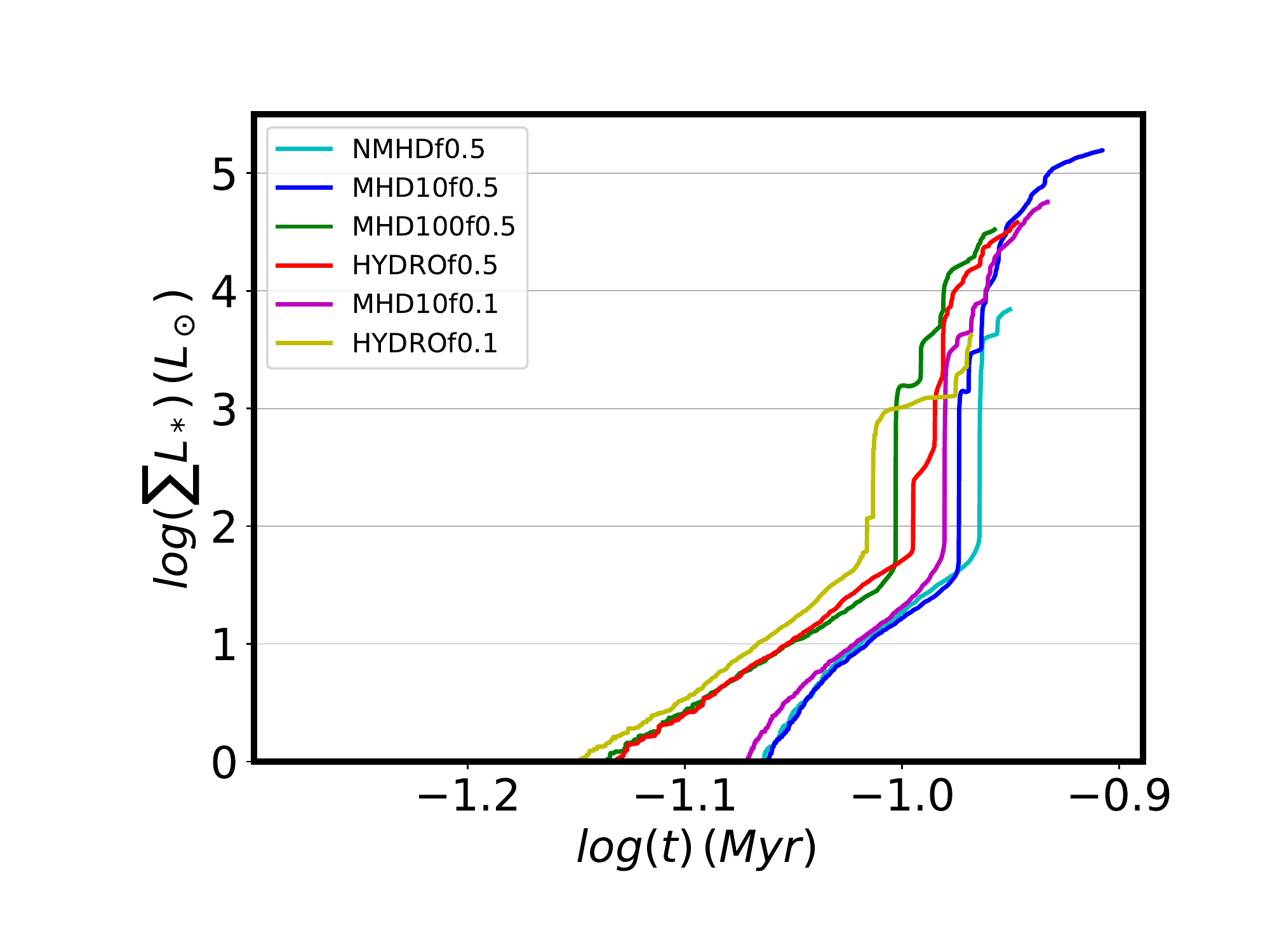}} 
\put(9,0){\includegraphics[width=8cm]{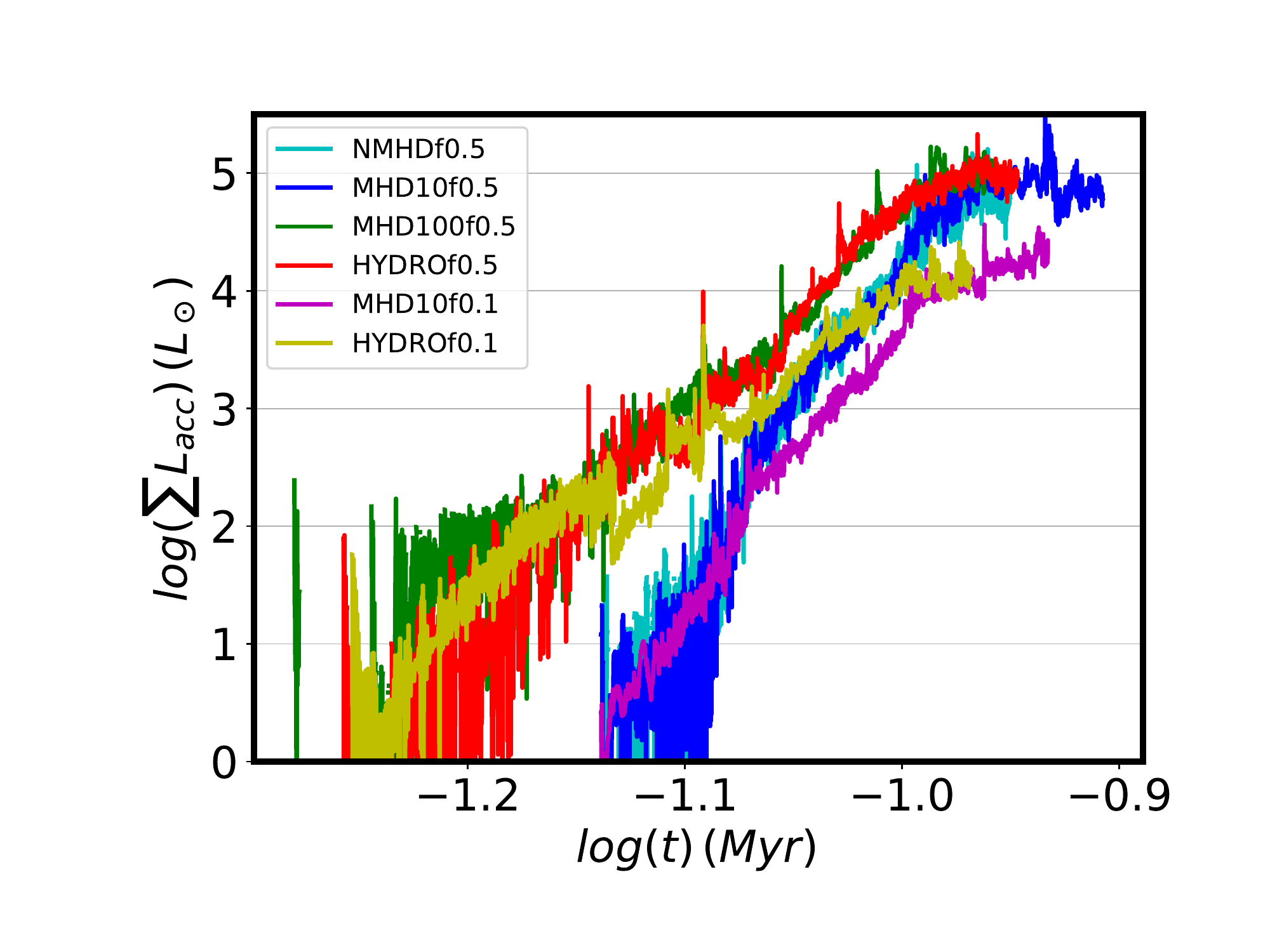}} 
\end{picture}
\caption{This figure complements Fig.~\ref{time_mass}.
Left panel  displays the total stellar luminosities and right one the total accretion luminosity.
}
\label{time_mass2}
\end{figure*}

\end{document}